\documentclass[12pt]{article}

\usepackage[
  letterpaper,
  top=1in,
  bottom=1in,
  left=0.8in,
  right=0.8in
]{geometry}

\usepackage{times}          
\usepackage[T1]{fontenc}
\usepackage[utf8]{inputenc}

\usepackage{setspace}
\usepackage{amsmath,amssymb,amsfonts}
\usepackage{amsthm}
\usepackage{bm}
\usepackage{mathrsfs}

\usepackage{verbatim}
\usepackage[hyphens]{url} 
\usepackage{hyperref}     
\usepackage[normalem]{ulem}
\usepackage{graphicx}
\usepackage{dcolumn}
\usepackage{xcolor}
\usepackage{caption}
\usepackage{xr}
\usepackage{fp}
\usepackage{tabularx}
\usepackage{algorithm}
\usepackage{algpseudocode}
\usepackage{multirow}
\usepackage{booktabs}
\usepackage[numbers,square,sort&compress]{natbib}

\hypersetup{
  colorlinks=true,
  linkcolor=black,
  citecolor=black,
  urlcolor=blue
}
\makeatletter
\g@addto@macro{\UrlBreaks}{\do\a\do\b\do\c\do\d\do\e\do\f\do\g\do\h\do\i\do\j\do\k\do\l\do\m\do\n\do\o\do\p\do\q\do\r\do\s\do\t\do\u\do\v\do\w\do\x\do\y\do\z\do\A\do\B\do\C\do\D\do\E\do\F\do\G\do\H\do\I\do\J\do\K\do\L\do\M\do\N\do\O\do\P\do\Q\do\R\do\S\do\T\do\U\do\V\do\W\do\X\do\Y\do\Z\do\1\do\2\do\3\do\4\do\5\do\6\do\7\do\8\do\9\do\0}
\makeatother

\newcount\boxheight
\newcount\boxwidth
\newcommand\testaspect[1]{%
  \setbox0=\hbox{#1}%
  \boxheight=\ht0\relax%
  \boxwidth=\wd0\relax%
  \FPdiv\theaspect{\the\boxwidth}{\the\boxheight}%
  \copy0%
}

\makeatletter
\renewcommand{\section}{\@startsection{section}{1}{\z@}%
  {-3.5ex plus -1ex minus -0.2ex}%
  {2.3ex plus 0.2ex}%
  {\normalfont\normalsize\bfseries}}
\renewcommand{\subsection}{\@startsection{subsection}{2}{\z@}%
  {-3.25ex plus -1ex minus -0.2ex}%
  {1.5ex plus 0.2ex}%
  {\normalfont\normalsize\bfseries\itshape}}
\makeatother

\title{\textbf{Design Principles for Reproducible Networks}}

\author{
  Jasper van der Kolk\textsuperscript{1,$\dagger$}
  \and 
  Cory Glover\textsuperscript{2,$\dagger$}
  \and Albert-L\'{a}szl\'{o} Barab\'{a}si\textsuperscript{1,2,3,*}
}

\date{}

\newcommand{\affiliations}{%
\begin{centering}
\textsuperscript{1} Department of Network and Data Science, Central European University, Vienna, Austria \\[2pt]
\textsuperscript{2} Network Science Institute, Northeastern University, Boston, MA, USA \\[2pt]
\textsuperscript{3} Department of Medicine, Brigham and Women's Hospital, Harvard Medical School, Boston, MA, USA \\[2pt]
\footnotesize($\dagger$ These authors contributed equally to this work.)\\
\footnotesize(\textsuperscript{*} Corresponding author: barabasi@gmail.com)\\[2pt]
\end{centering}
}

\begin{document}

\maketitle
\thispagestyle{empty}

\affiliations

\begin{abstract}

\begin{singlespace}
\noindent
\textbf{From protein complexes to electronic circuits, many natural and engineered systems function only if assembled in an exact, reproducible fashion. The structure of each of these systems can be understood as a network, yet network science lacks the mechanisms to consistently reproduce exact topologies, focusing instead on generating network ensembles. We introduce the framework of \emph{network design} where we encode the local constraints obeyed by a system’s building blocks in a design set, and derive the Unigraphical Design Theorem, which determines when these constraints guarantee reproducible assembly into a unique structure, a process we call \emph{unigraphical assembly}. For systems whose design sets do not specify a unique outcome, we identify \emph{guided assembly} as a second route to reproducibility, in which temporal ordering decomposes construction into unigraphical steps. Applying these results to 3,618 reproducible systems, including protein complexes, molecules, and robots, we classify those that undergo unigraphical assembly and those that require guided assembly. We further identify a diversity–redundancy boundary that explains how systems trade component variety for structurally interchangeable parts while retaining unique assembly. Finally, we experimentally test the theory using 3D-printed components to re-engineer generative construction sets into systems that assemble unigraphically into prescribed topologies. Network design thus reframes reproducibility as a mathematically testable property of real networks, opening a route to the rational engineering of complex systems.}
\end{singlespace}
\end{abstract}

\newpage


Many real-world networks are highly reproducible. The ribosome, for example, is a protein complex consisting of 82 proteins that assembles reliably from pairwise protein--protein interactions (PPIs) and maintains a highly conserved architecture across species~\cite{timsit2021evolution}. Likewise, the connectome of \textit{C.~elegans} is reproduced with remarkable fidelity across individuals~\cite{cook2019whole}. This reproducibility presents a fundamental puzzle for network science: how can local interactions among components repeatedly give rise to the same global architecture?

Network science has developed powerful tools to characterize the structure of real network, from the role of degree distributions~\cite{barabasi1999emergence,erdHos2024new} and clustering~\cite{watts1998collective} to communities~\cite{newman2004detecting} and motifs~\cite{milo2002network,vega2004fitness}, along with the generative mechanisms responsible for their emergence \cite{caldarelli2007scale,posfai2018talent,gershenson2007design,dorogovtsev2022nature,menczer2020first}. However, many natural and engineered systems do not merely produce networks with similar statistical features; they repeatedly produce exactly the same network, or a topology that is sufficiently similar to preserve its function. Further, the function of many systems depends on reproducibly reconstructing an exact network structure. In hemoglobin, for example, altered assembly can lead to sickle cell disease~\cite{mccavit2012sickle}; in neuronal circuits, wiring defects can impair synaptic specificity and disrupt locomotion~\cite{starich2009interactions}; and in electrical circuits, variations in wiring can produce device failure~\cite{haselmanFutureIntegratedCircuits2010}.

\begin{figure}[p]
    \centering
    \includegraphics[width=0.74\textwidth]{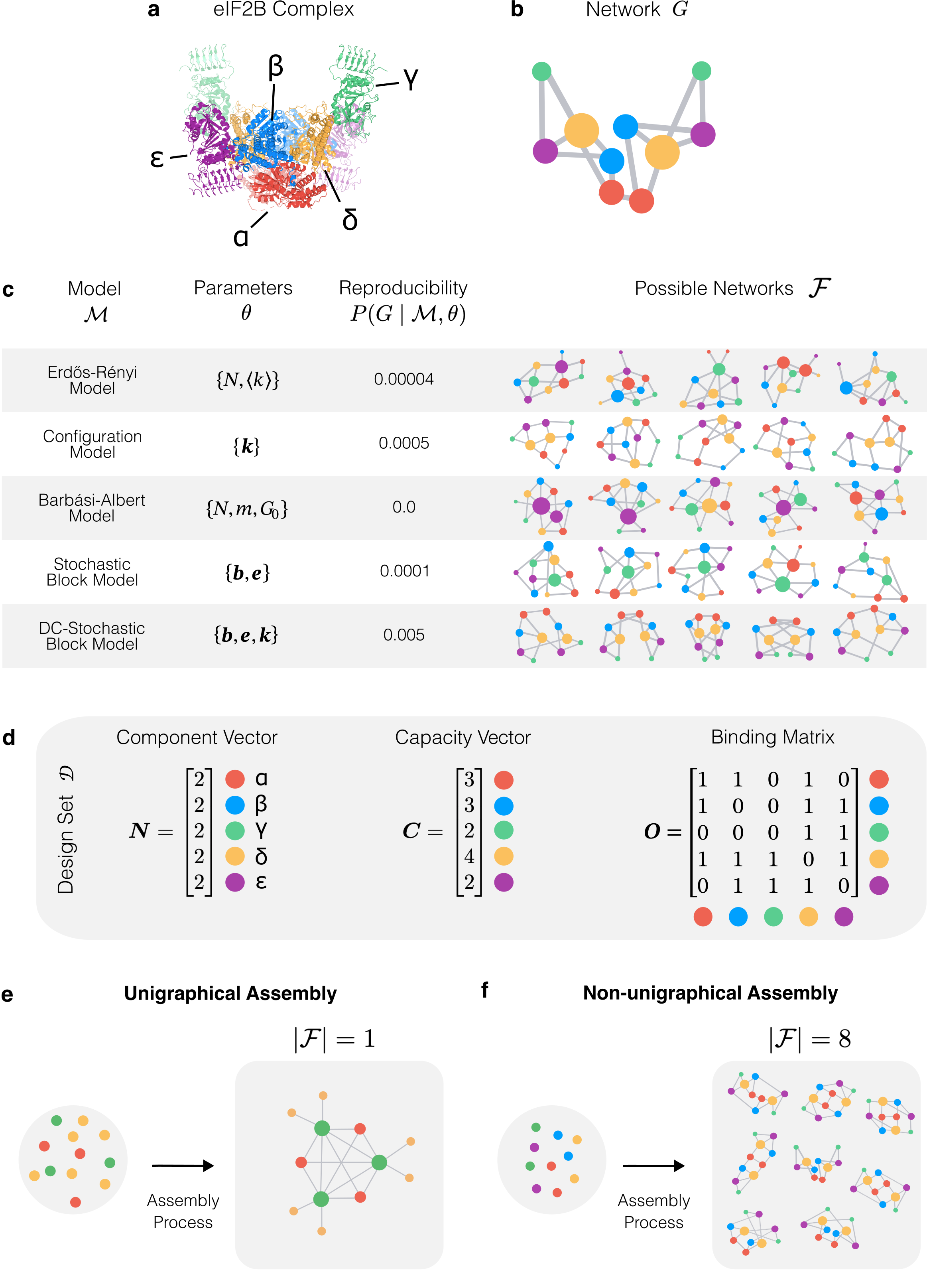}
    \caption{\footnotesize\textbf{The Network Design Framework.}
\textbf{a,b} The protein complex eIF2B, consisting of 10 proteins as shown in \textbf{a}, is formally described by the network shown in \textbf{b}, where each node is colored by its associated protein type.
\textbf{c} Using the established network science models, adapted to the constraints of the network in \textbf{b}, we find that the probability of producing the correct structure is rather small.
For example, fixing $N=10$ and $\langle k\rangle=2.8$, the Erd\H os--Rényi model recovers the network behind the protein complex with probability $0.00004$. Other canonical network models shown in the table are also unable to reliably reconstruct the network. \textbf{d} The Design Set and its three building blocks: the component vector $\boldsymbol{N}$, counting the number of each node type; the capacity vector $\boldsymbol{C}$, counting the maximum number of connections for each type; the binding matrix $\boldsymbol{O}$, which captures the pairwise binding rules of each node type.
\textbf{e,f} For each node set, we begin with a set of disconnected nodes and identify all possible networks without spurious connections and ensuring that all nodes are saturated. If only one network can be assembled from the design set $\mathcal{D}$, i.e., $|\mathcal{F}|=1$, as in \textbf{e}, we have unigraphical assembly. In contrast, we call it non-unigraphical assembly when the design set $\mathcal{D}$ is compatible with multiple distinct networks ($|\mathcal{F}|>1$) as in \textbf{f}.}
    \label{fig:framework}
\end{figure}

This distinction exposes a gap between statistical network modeling and reproducible assembly. Existing network models were not intended to construct a particular graph; rather, they are inherently stochastic, defining ensembles of graphs that share selected features, such as size, density, degree sequence or community structure \cite{bianconi2008entropy,orsini2015quantifying}. To illustrate this distinction, we ask whether canonical network models can recover the structure of the protein complex eIF2B (Fig.~\ref{fig:framework}a), represented as an abstract interaction network whose nodes represent $N=10$ proteins and whose edges capture $L=15$ pairwise binding interactions (Fig.~\ref{fig:framework}b). Starting with the Erd\H{o}s-Rényi model, and constraining the number of nodes and edges to $N=10$ and $L=15$, we find that only one of the more than 50,000 distinct, unlabeled networks the model generates matches the empirical wiring, corresponding to a recovery probability of $P \approx 4 \times 10^{-5}$ (Fig.~\ref{fig:framework}c). We also generated 50,000 networks using the Barabási--Albert (BA) model~\cite{barabasi1999emergence}, none of which recover the observed structure. The configuration model (CM)~\cite{bollobas1980probabilistic} and the stochastic block model (SBM)~\cite{holland1983stochastic} each yield $P \approx 10^{-4}$, while the more constrained degree-corrected DCSBM~\cite{karrer2011stochastic} achieves $P \approx 5 \times 10^{-3}$.
These results are not evidence of the failure of the existing modeling frameworks, rather they highlight a mismatch between the purpose of these models and the question posed by network reproducibility.

Here we propose that network reproducibility represents a previously unrecognized yet fundamental property of many real networks: their formation is driven by \emph{network design} principles, whereby predefined building blocks and local constraints restrict the space of possible final topologies. We formalize these principles by encoding the local constraints in a \emph{design set}, and ask under what conditions a design set specifies a unique network. Our key advance is the Unigraphical Design Theorem (UDT), which identifies when the local constraints encoded by a design set specify a unique final network, independent of assembly dynamics. For systems whose design sets do not specify a unique final topology, we also identify a second route to reproducibility in which the order of interactions between building blocks further constrain the final outcome.

We apply the resulting theoretical framework to a large dataset of real-world reproducible systems, including protein complexes, molecules, circuits, robots, furniture and image fragments, finding that a remarkably high fraction of real systems satisfy the formal conditions of the UDT, and identify many other systems whose unique outcome can be successfully achieved via guided assembly. We then demonstrate the practical potential of our approach by designing novel 3D-printed components that allow us to re-engineer construction sets into systems that can only be assembled into the prescribed topologies. By shifting the focus from statistical network generation to constraint-driven construction, network design offers a principled framework for understanding how complex systems reproducibly build functional topologies, and for engineering them in natural and artificial settings.

\clearpage
\section*{Formalizing Network Design: The Design Set}

To formalize network design, let us return again to the eIF2B complex composed of pairs of $\alpha$, $\beta$, $\gamma$, $\delta$ and $\epsilon$ proteins, which form a network of 5 distinct node types, each shown in different colors in Fig.~\ref{fig:framework}b. 
Each of these components interacts according to predefined pairwise binding rules, which we encode in the \emph{design set} $\mathcal{D}$ (Fig.~\ref{fig:framework}d), described below.

First, the \emph{component vector} $\mathbf{N}$ specifies the inventory of building blocks and their multiplicities (number of identical copies). 
As the eIF2B complexes contain two copies of each of the 5 protein types, we have $N_i=2$ for all $i$. 
Second, the \emph{capacity vector} $\mathbf{C}$ encodes the number of binding sites available to each node type. The $\alpha$, $\beta$ and $\epsilon$ proteins can bind to up to three partners simultaneously, hence $C_1=C_2=C_5=3$, while the $\gamma$ and $\delta$ chains can bind four and two proteins respectively, hence $C_3=4$ and $C_4=2$.
Third, the \emph{binding matrix} $\boldsymbol{O}$ specifies the allowed interactions between node types, along with the maximum number of simultaneous connections between them. For example, the $\alpha$-protein can bind to another $\alpha$-protein ($O_{11}=1$) and to the $\beta$ and a $\delta$ protein ($O_{12}=O_{14}=1$).

The three ingredients of the design set play distinct roles. The component vector ($\boldsymbol{N}$) fixes the inventory of available building blocks, while the capacity vector ($\boldsymbol{C}$) specifies how many connections each type can form. By contrast, the binding matrix ($\boldsymbol{O}$) adds a different kind of information: it encodes pairwise compatibility between component types. Such compatibility may for example arise from complementary protein interfaces~\cite{day2012binding}, genetically encoded neuronal matching rules~\cite{kovacs2020uncovering,barabasi2020genetic}, or engineered color and shape codes~\cite{jacobs2016self}. Thus, beyond specifying how many connections each component can make, the design set also specifies which connections are allowed and which are forbidden.

\clearpage

\section*{Unigraphical Design Theorem}  
To construct a network from a given design set $\mathcal{D}$, we allow nodes to assemble according to any procedure that respects the constraints encoded in $\mathcal{D}$.
Let $\mathcal{F}$ denote the set of all final network states (Fig.~\ref{fig:framework}e,f), subject to only 
two straightforward assumptions: (i) all links satisfy the interaction rules encoded in the binding matrix $\boldsymbol{O}$ (no spurious bindings) and (ii) each node has as many neighbors as allowed in $\boldsymbol{C}$ (saturation).

We deliberately define $\mathcal{F}$ in terms of the final outcome and disregard system specific assumptions such as assembly dynamics, binding energies, or the physical mechanism by which saturation is reached. This allows us to ask a more general question, agnostic to the exact nature of the system under study: does a design set $\mathcal{D}$ uniquely specify a single possible outcome $(|\mathcal{F}|=1$, Fig.~\ref{fig:framework}e), or is it compatible with multiple networks $(|\mathcal{F}|>1$, Fig.~\ref{fig:framework}f). 

\begin{figure}[p]
    \centering
    \includegraphics[width=0.74\textwidth]{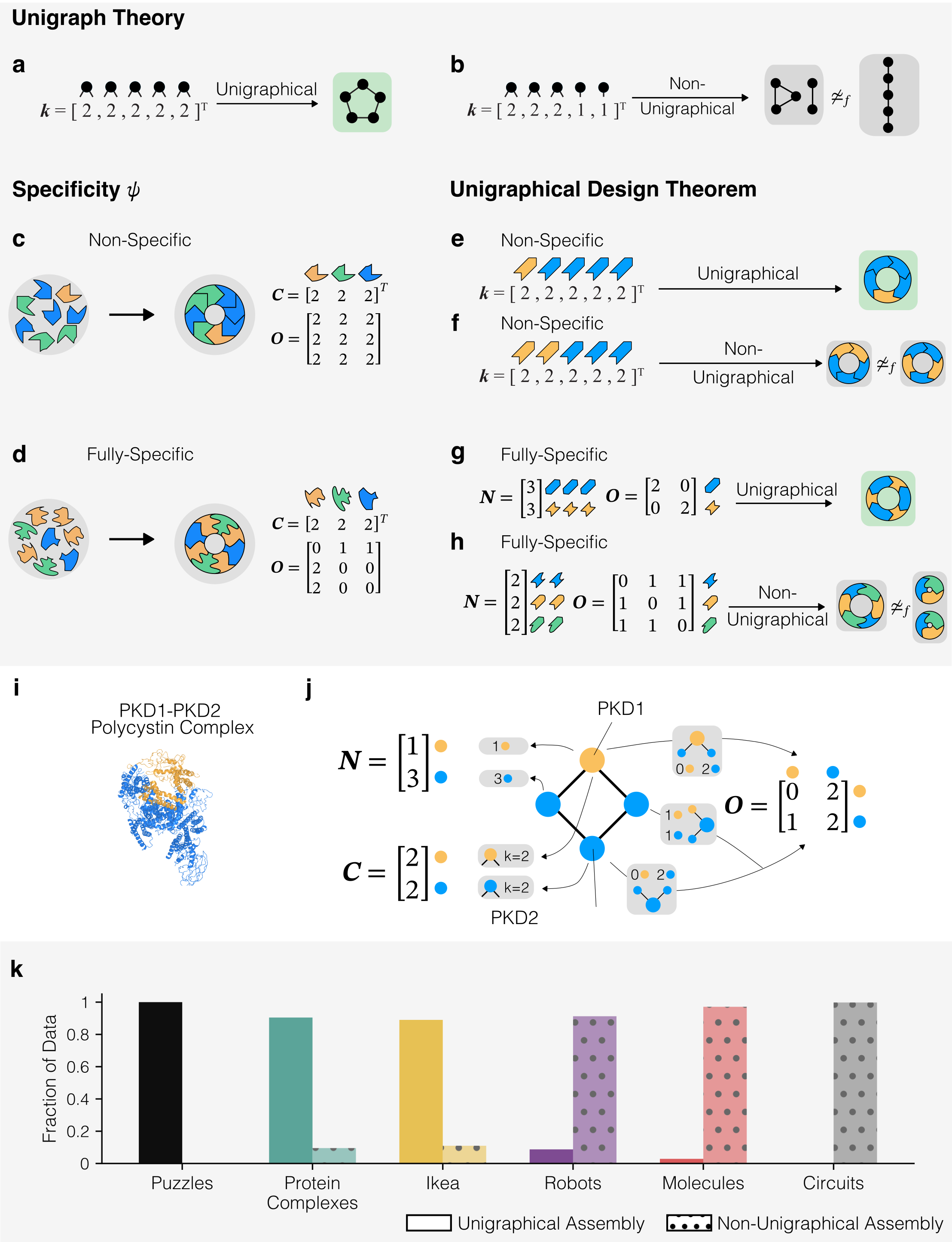}
    \caption{\footnotesize\textbf{Network Design via Unigraphical Assembly and the Unigraphical Design Theorem.}
\textbf{a,b} Unigraph theory~\cite{kleitman1975note} allows us to determine if a degree sequence is compatible with a unique graph.
\textbf{c,d} The specificity of a design set $\mathcal{D}$ tells us how restrictive the interactions are. If all nodes can connect to all others, $\mathcal{D}$ is non-specific, as shown in \textbf{c}. 
If each node can only connect to a predefined set of neighbors, $\mathcal{D}$ is fully specific, as illustrated in \textbf{d}.
\textbf{e,f} In network design, we distinguish different node types, shown in color in the figure. The distribution of the colors over the network can lead to different outcomes, even if the
underlying network stays fixed. We show examples of unigraphical \textbf{e}
and non-unigraphical \textbf{f} colorings for a non-specific system.
\textbf{g--h} For fully-specific systems, each node has a fixed set of
neighbors, reducing the amount of compatible networks. Here we show examples
of unigraphical \textbf{g} and non-unigraphical \textbf{h} design sets.
\textbf{i,j}  To extract the design set of the PKD1-PKD2 Polycystin protein
complex we first extract the component vector by counting the number of nodes
of each type (3 blue and 1 yellow). Next, the capacity vector is identified by
looking at the degrees of each node type (2 for each node). Finally, the
binding matrix is extracted by setting the $O_{ij}$ entry to the maximum
number of connections a node of type $i$ has with nodes of type $j$.
\textbf{k} We apply the Unigraphical Design Theorem and numerical tools to our 3,618 design sets, allowing us to classify each network as either `unigraphical assembly' (filled) or `non-unigraphical assembly' (dotted).}
    \label{fig:theory}
\end{figure}

To identify the precise mathematical conditions that guarantee a single network, we draw on the graph-theoretic framework of \emph{unigraphs}~\cite{kleitman1975note,johnson1975simple}. A degree sequence $\boldsymbol{k}$ is called \emph{unigraphical} if it has exactly one graph realization~\cite{kleitman1975note,johnson1975simple}. For example, the only graph we can generate from the sequence $\boldsymbol{k} = [2,2,2,2,2]^T$ is the 5-cycle (Fig.~\ref{fig:theory}a). By contrast, the sequence $\boldsymbol{k} = [2,2,2,1,1]^T$ is compatible with two distinct graphs, and is therefore non-unigraphical (Fig.~\ref{fig:theory}b). Classical unigraph theory therefore asks when a degree sequence has exactly one graph realization. 
Our goal is to extend the concept of unigraphicality from degree sequences to design sets, where not only the degree sequence but also node type and pairwise binding rules encoded in the binding matrix $\boldsymbol{O}$ define the set of possible graphs. We therefore call a design set $\mathcal{D}$ \emph{unigraphical} if it is compatible with exactly one network ($|\mathcal{F}|=1$), implying that the components must reproducibly assemble into a unique topology. Hence, \emph{unigraphical assembly} implies that any assembly process that respects the design set and reaches saturation can produce only one final graph, independent of the dynamics of assembly or the details of the specific system under study.

Our first key result is the \emph{Unigraphical Design Theorem} (UDT), which allows us to exactly determine whether a design set is compatible with a single network (unigraphical assembly) or if it can generate multiple outcomes. 
Central to this theorem, whose proof is provided in Supplementary Information III, is the design set's specificity $\psi$ \cite{jacobs2016self,zeravcic2014size,hubl2025accessing}.
A node is non-specific ($\psi=0$) if it can connect to any other node in the system (Fig.~\ref{fig:theory}c). This is the case for molecules, where connections depend on valence rather than the identity of the constituent atoms. 
In contrast, a node is fully specific ($\psi=1$) if each of its binding sites is only compatible with one node type (Fig.~\ref{fig:theory}d), a situation often encountered in proteins, which require complementary 3D facets to bind.

The Unigraphical Design Theorem provides the exact conditions under which a design set $\mathcal{D}$ is unigraphical for both non-specific and fully specific design sets. For a non-specific $(\psi=0)$ design set to be unigraphical, it must have a unigraphical capacity vector (degree sequence)~\cite{koren1976sequences} and must not admit node swaps that lead to different network colorings — a condition that arises because our nodes are labeled, a feature absent from classical unigraph theory. 
Therefore, even if the degree sequence realizes a unique graph, a design set can fail to assemble consistently if two non-equivalent colorings are possible. This is illustrated in Fig.~\ref{fig:theory}e,f, where the 5-cycle with one green node and 4 blue nodes is unigraphical (Fig.~\ref{fig:theory}e), as opposed to the design set with two green nodes and 3 blue nodes (Fig.~\ref{fig:theory}f), which does not lead to unigraphical assembly as it allows for two different colorings. 

By contrast, in fully specific systems ($\psi=1$), thanks to the more restrictive binding rules, it is often impossible to interchange nodes of different types. For example, in Fig.~\ref{fig:theory}g the blue and yellow nodes cannot be interchanged, hence, the cycle where they alternate is the only compatible network. However, some fully specific design sets are non-unigraphical, as illustrated in Fig.~\ref{fig:theory}h where either one large cycle or two small cycles can be realized. Determining unigraphicality of a fully specific system requires new machinery that decomposes the graph into subgraphs related through the binding matrix to which classical unigraph theory~\cite{johnson1975simple,koren1976sequences} can be applied. 
The exact procedure is provided in SI III, where we also show how to evaluate the unigraphicality of semi-specific networks ($0<\psi<1$).

These exact results raise a fundamental question: can the Unigraphical Design Theorem explain the assembly of real systems? To address this, we started by identifying six classes of systems that are known to form reproducible networks: (1) 582 manually curated protein complexes from the EMBL-EBI Complex Portal \cite{balu2025complex}; (2) 1321 stable organic molecules from the PubChem database \cite{kim2023pubchem}; (3) 949 electronic circuits representing analog devices, documented in the LTSpice software \cite{broll2023ltspice}; (4) 572 robots capable of traversing complicated terrains whose construction rules are provided in Ref.~\cite{zhao2020graphgrammar}; (5) 99 Ikea furniture sets instruction manuals compiled in~Ref.~\cite{zhang2025manual}; (6) 95 fragmented images, used in computer vision to train the assembly of shattered frescoes and torn-up photos~\cite{tsesmelis2024re,li2019hierarchical}.

For each of these 3,618 networks, we were able to identify the corresponding design set using a set of methods outlined in Fig.~\ref{fig:theory}i,j and discussed in SI I. We then applied the UDT to each design set asking if any of these satisfy the stringent mathematical conditions of unigraphical assembly. To our surprise, we find that despite the diversity of the systems and the many auxiliary constraints they obey individually, 813 (23\%) of the design sets are only compatible with a single outcome, $|\mathcal{F}|=1$, demonstrating the ability of the network design framework to explain their reproducible assembly. Specifically, we find that the vast majority of protein complexes (93\%) and Ikea furniture sets (89\%), as well as all fragmented images, obey design constraints strict enough to specify a unique final network (Fig.~\ref{fig:theory}k). For each of these systems, function requires a unique outcome, hence reproducibility is already a property of the underlying network structure. 

At the same time, the UDT also reveals that the vast majority of robots (93\%), molecules (97\%) and circuits (100\%) have design sets compatible with multiple outcomes. This raises a central question: if the remaining 2,805 networks are not uniquely specified by their design sets, how do they achieve the reproducible assembly, essential for their function?

\section*{Guided Assembly}  

\begin{figure}[p]
    \centering
    \includegraphics[width=1\textwidth]{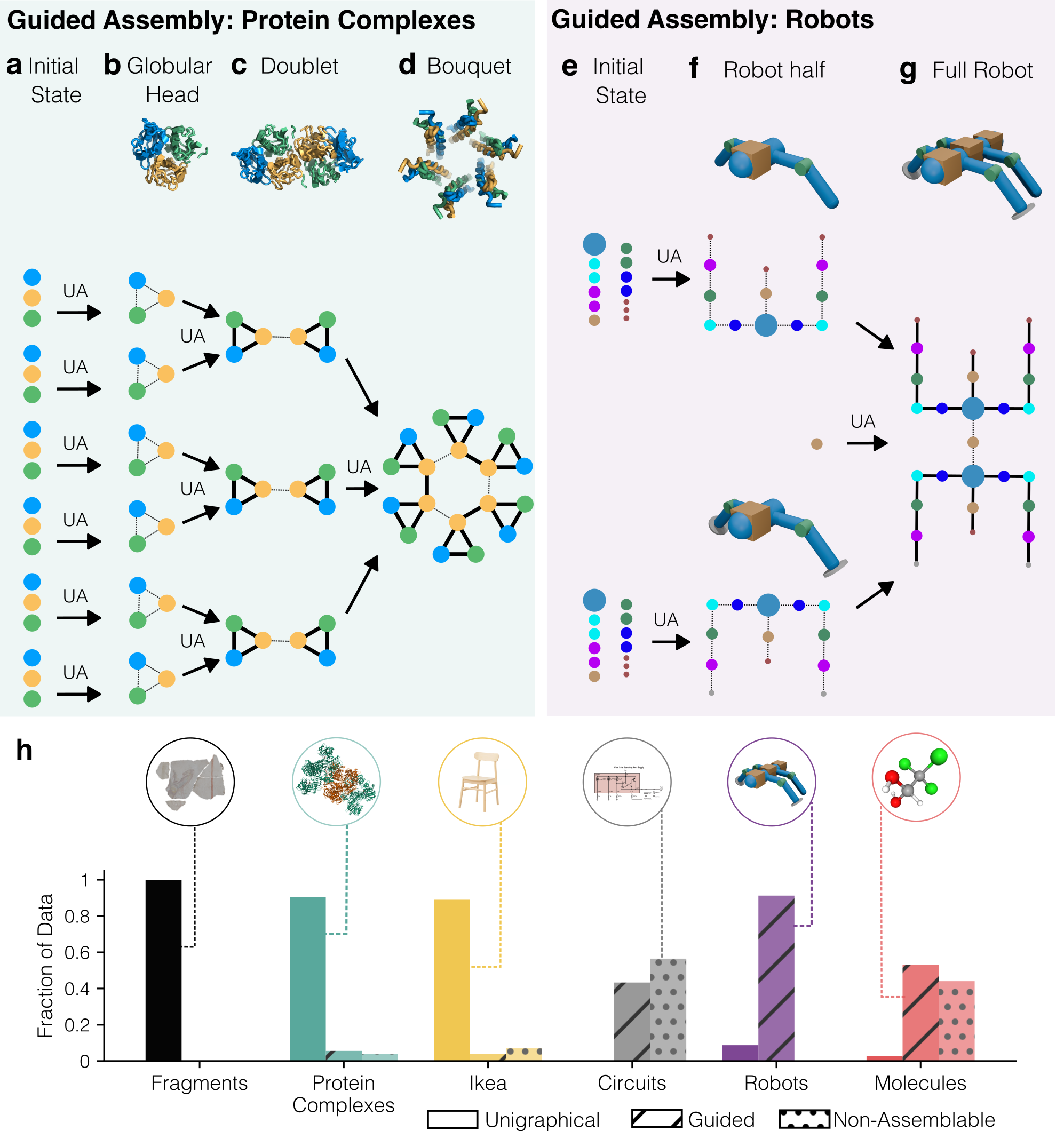}
    \caption{\footnotesize\textbf{Network Design via Guided Assembly.}
\textbf{a--d} There is experimental evidence~\cite{thielens2017c1q} that the protein complex C1q is constructed in three distinct steps:
First, the globular head (\textbf{b}) is unigraphically assembled (UA) from the three proteins shown in \textbf{a}. Next, two globular heads are combined to form a doublet (\textbf{c}). Finally, three doublets are connected to form the final structure (\textbf{d}).
\textbf{e--f} We use the Guided Assembly Propositions to understand how to build a robot with a fixed outcome, given that according to the UDT, its design set is compatible with two possible outcomes. We first construct the two halves of the robot through unigraphical assembly (\textbf{e--f}). Next we connect the two halves to form the final robot (\textbf{g}). Note that in each step of the process,  unigraphical assembly applies.
\textbf{h} For each real network in our dataset we ask if it can be built through unigraphical assembly (full color) or guided assembly (shaded color), or if it cannot be designed by either of these methods (dotted color). We find that a fraction of circuits, robots and molecules, which are not subject to unigraphical assembly, can be assembled using guided assembly.}
    \label{fig:guided}
\end{figure}

To understand the assembly of the remaining 2,805 networks, we note that many of them are built in a modular way. For example, robots are often built from modular subnetworks and IKEA furniture comes with instruction manuals delineating a sequence of assembly steps. Furthermore, sequential assembly is not a unique feature of engineering, but also appears in nature, documented when complexes are built from individual proteins~\cite{ahnert2015principles}. Consider the protein complex C1q, a key component of the innate immune system, of which $O(10^{14})$ copies are accurately reproduced within the human body \cite{kohler1972metabolism}. According to the UDT, the design set of C1q, describing the binding of its 18 proteins, is compatible with 52 topologically distinct graphs.
Experimental evidence, however, has elucidated how the C1q complex avoids this multiplicity~\cite{thielens2017c1q}: (i) First, the protein chains C1QA, C1QB, and C1QC are expressed and allowed to assemble into a globular head (Fig.~\ref{fig:guided}b), forming a triple helix of collagen-like strands. 
In graph theoretic terms, this corresponds to the formation of a triangle between C1QA, C1QB, and C1QC. 
(ii) Next, a bond forms between C1QC proteins, allowing pairs of triangles to form doublets (Fig.~\ref{fig:guided}c). 
(iii) Finally, three such doublets assemble into the final bouquet structure (Fig.~\ref{fig:guided}d).

While the design set of the C1q complex is compatible with 52 distinct network outcomes, hence is not unigraphical, we find that each of the three intermediate steps (i)-(iii) in Fig.~\ref{fig:guided}a-d is unigraphical in isolation. 
Formally, this means that the construction of the C1q complex perfectly aligns with an assembly protocol that takes advantage of unigraphical assembly limited to a subset of components, ultimately ensuring the reproducible formation of the target network \cite{thielens2017c1q}.

Inspired by this example, next we introduce the \emph{guided assembly} framework, which temporally constrains the assembly process by forcing the building blocks to arrive in a specific order and, most importantly, requires that each step of the design protocol is subject to unigraphical assembly.
In SI IV we formalize our results in the form of three \emph{Guided Assembly Propositions} that cover three different types of graphs: (i) trees, (ii) networks with unigraphical 2-cores and (iii) several ring graphs. These propositions, combined with numerical methods, allowed us to see if any of the real design sets is amenable to guided assembly. We find that of the 2,805 networks that are not capable of unigraphical assembly, more than half (1,661) can generate unique outcomes following a guided assembly procedure. The majority of these networks correspond to robots and circuits, but also to molecular systems produced by chemical processes (Fig.~\ref{fig:guided}h). For example, the design set of the tree-like robot in Fig.~\ref{fig:guided}e-g is compatible with $|\mathcal{F}|=2$ outcomes. The off-target network is avoided if we first assemble the two halves, which are unigraphical in isolation (Fig.~\ref{fig:guided}f), and then bring these two halves together to form the final structure (Fig.~\ref{fig:guided}g).

IKEA furniture helps us illustrate the key distinction between unigraphical and guided assembly. Each furniture set is accompanied by an instruction manual, suggesting that they are better suited for guided assembly. Yet, we find that 89\% of IKEA sets are compatible with only one final structure ($|\mathcal{F}|$=1), meaning that they could, in principle, be assembled without a manual. The purpose of the instructions, therefore, is not to ensure assembly, but to reduce the assembly time by avoiding incorrect intermediate choices. Only 4 of the 99 furniture sets have multiple outcomes that are avoided by guided assembly.

Taken together, of the original 3,618 networks, we can explain the reproducibility of 813 by unigraphical assembly and another 1,661 can be built using guided assembly. That means that only 1,144 networks, mainly circuits and molecular systems, are compatible with multiple outcomes.
We label these sets as non-assemblable, despite being known to be capable of producing unique outcomes, raising the question, how are they assembled?

Many of these systems are subject to additional, system-specific constraints that go beyond the pairwise compatibility encoded in the current design set $\mathcal{D}$. For example, in some protein complexes spatial constraints, like access to a specific binding site, can further reduce the space of compatible graphs \cite{hlavacek1999steric}, chaperone proteins and protein cotranslation can guide protein assembly to build complexes deemed non-assemblable by our methods \cite{hendrick1995role,mallik2026twisted}. 
Additional assembly constraints also arise in physical networks and metamaterials~\cite{dehmamy2018structural,posfai2024impact, damasceno2012predictive,wytock2026irregular,vskrbic2021building}, where restrictions induced by volume exclusion are prevalent or when kinetics guide the components towards specific target structures, like in colloids~\cite{hubl2025accessing,jacobs2015rational}. In the end, only 31\% of the studied systems require additional system-specific constraints to explain their reproducible assembly. Hence our key result is the finding that for 69\% of systems, characterized by a remarkable functional and structural diversity, unique outcomes can be accurately predicted through the mathematically exact network design framework, indicating that their design set can help explain their reproducibility.

\section*{Network Characteristics that Define Assembly}

\begin{figure}[p]
    \centering
    \includegraphics[width=1\textwidth]{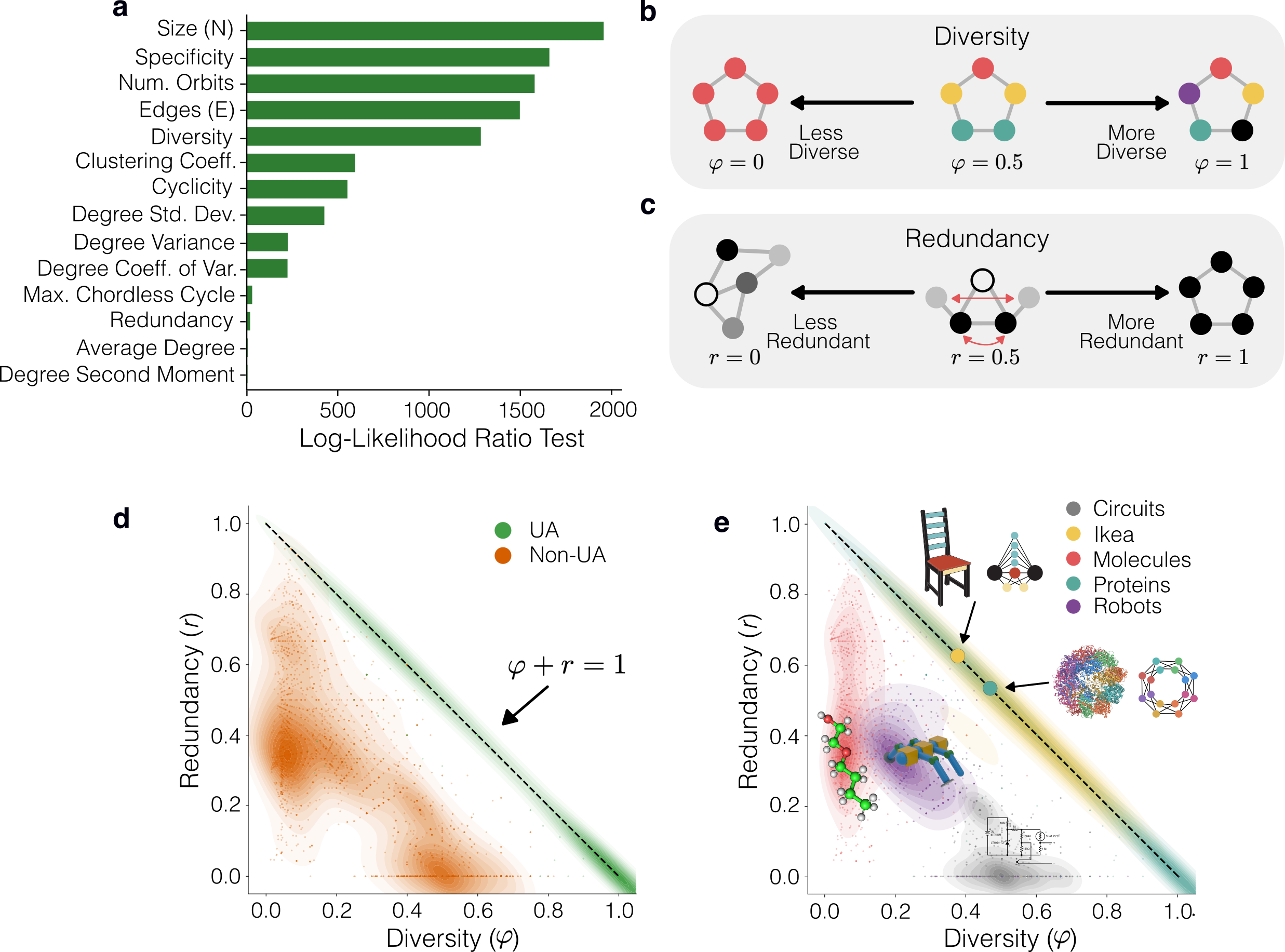}
    \caption{\footnotesize\textbf{The Diversity-Redundancy Boundary}
\textbf{a} Logistic regression predicts the impact of 15 network characteristics on unigraphical assembly.
\textbf{b} Diversity measures the number of distinct node types in the network, denoted by color: $\varphi=0$ if all nodes are the same type and $\varphi=1$ if all nodes are different colors. We also show an intermediate level of diversity $\varphi=0.5$, where some node types are repeated.
\textbf{c} Redundancy measures how structurally interchangeable nodes are, where two nodes are interchangeable if they play the same topological role in the network. Interchangeable nodes are denoted by shading. For $r=0$ no nodes are interchangeable, for $r=0.5$ two pairs of nodes can be interchanged and for $r=1$ all nodes are interchangeable.
\textbf{d} We placed the 3,618 design sets on the diversity-redundancy diagram, marking in orange systems that do not assemble unigraphically, and with green those that do. We find that most unigraphical assembling networks (green nodes), fall near the line $\varphi + r = 1$.
\textbf{e} When colored by dataset, we find that unigraphical systems such as puzzles, protein complexes such as TRiC, and IKEA furniture such as the
Kaustby chair fall near the $\varphi+r=1$ line. By contrast,  systems do not undergo unigraphical assembly, such as molecules, circuits and robots, form clusters beneath this line.}
    \label{fig:analysis}
\end{figure}

While the theorems we introduced above help us identify systems capable of unigraphical or guided assembly, they do not explain why some networks can assemble and others do not. We therefore next ask if there are structural features that make a design set sufficiently restrictive to specify a unique network. We use logistic regression to examine the role in unigraphical assembly of 15 network and design-set properties (Fig.~\ref{fig:analysis}a), ranging from network size ($N$) and average degree $(\langle k\rangle)$ to specificity ($\psi$) and diversity ($\varphi$). We find that the single most predictive property is network size~(Fig.~\ref{fig:analysis}a), indicating that smaller networks are more likely to be unigraphical compared to larger ones. Yet, size is an imperfect classifier, with a McFadden $R^2$ of $\approx 0.6$, reflecting the fact that most large protein complexes and all large fragmented images are in fact capable of unigraphical assembly. However, we find that two characteristics of the design set, namely diversity and redundancy (Fig.~\ref{fig:analysis}b,c), together reach a McFadden $R^2\approx0.67$, prompting us to ask what role these features play in assembly.

Diversity $\varphi$ measures how many distinct node types a network has (colors in Fig.~\ref{fig:analysis}b): We have $\varphi=0$ when all nodes are identical and $\varphi=1$ when every node is unique.
Redundancy (or symmetry), $r$, is a network property capturing how interchangeable the nodes are within a given network structure~\cite{ball2018symmetric} (shading in Fig.~\ref{fig:analysis}c): 
two nodes are redundant if swapping them leaves the network structure unchanged. We have $r=0$ when no two nodes can be swapped and $r=1$ when all nodes are interchangeable. For example, the network in Fig.~\ref{fig:analysis}d has two interchangeable pairs of nodes (grey and black), hence $r=0.5$.

Prompted by the high predictive power of $\varphi$ and $r$, we placed all real systems in a $(\varphi,r)$ diagram (Fig.~\ref{fig:analysis}d), coloring systems capable of unigraphical assembly in green and systems that are not unigraphical in orange. We find a surprising separation between the two classes: 93\% (758) of the unigraphical networks are in the immediate vicinity of the $\varphi+r=1$ boundary, characterized by $\varphi+r>0.8$.
At the same time, 92\% (2573) of the non-unigraphical networks are far from the boundary, scattered in the plane of the phase diagram, characterized by
$\varphi+r\leq0.8$. Further robots, molecules and circuits form distinct clusters within the diagram (Fig.~\ref{fig:analysis}e).

To understand the origin of this unexpected segregation in the $(\varphi,r)$-space, we proved a corollary of the Unigraphical Design Theorem (SI V), predicting that any fully specific network ($\psi=1$, see Fig.~\ref{fig:theory}d) that assembles unigraphically must be on the diversity-redundancy boundary $\varphi+r=1$. 
While the reverse is not true (lying on this boundary does not necessarily guarantee unigraphical assembly) we find that 92\% of the real systems that lie on this boundary unigraphical.  

The fact that components with maximal diversity ($\varphi=1$, the rightmost point of the diagram) and full specificity lead to a unique outcome was recognized previously \cite{hubl2025accessing,zeravcic2014size,jacobs2015rational}.
However, we find that 48\% of the natural and engineered systems that are capable of unigraphical assembly are not fully diverse ($\varphi\neq1$), but are scattered on the $\varphi+r=1$ boundary.
This leads to our final major insight, encoded by the diversity-redundancy boundary: a formal derivation of how to reduce diversity via network-based redundancy, and still maintain reproducibility. 
For example, we find that complexes reuse multiple copies of the same protein by assigning them structurally identical roles (Fig.~\ref{fig:analysis}e), remaining on the $\varphi+r=1$ boundary. 
It also formalizes how IKEA engineers duplicates parts: they ensure that the network role of the repeating components is interchangeable. For example, the Kaustby Chair (Fig.~\ref{fig:analysis}e) has low diversity ($\phi=0.37$), reflecting the use of repeated components, but high redundancy ($r=0.63$), because those repeated components occupy interchangeable network roles, keeping the chair exactly on the diversity-redundancy boundary and ensuring unigraphicality.
In other words, the diversity-redundancy boundary offers a principled theoretical framework on how to reduce the complexity of the design process via redundancy. 

\section*{Experimental Realization and Inverse Design} 
 
\begin{figure}[p]
    \centering
    \includegraphics[width=0.93\textwidth]{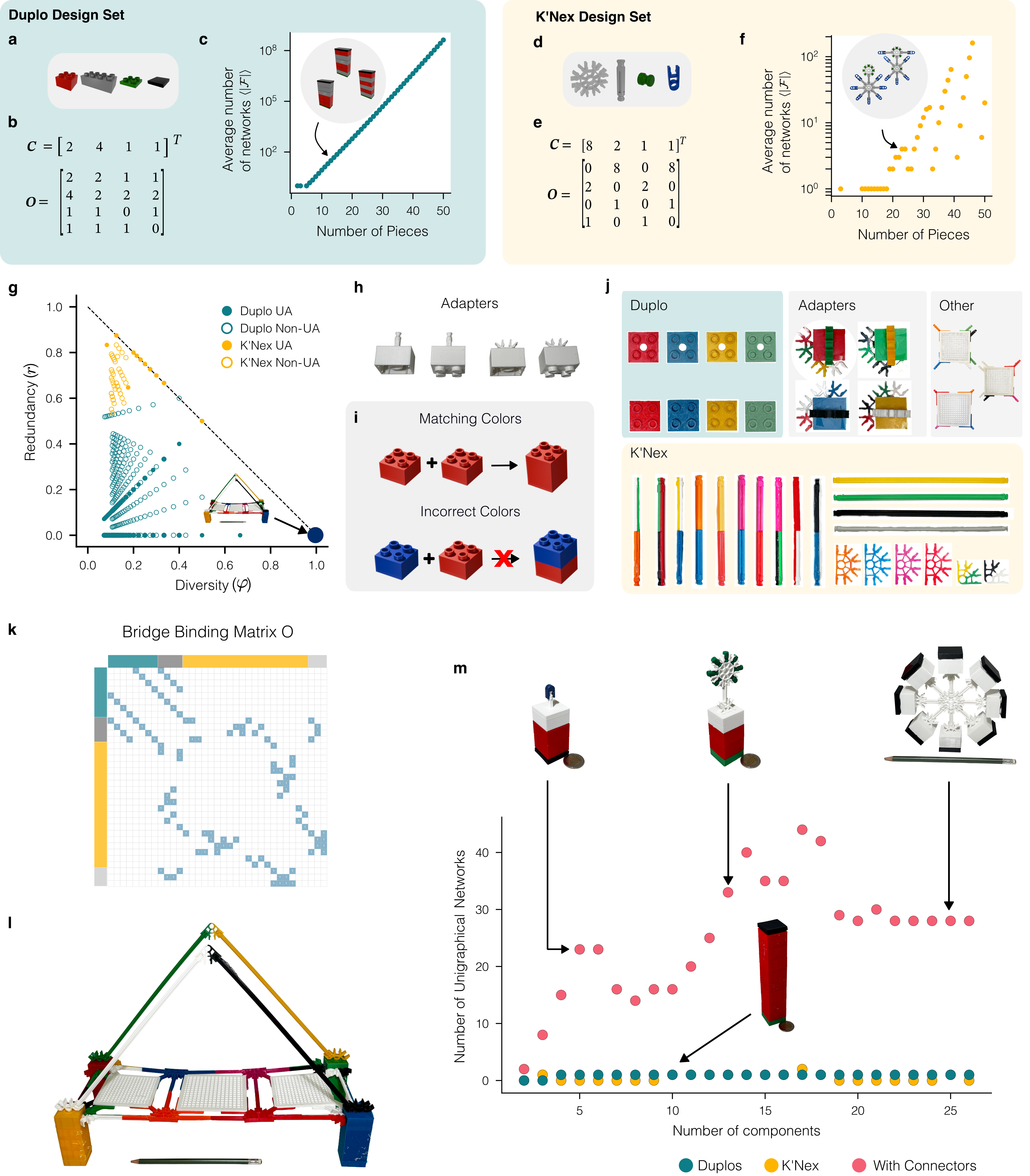}
    \caption{\footnotesize\textbf{Designing Unigraphical Networks}
\textbf{a--f} For Duplo and K'Nex, two well-known construction sets, we extract the associated
design set shown in \textbf{b} and \textbf{e}. \textbf{c,f} With an increasing number of pieces, both construction sets can build an exponentially growing number of possible structures.
\textbf{g} The diversity-redundancy plot indicates that both Duplo and K'Nex constructions fall well below the diversity-redundancy boundary, explaining why they are unable to undergo unigraphical assembly.
\textbf{h--l} By combining both toy sets using adapters \textbf{h}, we can create networks with higher diversity \textbf{j}. By also enforcing that connecting nodes are of the same color as illustrated in \textbf{i}, we can create 35 pieces shown in \textbf{j}, which connect to each other via the binding matrix $\boldsymbol{O}$ shown in \textbf{k}. The design set of \textbf{j} and \textbf{k} is compatible with a unique outcome: the unigraphical bridge shown in \textbf{l}.
A set of at most 11 different building blocks (3 Duplo, 4 K'Nex and 4 connectors), where we lift the color restriction, is able to generate many structures with a unique outcome, as shown in \textbf{m}.}
    \label{fig:toys}
\end{figure}

A critical test of the network design framework is whether it empowers us to design novel structures with unique outcomes.
We demonstrate this in practice using two construction sets, Duplo (Fig.~\ref{fig:toys}a) and K'nex (Fig.~\ref{fig:toys}d), each with its known design sets (Fig.~\ref{fig:toys}b,e). 
Construction sets are known for their generative capacity, allowing us to build many different structures.
Indeed, we find that the average number of structures we can construct using either Duplo or K'Nex systems grows at least exponentially with the number of pieces (Fig.~\ref{fig:toys}c,f, SI VI).
The low diversity (only four distinct components) and low specificity (multiple binding partners for each component) places the structures we can build using these sets far from the diversity-redundancy boundary (empty circles in Fig.~\ref{fig:toys}g), making unigraphical assembly impossible in most cases.
Only a few simple structures, like towers, are capable of unigraphical assembly (full symbols in Fig.~\ref{fig:toys}g), prompting us to ask, can we rely on the design theorems we introduced above to build larger and more complex structures with unique outcomes using these construction sets?

Inspired by our finding that unigraphical structures tend to lie on the $\varphi+r=1$ diversity-redundancy boundary, we decided to increase the diversity $\varphi$ of the construction system via two simple modifications: (i) adding four new 3D-printed \emph{adapters} that are designed to connect Duplo blocks to K'nex components (Fig.~\ref{fig:toys}h) and (ii) imposing a binding rule stating that only components of the same color are allowed to directly connect (Fig.~\ref{fig:toys}i). 

To illustrate the impact of these changes, consider the 35 pieces shown in Fig.~\ref{fig:toys}j subject to the binding matrix in Fig.~\ref{fig:toys}k. Due to the exponential growth with $N$ (Fig.~\ref{fig:toys}c), 35 Duplo blocks would, on average, be compatible with at least $\sim 10^5$ different structures. 
Yet, as we can prove using the Unigraphical Design Theorem (SI VI), the only structure we can build using all 35 pieces in Fig.~\ref{fig:toys}j subject to the binding matrix shown in Fig.~\ref{fig:toys}k is the bridge shown in Fig.~\ref{fig:toys}l.
In other words, by increasing diversity from $\varphi\approx0.11$ to $\varphi=1$ and by making the bridge fully specific ($\psi=1$), we successfully moved a large structure to the diversity-redundancy boundary, ensuring its unigraphical assembly.

Using adapters leads to an increase in the number of unigraphical structures, even without the color constraint.
We find that the number of unigraphical networks built with $N$ of the 11 distinct building blocks (4 K'Nex, 3 Duplo blocks and 4 connectors) is much higher than what can be achieved using only Duplos or K'Nex pieces (Fig.~\ref{fig:toys}m). For example, with 25 pieces we can now construct 28 distinct unigraphical systems, like the wheel shown in Fig.~\ref{fig:toys}m, while Duplos can only produce a single system with a unique outcome, the tower of height $25$.

\section*{Discussion}

Network design shifts the focus from a statistical description of real networks, a methodological perspective that currently dominates network science, to the constraints that govern the reproducibility of complex systems. Unlike traditional network models, which generate ensembles of possible configurations, the design framework identifies the conditions under which unique, reproducible networks can form. 
Our main finding is that reproducibility is not incidental, but arises from constraints encoded in the system’s building blocks.

Our key technical advance is the identification of two complementary and mathematically groun\-ded mechanisms, each capable of encoding reproducible outcomes: unigraphical assembly, which occurs when the design set is sufficiently restrictive, and guided assembly, which enforces uniqueness through temporally ordered interactions. Together, these two mechanisms provide a unified framework for understanding how systems reliably generate unique structures from a set of pre-determined components. 

Importantly, the network design principles identified here are not only descriptive but generative. Indeed, we were able to demonstrate how to use them to realize physical systems that assemble consistently into unique structures, opening the door to the rational design of reproducible networks in both natural and artificial settings.
 
Recent advances in statistical physics and chemistry~\cite{whitelam2015statistical,hormoz2011design,freeman2018reversible,murugan2015multifarious,osat2023non}, including numerical searches over interaction rules that favor a specified target~\cite{russo2022sat,bohlin2023designing},
have made initial steps towards moving self-assembly away from a focus on lattices to more complex structures such as polymer networks and heterogeneous, multi-component systems. 
The network design principles could inform these efforts as well: by abstracting away detailed physical interactions and focusing instead on connectivity constraints, we offer an analytically tractable framework for predicting assembly outcomes that are potentially applicable to a wide range of real systems, from molecular complexes to synthetic materials. 

This abstraction may be particularly relevant for the growing experimental interest in programmable self-assembly, spanning systems from DNA tiles~\cite{evans2017physical} to colloids~\cite{huang2024colloidal}. Many programmable colloidal systems rely on full diversity and specificity; and when this limit is not feasible in practice, they rely on optimization procedures~\cite{jacobs2015self,jacobs2016self} to reduce diversity while maintaining self-assembly, taking advantage of the system symmetries \cite{hubl2025accessing, ahnert2010self,tkachenko2026structural}. The network design principles offer a theoretical framework to extend self-assembly to systems that are not fully diverse. Indeed, as we show in Fig.~\ref{fig:toys}, deliberate changes in network redundancy can allow even non-diverse systems to self-assemble. 

Network design also connects naturally to the exploration of viable chemical compound space~\cite{dobson2004chemical}. Building block encodings have been used to define minimal representations for structures whose uniqueness is presupposed~\cite{tkachenko2026structural,ahnert2010self}. Assembly Theory~\cite{marshall2017probabilistic,sharma2023assembly} has recently used combinatorial arguments to define complexity measures as potential biosignatures. Network design principles have a different but complementary goal: to explore system-specific outcomes compatible with the inherent interactions formally encoded in the design set and could potentially offer new measures of network complexity that bridge combinatorial structure and physically realizable assembly. 

Our results leave room for future work. For example, here we deliberately focused on final assembly states and did not account for the dynamics of the assembly process. Dynamics can enhance assembly efficiency~\cite{gartner2024design,hubl2025accessing,anderson2020hoomd} and help a system avoid kinetic traps~\cite{murugan2015undesired,graziano2026designing}, 
hence incorporating assembly kinetics into the network design framework may enable more efficient exploration of accessible configurations, potentially expanding the space of reproducible systems. 

Relatedly, requiring a single outcome ($|\mathcal{F}|=1$) may be too strict for some natural systems. A probabilistic treatment can relax this by assigning binding affinities to node pairs~\cite{hubl2026polyhedral} or by making the binding matrix inherently probabilistic~\cite{barabasi2020genetic}, so that a design set specifies a distribution over likely assemblies. So far we also assumed that the number of building blocks is fixed through the component vector. In practice, component supply is often imbalanced or fluctuating, and incorporating this constraint into the design set is a natural next step.

Finally, not all networks reproduce. For example, many large social, infrastructural and ecological networks are open-ended systems, whose nodes and links are continuously added, removed or rewired. In these systems, the reproducibility of a particular structure is often less important than the statistical, dynamical or functional properties of the ensemble: a social network must support many possible friendship patterns, the internet needs to allow alternative routing paths, and a machine-learning architecture must tolerate multiple weight configurations with comparable performance. For such systems, traditional network models remain the natural language, as they capture variability, adaptation and population-level regularities rather than exact reproducibility.

This distinction suggests a broader division in network science. Some networks are \emph{generative}: their function lies in maintaining many possible configurations, allowing exploration, adaptation or robustness to changing conditions. Others are \emph{constructive}: their function requires that components assemble into a specific architecture, making reproducibility itself a design constraint. Network design applies to the latter class - it does not replace ensemble-based network science, but complements it by identifying the conditions under which networks cease to be statistical objects and become reproducible constructions. Understanding the conditions under which such reproducibility is possible therefore represents a new chapter for network science and is essential for a deeper understanding of complex systems.

\newpage


\subsection*{Data Availability}

The networks analysed during this study are available at
\url{https://github.com/Barabasi-Lab/NetworkDesign}.

\subsection*{Code Availability}

The code used to determine unigraphicality is publicly available at \url{https://github.com/Barabasi-Lab/NetworkDesign}. All other custom computer code and scripts generated during the current study are available from the corresponding author upon reasonable request.

\section*{Acknowledgments}

We thank Dániel Barabási for his insightful thoughts and discussions on the brain and other self-assembling networks and Albert T. Liu and Sungwan Park for insights on self-assembly, that inspired this project. We also thank Tina Rosado for her help with figure design and the photography of the construction sets. A.-L.B. and C.G. were supported by the National Science Foundation (NSF) award No.2243104 – COMPASS. A.-L.B. and J.K. acknowledge support from the European Union’s Horizon 2020 research and innovation programme No. 810115 – DYNASNET. J.K. acknowledges support from the AccelNet-MultiNet program, a project of the National Science Foundation (Awards No. 1927425 and No. 1927418). 

\section*{Author Contributions}
All authors conceived the study, contributed to the research, and co-wrote the manuscript. J.K. and C.G. contributed to data collection and analysis, as well as the development of analytical and numerical tools.

\section*{Competing Interests}

A.-L.B.~is the founder of Scipher Medicine, a company that explores the role of networks in
health. J.K.~and C.G.~declare no competing interests.

\bibliography{ref.bib}



\end{document}



\maketitle
\thispagestyle{empty}

\affiliations

\setstretch{2}
\setlength{\parskip}{0pt}

\tableofcontents

\clearpage
\section{Datasets}
\label{si:datasets}

Many real systems are capable of consistent reproduction, a feature that is often essential for their function. In this work, we focus on networks built from interacting building blocks that must reproducibly build the same network over and over in order to maintain their function. For this purpose we collected data from real systems that are expected to be reproducible like protein complexes, molecules, circuits, robots, Ikea furniture and image fragments. These systems are ideally suited for our aims as they (1) are made of clearly defined building blocks, (2) interact pairwise and (3) require specific final connectivity patterns to function. In Tab.~\ref{tab:group-summary} we summarize the relevant properties of each of the 6 systems studied. 

\begin{table}[htbp]
  \centering
  \caption{Structural summary of each network group. For each group we report
  what the nodes and edges represent, the number of networks analyzed, the size range $N$, the range of mean
  degrees $\langle k\rangle$ and the range of diversity and redundancy, along
  with the range of the maximum chordless cycle length.}
  \label{tab:group-summary}
  \footnotesize
  \setlength{\tabcolsep}{4pt}
  \begin{tabular}{@{}l c c c c c c@{}}
    \toprule
     & \textbf{Proteins} & \textbf{Molecules} & \textbf{Circuits} & \textbf{Robots} & \textbf{Ikea} & \textbf{Fragments} \\
    \midrule
    \textbf{Nodes} & Proteins & Atoms & Components/Wires & Modules & Pieces & Fragments \\
    \textbf{Edges} & Physical Bonds & Covalent Bonds & Connections & Connections & Connections & Proximity\\
    \midrule
    \textbf{\# Networks}                & 608         & 1321        & 949         & 572          & 100          & 95         \\
    \midrule
    \textbf{Size $N$}                  & 3--42       & 3--150      & 3--199      & 5--80        & 2--24        & 3--72      \\
    \textbf{Avg. degree $\langle k\rangle$}   & 0.29--8.08 & 1.33--2.36  & 1.16--2.81  & 1.6--1.98    & 1--4.21      & 0.86--4.2\\
    \textbf{Diversity $\varphi$}            & 0.077--1   & 0.0198--1   & 0.131--1    & 0.1--1       & 0.17--1     & 1--1       \\
    \textbf{Redundancy $r$}             & 0--0.96    & 0.04--0.94 & 0--0.92    & 0.14--0.79 & 0.125--0.85 & 0--0.71   \\
    \textbf{Larges cycles} & 0--12      & 0--26       & 0--24       & 0--0         & 0--12        & 0--39      \\
    \bottomrule
  \end{tabular}
\end{table}

\subsection{Protein Complexes}

The protein complexes studied in this work are extracted from the Complex Portal \cite{balu2025complex}, an encyclopedic resource of macromolecular complexes spanning multiple model organisms. In our study, we focus on humans (Homo sapiens, Taxonomy ID 9606), mice (Mus musculus, Taxonomy ID 10090) and yeast (Saccharomyces cerevisiae, Taxonomy ID 559292).

A complex is defined as a stable set of interacting proteins that can be co-purified and has been shown to exist as an isolated, functional unit \textit{in vivo}. Any interacting non-protein molecules such as small molecules and nucleic acids may also be included if they are essential, functional components. The Complex Portal provides different confidence scores to the curated complexes based on the level of experimental evidence available. In this study, we only analyze complexes with the highest, 5-star, confidence scores. 

Data from the Complex Portal can be downloaded as a set of JSON files. These files are organized into a set of \textit{interactors}, i.e. the proteins and small molecules, as well as a set of \textit{interactions}. This allows us to extract the simple, labeled graphs that form the basis of our analysis. A link exists between two vertices whenever binding regions or residues within proteins are known to directly interact within the complex. A detailed explanation of the JSON data files can be found at \cite{balu2025complex}.

Because networks that are sufficiently small ($N\leq 2)$ or sufficiently sparse ($L\leq 1$) will always trivially unigraphically assemble, we remove these complexes. We are then left with 582 complexes of which 369 correspond to humans, 22 to mice and 191 to yeast.  

\textbf{Extracting the design set $\mathcal{D}$ of protein complexes: }The design set of the protein complex is extracted empirically. First, the component vector $\boldsymbol{N}$ is extracted by counting how many copies of each protein or small molecule is present in the network. Then, the capacity vector $\boldsymbol{C}$ is found by counting the maximum number of neighbors each node type has. In general, vertices of the same type tend to have the same capacity within a network. Only in 6.0\% of networks do vertices of the same type have varying degrees. These cases cannot be captured by the analytic methods developed in this work and must be treated numerically. They are therefore not included in the 582 networks studied in this work. Finally, an entry of the binding matrix $O_{uv}$ is extracted by counting how many neighbors of type $v$ the vertices of type $u$ have, and taking the maximum value. For example, if a network contains two vertices of type $u$ that have 2 and 5 neighbors of type $v$ respectively, we set $O_{uv}=5$. The rationale for this choice is that if there exists at least one node of type $u$ with 5 binding sites of type $v$, then we would expect all vertices of type $u$ to have the potential to make 5 connections with vertices of type $v$.

Note that extracting the binding matrix this way is very conservative and may lead to under-counting the number of potential neighbors of a certain type. In our previous example, there might be 6 binding sites from type $u$ to type $v$, even though only 5 are manifest in the studied complex. Similarly, a zero in the binding matrix implies that two proteins are not compatible, when in reality it only means that they do not interact in the specific complex under study. It must, however, be remarked that even if two proteins could potentially interact, their dimer might not be stable. In this sense, their interaction would only be transient. As we are only interested in the final, asymptotic behavior of the different complexes, keeping only those edges actually present in the complexes obtained from the Complex Portal is reasonable as they likely represent the stable interactions. 

To further justify the validity of our extraction procedure of the binding matrix, we perform several checks specifically on the zeros of the binding matrix, looking at two different protein-protein interaction datasets (PPIs) to find potential protein type pairs missing in the complexes.

\begin{figure}
    \centering
    \includegraphics[width=\textwidth]{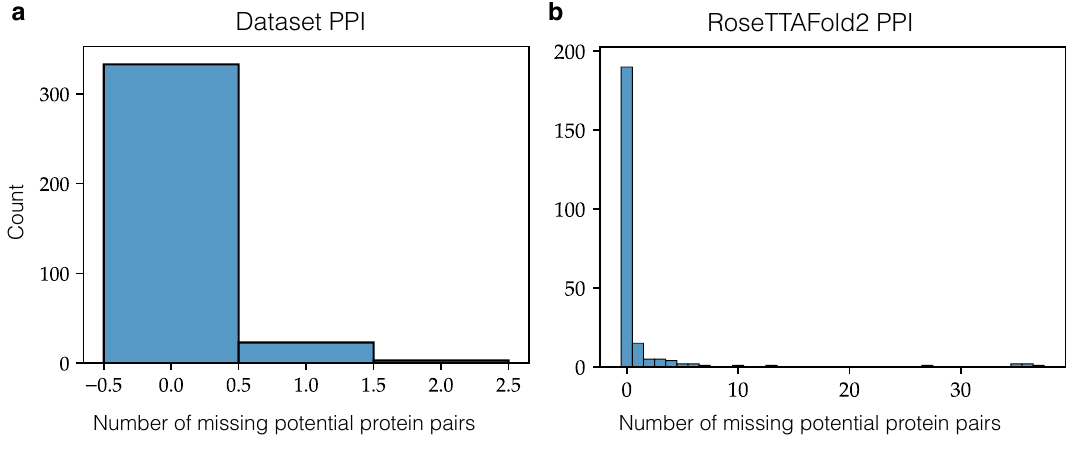}
    \caption{The number of missing proteins pairs per complex, i.e., the potentially incorrect zeros in the $O$-matrix, based on two PPIs. In \textbf{a}, the results of the PPI constructed from our data set is shown, whereas in \textbf{b}, the results of the PPI given in Ref.~\cite{zhang2025predicting} is shown.   }
    \label{fig:justifying_O_protein}
\end{figure}

First, we construct a dataset-wide protein-protein interaction dataset, focusing specifically on the class of human complexes. This is done by creating a network where the node set represents all unique proteins present in the human complexes and where edges are placed if two proteins interact in \emph{any} of the complexes. Note that we exclude small molecules from this analysis. For each complex, we check whether there exist a non-edge which is manifest in the PPI. We find that only 5.4\% of complexes fall into this category, underlining the effectiveness of our extraction method. In Fig.~\ref{fig:justifying_O_protein}a we show the distribution of the number of potentially missing pairs per complex.

Our complex-based PPI only contains those interactions present in complexes with the highest level of experimental evidence, and therefore might not be comprehensive. For this reason we repeat the same analysis with a pairwise PPI constructed in Ref.~\cite{zhang2025predicting} where information from several protein interaction databases as well as structural DDI (domain-domain interaction) information from AlphaFold~\cite{varadi2022alphafold} is combined to predict PPIs among the 200 million possible human protein pairs. We filter predicted interactions such that all predicted precision values are over 90\%.

We perform a similar analysis to the one applied to the dataset-wide PPI, and find that, of the 248 complexes where all constituent proteins are present in the pairwise PPI (63\% of the human complexes in the dataset), only 18\% contain a disconnected pair of protein types that could potentially bind according to the pairwise PPI. In Fig.~\ref{fig:justifying_O_protein}b we show how many potential protein pairs are not present per complex, noting that this number is generally low. 


\subsection{Molecules}
The molecular data is collected from the PubChem database of the National Institute of Health (NIH). We specifically select three subsets of molecules using the PubChem Classification Browser:
\begin{itemize}
    \item \textbf{WHO ATC:} The World Health Organization Anatomical Therapeutic Chemical classification system classifies active substances based on the organ or system they act on and their therapeutic, pharmacological and chemical properties. We download (10/02/2025) all 4383 molecules that have been classified in this scheme. 
    \item \textbf{The Natural Product Atlas:} This dataset provides open access coverage of bacterial and fungal natural products \cite{poynton2024natural}. The classification scheme is based on which species produces which chemical. We download (09/10/2025) all 8795 molecules that consist of at most 20 atoms, not including hydrogen. In other words, we select those molecules with a heavy atom count of at most 20. 
    \item \textbf{CPDat:} The EPA Chemical and Products Database is a resource for exposure-relevant data on chemicals in consumer products. It maps a large set of chemicals to a set of terms categorizing their usage and function in consumer products. We download (09/10/2025) all 9690 molecules classified using this scheme available in PubChem. 
\end{itemize}

After filtering out duplicates between the three datasets, we obtain the molecules' International Chemical Identifier (InChI) from the CSV files downloaded from PubChem. Using RDKit, an open-source toolkit for cheminformatics \cite{l2010rdkit}, we are able to turn these identifiers into networks, where vertices are atoms and edges are covalent bonds. Note that hydrogens were explicitly added before graph construction. Because we restrict our analysis to simple (unweighted) graphs without bond multiplicity, we exclude molecules containing double, triple, or aromatic bonds. Multi-component species such as salts or ion pairs, which are not connected by covalent bonds, appear as disconnected components in the molecular networks. As this is not interesting from a network design point of view, we exclude molecules whose networks consist of multiple disconnected components. To focus on standard neutral organic bonding patterns, we excluded molecules containing formally charged or hypervalent atoms, retaining only species in which atoms exhibit their most common valence configurations. Finally, as was done in the case of protein complexes, we remove networks with less than 3 vertices or less than 2 edges. This leaves us with a total of 1,321 molecules (90 WHO ATC, 258 Natural Product Atlas and 973 CPDat).

\textbf{Extracting the design set $\mathcal{D}$ of molecules: }For each molecule, we construct the component vector $\boldsymbol{N}$ by counting the number of copies of each of the atoms present. As a result of our preprocessing procedure, the capacity vector $\boldsymbol{C}$ can be read off from the network by counting the number of neighbors each atom has. 

Finally, we treat atomic interactions as valence-constrained but otherwise non-specific, meaning that we do not impose explicit pairwise compatibility restrictions between atom types. This results in a binding matrix $\boldsymbol{O}$ where each entry $O_{uv} = C_u$, determined solely by the valence capacity of atom type $u$.

We adopt this assumption because covalent bonding in neutral main-group organic molecules is primarily governed by valence constraints rather than highly specific pairwise interaction rules \cite{weininger1988smiles}. Within this restricted chemical space (single bonds, no formal charges, typical valence states), many atom types can in principle bond to a range of other species, and the absence of a bond in a particular molecule does not imply fundamental incompatibility.

We note that certain atom pairs (e.g., O–O single bonds) are energetically disfavored or comparatively unstable. Incorporating such restrictions would reduce the number of compatible network topologies and therefore further constrain assembly. By not imposing these additional chemical constraints, we obtain a maximally permissive binding matrix, making this a conservative modeling choice.

\subsection{Circuits}
Circuits are made of components (resistors, capacitors, insulators, etc.) connected to each other by wires.
A circuit can be represented as a bipartite network where each component and wire is considered a node and a link exists if a wire connects to a particular component. A wire can connect to several components. Note that, in this bipartite framework, degree-one vertices are generally considered `wires'. One example of degree-one vertices is given by the grounds.
We define the node types based on the component or wire type as well as the number of connections they can make. For example, two wires that connect to different numbers of components are considered as being different types.

We gather 949 circuits taken from Analog Devices' LTSpice collection \cite{broll2023ltspice,ltspice}, 82 correspond to this collection's demos and 867 correspond to its examples.
Each circuit is encoded in an LTSpice netlist file.
The first portion of the netlist file contains relevant component information.
The rest is commands necessary to simulate the circuit which can be ignored for our purposes.
In the component portion, each line represents a distinct component.
The first entry is the name and type of the component and the subsequent entries on the line indicate the wire connections the component has.
The line of each component will vary in length depending on the component type and the amount and types of connections it can make.
Additional information may be given on each line about the voltage and other circuit requirements for the component.
A detailed explanation of netlist files can be found in the documentation of any LTSpice circuit simulator.

Using these LTSpice netlist files, we are able to extract the associated network and collection of node types for a given circuit.

\textbf{Extracting the design set $\mathcal{D}$ of circuits: }The design set of the circuit is extracted as follows.
The component vector $\mathbf{N}$ is found by counting the number of components and wires for each distinct type. 
The capacity vector $\mathbf{C}$ is found by counting the degree of each distinct node type.
This will be consistent as wires with different degrees are considered distinct node types.
The binding matrix $\mathbf{O}$ is conservatively estimated to be fully non-specific, while respecting the bipartite structure of the network. This means that any component can connect to any wire-like node, while components cannot be connected to other components and wires cannot be connected to other wires. 

\subsection{Robots}
Robots can be modular, built up from individual parts. One way to describe how this process takes place is with graph grammars \cite{zhao2020graphgrammar,klavins2004graphgrammar}. Representing the robot as a graph, a grammar provides a set of rules on how a subgraph $S$ can be replaced with another subgraph $S'$. Zhao \textit{et al.} use this framework to automate optimizing a robot's design for different types of terrain. As the set of rules can be applied iteratively, a small set of rules can create a large space of possible designs. We leverage this fact to generate a dataset of modular robots from the graph grammar provided in Fig.~3 and 4 of Ref.~\cite{zhao2020graphgrammar}. 

The grammar is provided in a word document containing a list of rules, each consisting of a subgraph L (the subgraph to be replaced) and a subgraph R (the subgraph to be inserted). Robots always start off with a single start symbol $S$. This can then be replaced by the central \textit{head-body-tail} structure. Next, this body can either be lengthened to \textit{head-body-body joint-body-tail}, or limbs can be attached to the body component. Note that limbs are always attached symmetrically, i.e., they are attached two at a time. Appendages can also be attached to either the limbs or head/tail components. Finally, limbs and appendages can both be ended either with or without a wheel. 

In order to create robots with different levels of complexity, we fix the number of body lengthening steps that can be performed. In other words, we begin to build each robot using a ``start'' node, then create the basic body plan, and then successively perform zero, one, two or three body lengthening steps. We then iteratively apply randomly chosen applicable rules until no more rules can be applied. This leads to 1000 robots - 250 robots for each body length - each uniquely labeled by the sequence of rules that formed them. Importantly, a unique rule order does not imply a unique robot. We therefore filter out duplicated robots. This procedure produces 572 robots, 61 with body length 1, 130 with body length 2, 173 with body length 3 and 208 with body length 4. There are more robots of longer length as having more body components increases the number of possible robots, thereby reducing the number of duplicates in the sample.  

\textbf{Extracting the design set $\mathcal{D}$ of robots: }The component vector $\boldsymbol{N}$ is extracted by counting the number of each type of components in the graphs. Next, due to the nature of the grammar rules, each component has a predefined degree as well as a predefined set of possible neighbors, making the capacity vector $\boldsymbol{C}$ and binding matrix $\boldsymbol{O}$ exact. 

\subsection{Ikea Furniture}
Executing complex assembly tasks based on the information present in abstract instruction manuals poses a significant challenge for robots. Much work in computer vision and robots has been performed to solve the problem of translating the visual information present in instruction manuals into a form that robots can use to assemble the final structure \cite{wang2022ikea,zhang2025manual,tie2025manual2skill}. One popular dataset to do so is given by IKEA instruction manuals, as they are human designed and readily available. 

One common technique used is to convert the final object into a graph \cite{tie2025manual2skill}, where assembly parts are vertices and physical connections are edges. Most datasets contain a significant amount of extraction errors, where instruction manuals are not properly translated into graphs. For this reason we extract the networks by hand, choosing the set of instruction manuals explored in \cite{tie2025manual2skill}. After filtering out very small networks ($N<2$ or $E<1$), we are left with 99 networks- 8 benches, 54 chairs, 4 desks, 19 tables, 3 shelves and 11 miscellaneous items. 

\textbf{Extracting the design set $\mathcal{D}$ of Ikea furniture: }We extract the component vector $\boldsymbol{N}$ by counting how many copies of each node type are present in the graph and the capacity vector by measuring the capacity of each node type. Our data reveals no components of the same type having different degrees maintaining consistency in the capacity vector. Finally, we extract the entries $O_{uv}$ of binding matrix $\boldsymbol{O}$ manually by identifying connection rules between components using the instruction manual.

\subsection{Image Fragments}
From a network design perspective, typical jigsaw-puzzles do not present a particularly interesting class of systems due to their regular nature - they can generally be described as 2D square lattice. As all puzzle pieces are different, both due to the images printed on them as their shape, the only free variable in a jigsaw-puzzle dataset would be the puzzles' size. To obtain more interesting test cases, we use two datasets of \emph{arbitrarily cut} images, (1) RePAIR \cite{tsesmelis2024re} and (2) GVCPuzzles \cite{JigsawNet18,li2019hierarchical}, where pieces have different sizes and shapes. 

The \textbf{RePAIR} dataset consists of 2D and 3D scans of fragments of a fresco, destroyed during a World War II bombing at the Pompeii archaeological park. The fragments are also eroded and have missing pieces with irregular shapes and different dimensions. Ground truth assemblies are based on manual puzzle solving by local archeologists. 

The \textbf{GVCPuzzles} dataset consists of photos shredded using a computer algorithm based on three parameters: (1) puzzle complexity (number of cuts to generate), (2) randomized cutting orientation and (3) perturbations along the cutting curve. Photos are scraped from the website Pexels. 

For each puzzle we have a ground truth file which contains the $(x,y)$-coordinates of the center of mass of each of the pieces as well as the piece's proper rotation. Additionally, there is an image file of each of the puzzle pieces. We can use these images to extract the contours of these pieces, after which they are placed in the corrected position and oriented according to the ground truth file. We then extract a contact graph, where vertices are puzzle pieces and an edge is placed between two pieces if their contours touch at any point. After removing very small graphs ($N<3$ or $E<2$), this procedure produces 95 graphs in total (80 RePair and 15 GVCPuzzles).

\textbf{Extracting the design set $\mathcal{D}$ of image fragments: }As all networks are maximally diverse due to the unique shapes of the constituent fragments, the component vector $\boldsymbol{N}$ is given by a vector of ones of size $N$. Similarly, the capacity vector $\boldsymbol{C}$ is given by the degree sequence of the network. The high complexity of the fragments' contours makes the system fully specific - all binding sites are compatible with only a single, unique fragment. Therefore, the binding matrix $\boldsymbol{O}$ equals the adjacency matrix of the network. 

\clearpage
\section{Standard Network Models}
Before we turn to network design, we must quantify to what degree standard network models are able to consistently reproduce the systems introduced above. Of course, we know that the networks produced by them are not reproducible, as these network models were never designed for that purpose. To make this assumption plausible, in the main text we quantify how likely it is that the Er\H{o}s-Rényi model, the configuration model, the Barabási-Albert model, the stochastic block model or its degree corrected variant reproduce the protein complex eIF2B. To this end we uniformly draw $50,000$ networks for each of these models, after which we remove isomorphic duplicates. We then estimated probability of generating eIF2B, given by the total number of times it was produced divided by the total number of unique graphs produced by the models. Note that this value represents an upper bound, the true probability will be lower. First, in order for us to exhaustively enumerate all networks compatible with a model, the number of draws would need to be increased. Second, we do not take node type distribution into consideration when evaluating isomorphism. However, as the goal of this section is solely to motivate the need for a new network framework for consistent reproduction, we consider the upper bound sufficient. 

We summarize the relevant values used in this experiment in Table~\ref{tab:model-recovery}.

\begin{table}[htbp]
  \centering
  \caption{Recovery of the eIF2B interaction network by canonical network models.
  For each ensemble we draw $50{,}000$ samples and report the number of distinct
  unlabeled graphs, how often the empirical network $G$ is recovered, and the
  resulting recovery probability $P(G)$.}
  \label{tab:model-recovery}
  \begin{tabular}{@{}l l S[table-format=5.0] S[table-format=5.0] S[table-format=2.0] S[table-format=1.5]@{}}
    \toprule
    \textbf{Model} & \textbf{Parameters} & {\textbf{Total}} & {\textbf{Distinct}} & {\textbf{$G$ recovered}} & {\textbf{$P(G)$}} \\
                   &                     & {\textbf{generated}} & {\textbf{unlabeled graphs}} & {\textbf{(count)}} & \\
    \midrule
    Erd\H{o}s--R\'enyi   & $N, \langle k\rangle$        & 50000 & 40486 & 2  & {0.00004} \\
    Configuration model  & $\boldsymbol{k}$        & 50000 & 2197  & 23 & 0.0005 \\
    Barab\'asi--Albert   & $N, m, G_0$   & 50000 & 17933 & 0  & 0 \\
    Stochastic block     & $\boldsymbol{b}, \boldsymbol{e}$        & 50000 & 9945  & 5  & 0.0001 \\
    Degree-corrected SBM & $\boldsymbol{b}, \boldsymbol{e}, \boldsymbol{k}$     & 50000  & 122  & 268 & 0.005 \\
    \bottomrule
  \end{tabular}
\end{table}

\clearpage
\section{Unigraphical Assembly}

In this section we describe two methods of determining the unigraphicality of a design set, i.e., when the number of compatible graphs $|\mathcal{F}|=1$. First, we introduce the Unigraphical Design Theorem (UDT), which allows us to determine unigraphicality analytically for most networks in our dataset. Second, for those networks with structures that cannot be directly resolved by the UDT, we introduce numerical tools that enumerate all compatible graphs, allowing us to determine $|\mathcal{F}|$.

\subsection{The Unigraphical Design Theorem}

We begin with a series of definitions formalizing the framework of network design.
Note that throughout our definitions, both the terms vertices and vertices as well as the terms edges and links may be used interchangeably.
To avoid confusion, the labeling $i,j$ will generally refer to node indices and the labeling $s,t$ will generally refer to node types.

\begin{definition}
    A design set is an ordered triple $\mathcal{D}=(\boldsymbol{N},\boldsymbol{C},\boldsymbol{O})$ comprising
    \begin{enumerate}
        \item the component vector $\boldsymbol{N}$ where $N_s$ is the number of vertices of type $s$,
        \item the capacity vector $\boldsymbol{C}$ where $C_s$ is the maximal degree of vertices of type $s$, and
        \item the binding matrix $\boldsymbol{O}$ where $O_{st}$ is the maximum number of connections a node of type $s$ can maintain with a node of type $t$.
    \end{enumerate}
\end{definition}

In this paper, we generally worked with `typed', or `colored', graphs, where each node has a certain building block type associated to it. These can represent different proteins, different atoms or different electronic components. In Fig.1b of the main text, for example, nodes are colored by the type of protein they represent. In graph theory, typed graphs are not only determined by the set of vertices $V$ and edges $E$, but also by the type of each of the nodes. We formalize this information as follows: 

\begin{definition}
    A typed graph is an ordered triple $G=(V,E,f)$, where $V$ is a set of vertices, $E\subseteq\{\{x,y\}\mid x,y\in V,x\neq y \}$ a set of edges and $f$ is a type function $f:V\rightarrow \{1,...,M\}$ from the set of vertices $V$ of a graph to set of type indices. 
\end{definition}

\begin{definition}
    The set of vertices of type $s$ is denoted by $\theta_s=\{i\in V\mid f(i)=s\}$.
\end{definition}

We next formalize what it means for a network to be a realization of the design set. This will allow us to connect the local rules of the design set with specific network topologies and define the set of compatible networks $\mathcal{F}$. In the main text, we stated that a network $G$ is a realization of a design set $\mathcal{D}$ if (1) its links respect the binding matrix and (2) all nodes are at capacity. Here, we formalize these two conditions and add the rather trivial condition that the network must contain as many nodes of each type as given in the component vector $\boldsymbol{N}$.

\begin{definition}
    A typed graph $G=(V,E,f)$ is a realization of a design set $\mathcal{D}$ if 
    \begin{enumerate}
        \item for all $s$, the number of vertices of type $s$ is $N_s$ where $N_s$ is the $s^{th}$ entry of the component vector, i.e., $N_s=|\theta_s|$,
        \item for all $s$, the degree of vertices of type $s$ is given by $C_s$, i.e., $\forall i\in\theta_s, d_i= C_s$ where $d_i$ is the degree of node $i$, and
        \item the binding rules defined in the O-matrix are respected by all vertices, i.e., $|\{j \in nb(i)\mid f(j)=t\}|\leq O_{f(i)t}$ $\forall i\in V$,$\forall t\in\{1,...,M\}$ where $nb(i)$ is the set of neighbors of node $i$. 
    \end{enumerate}
    We denote the set of all unlabeled realizations with $\mathcal{F}$ and the set of all labeled realizations with $\widetilde{\mathcal{F}}$. 
\end{definition}

Note that for the formal proofs in this section of the supplementary information we will mainly work with labeled graphs and therefore with the set $\widetilde{\mathcal{F}}$. However, for real-world applications the node labeling generally does not play a role, making $\mathcal{F}$ the relevant quantity to consider. 

As stated in the main text, the precise form the Unigraphical Design Theorem takes depends on the specificity of the design set.
We now formally define specificity, as introduced in Fig.2c,d of the main text, including the special cases of non-specific and fully specific design sets. 

\begin{definition}
    A design set where $O_{st} = C_s$ $\forall s,t$ is called fully non-specific.
\end{definition}

This definition quantifies how in the non-specific limit, the binding matrix $\boldsymbol{O}$ is fully determined by the capacity vector $\boldsymbol{C}$. Consequently, the binding matrix adds no additional restrictions to the system.
Instead, the system is solely governed by the pair $\{\boldsymbol{N},\boldsymbol{C}\}$.

\begin{definition}
    A design set where $\sum_t O_{st} = C_s$ $\forall s$ is called fully specific.
\end{definition}

In contrast to the non-specific case, for fully specific systems the capacity vector $\boldsymbol{C}$ is fully determined by the binding matrix $\boldsymbol{O}$. Consequently, the capacity vector adds no additional restrictions and the system is effectively governed by the pair $\{\boldsymbol{N},\boldsymbol{O}\}$. 

\begin{definition}
    A design set that is not fully specific nor non-specific is called semi-specific.
\end{definition}

We say a system is semi-specific if it is neither non-specific nor fully specific.
For semi-specific design sets, all three ingredients of the design set, $\{\boldsymbol{N},\boldsymbol{C},\boldsymbol{O}\}$, are relevant.

In order to interpolate between the two extremes of specificity, we define the specificity of a design set $\mathcal{D}=\{\boldsymbol{N},\boldsymbol{C},\boldsymbol{O}\}$ as
\begin{equation}
    \psi = \frac{1}{N}\sum_s N_s\left(1-\frac{\sum_tO_{st}-C_s}{C_s(M-1)}\right).
\end{equation}

We must now define what it means for a design set to be unigraphical. To this end we first reproduce here the classical definition of unigraphical degree sequences:

\begin{definition}[Johnson 1975 \cite{johnson1975simple}, Koren 1976 \cite{koren1976sequences}]\label{def:unigraph}
    A degree sequence $\boldsymbol{d}$ is unigraphical if for all $G,G'$ with degree sequence $\boldsymbol{d}$, $G\cong G'$.
    A unigraph is any realization of a unigraphical degree sequence. 
\end{definition}

We use the algorithm introduced in Ref.~\cite{borri2011recognition} to determine if a graph is unigraphical or not. This algorithm uses the decomposition theorem proved in Ref.~\cite{tyshkevich2000decomposition}, which states that any unigraph can be decomposed into a series of \emph{split} graphs (graphs where one can decompose the vertex set $V=V_K\cup V_S$ such that the induced subgraph on $V_K$ is complete and the induced subgraph on $V_S$ is empty) and a single special graph. Using this decomposition, the algorithm can determine unigraphicality in linear time.

We employ the language of Definition.~\ref{def:unigraph} to make precise the concept of unigraphicality in design sets. 

\begin{definition}\label{def:type_preserving_isomorphism}
    A design set $\mathcal{D}$ is considered unigraphical if $G\cong_fG'$ $\forall G(V,E,f),G'(V',E',f')\in\widetilde{\mathcal{F}}$.
    This means that there exists some function $\phi:V\rightarrow V'$ such that $\{i,j\}\in E\Leftrightarrow\{\phi(i),\phi(j)\}\in E'$ $\forall i,j\in V$ and $f'(\phi(i))=f(i)$ for all $i\in V$. Such a mapping is called a type preserving isomorphism.
\end{definition}

In other words, this means that all compatible networks in the labeled set $\widetilde{\mathcal{F}}$ are isomorphic, i.e., there is only one possible state, where the isomorphism must respect node type. 

The type preserving isomorphism also allows us to formally relate the two sets $\mathcal{F}$ and $\widetilde{\mathcal{F}}$. We say that $\mathcal{F}=\widetilde{\mathcal{F}}/\cong_f$, where each element of $\mathcal{F}$ is an equivalence class of labeled typed graphs under type preserving isomorphism. If all elements in $\widetilde{\mathcal{F}}$ are type-preserving isomorphic, there is only one equivalence class and $|\mathcal{F}|=1$. Additionally, we say that a design set being unigraphical is equivalent to saying that the graph $G\in\mathcal{F}$ unigraphically assembles. 

With these definitions, we have the necessary terminology to derive the conditions for a design set  $\mathcal{D}$ to be unigraphical, i.e., $|\mathcal{F}|=1$.
We collect these conditions in a theorem that we call the \textbf{Unigraphical Design Theorem} (UDT). Importantly, we assume in the following that $|\mathcal{F}|\neq 0$, i.e., that there is at least one graph that is compatible with the design set. For our purposes, this is a reasonable assumption, as we work with real graphs from which we extract the design set. Then, $\mathcal{F}$ will at least contain the empirical graph. However, if one were to start from a generic design set, one would first need to make sure that $\mathcal{D}$ is graphical~\cite{tripathi2010short,erdos1960graphs,gale1957theorem,ryser1957combinatorial}.

Because a design set's ability to unigraphically assemble is intrinsically related to its specificity, we must work by cases.
We begin by demonstrating the conditions under which a fully non-specific graph unigraphically assembles.

\subsubsection{Unigraphical Design Theorem for Non-Specific Design Sets}

In non-specific design sets, the binding rules between node types are irrelevant as all nodes can connect. Therefore, the possible connectivity patterns are fully determined by the degree sequence, as is the case in classical unigraph theory. However, having different node type the set of possible final states. As an example take the degree sequence $\boldsymbol{d}=(2,2,2,2,2)$ given in Fig.~2a of the main text. Classically, this sequence is unigraphical, but in the case of $\boldsymbol{N}=[2,3]^T$, two different final states are possible (Fig.~2f of the main text). These two final states will have the same underlying connectivity pattern, but different type-distributions (a five-cycle with type distribution (A,A,A,B,B) versus a five-cycle with (A,B,A,B,A)). In the following we show that a non-specific design set is unigraphical if and only if (1) its degree sequence is unigraphical and (2) interchanging any two types, respecting the capacity vector, leads to same distribution of types on top of the graph. The second condition must hold because in a non-specific system, any two nodes with the same degree could end up in the each others location. Even though such a swap leaves the underlying topology unchanged, the new distribution of types might lead to impaired network function. Thus, such a swap must leave the \emph{colored} graph invariant, making all nodes interchangeable without changing the global structure.

\begin{theorem}[\textbf{Unigraphicality of Non-Specific Graphs}]
    Let $\mathcal{D}$ be a fully non-specific design set, and let $\boldsymbol{d}=(d_1,d_2,...,d_n)$ be the corresponding degree sequence. Then, $\mathcal{D}$ is unigraphical with an associated unigraph $G_\mathcal{D}=(V,E,f)\in \widetilde{\mathcal{F}}$  if and only if the following conditions hold:
    \begin{enumerate}
        \item the degree sequence $\boldsymbol{d}$ is unigraphical and
        \item for all $i,j\in V$ such that $d_i=d_j$, letting $f'$ denote the function obtained from $f$ by swapping the values at $i$ and $j$ (i.e, $f'(i)=f(j)$, $f'(j)=f(i)$ and $f'(k)=f(k)$ $\forall k\neq i,j$), then $\exists\pi\in Aut(G_\mathcal{D})$ such that $f\circ \pi=f'$.
    \end{enumerate}
        
\end{theorem}

\begin{proof}
    $(\Rightarrow)$\\
    We first turn to condition (1). By contrapositive, assume that $d$ is not unigraphical. Then there exists some $G'\not\cong G_\mathcal{D}$ with degree sequence $d$. Because the system is fully non-specific, the same type function $f$ used for $G_{\mathcal{D}}$ can be used for $G'$.
    Therefore $G'$ is a realization of $\mathcal{D}$ however $G'\not\cong_fG_\mathcal{D}$ and therefore $\mathcal{D}$ is not unigraphical. By contrapositive then if $\mathcal{D}$ is unigraphical, then $d$ must be unigraphical.

    For condition (2) assume $G_\mathcal{D}(V,E,f)$ is in $\widetilde{\mathcal{F}}$, i.e, is a realization of the design set. Now, take any two vertices $i,j\in V$ with the same degree $d_i=d_j$ and define $f'(i)=f(j)$, $f'(j)=f(i)$ and $f'(k)=f(k)$ $\forall k\neq i,j$. Because the design set is non-specific, the graph $G'=(V,E,f')$ must also be in $\widetilde{\mathcal{F}}$. Because we know that $\mathcal{D}$ is unigraphical, there must exist a colored isomorphism  between $G_\mathcal{D}$ and $G'$. Because they share the same node and edges set, this isomorphism is by definition an automorphism $\pi:V\rightarrow V$. By the definition of a colored isomorphism, $f\circ \pi=f'$.  
        
    $(\Leftarrow)$\\
    In order for $\mathcal{D}$ to be unigraphical, any two of its realizations must be isomorphic under the type function $f$. Because any realization of $\mathcal{D}$ has the same degree sequence by the capacity vector, condition (1) gives that any two realizations must be isomorphic as graphs with no type. We must then only prove that condition (2) is sufficient for the type graph case. We note that any type function $f'$ can be related to the original type function $f$ through a permutation on the node indices. Next, it is known that any permutation can be written as a product of transpositions. Thus, we can generate a sequence of graphs from $G'=(V,E,f')$ to $G_\mathcal{D}=(V,E,f)$ that all differ by exactly one transposition of the type function $(V,E,f')\rightarrow(V,E,f_0)\rightarrow(V,E,f_1)\rightarrow ...\rightarrow(V,E,f_l)\rightarrow(V,E,f)$. Note that these transpositions must be between vertices of the same degree in order to satisfy the conditions set by the capacity vector $\boldsymbol{C}$. Now, by Def.~\ref{def:type_preserving_isomorphism}, condition (2) states that $G'=(V,E,f')\cong_f(V,E,f_0)\cong_f(V,E,f_1)\cong_f ...\cong_f(V,E,f_l)\cong_f(V,E,f)=G_\mathcal{D}$, and thus, by the transitive property of graph isomorphisms, this implies that $G'\cong_f G_\mathcal{D}$.
\end{proof}

\subsubsection{Unigraphical Design Theorem for Fully Specific Design Sets}\label{sec:fullyspecificUDT}

For fully specific design sets, node types not only define a graph coloring but also restricts the set of possible node interactions, effectively decreasing the set of possible connectivity patterns. Consequently, the binding matrix determines the set of final states rather than the capacity vector. This requires the development of new machinery, which we will introduce in the following.

In order to prove the unigraphicality of fully specific graphs, we must establish two more definitions. First, we note that, due to the fully specific nature of the design set, all vertices of type $s$ must have the same number of neighbors of type $t$, for all $s$ and $t$. This gives rise to two observations: (1) the induced subgraph of all vertices of type $s$ must be a regular graph and (2) the subgraph of all vertices of type $s$ and $t$ and the edges between these types must be a biregular graph. Using this we can define the \emph{(bi)regular decomposition} $\mathcal{S}$. 

This (bi)regular decomposition breaks the original graph into subgraphs made from the pairs of node types.
For example, the graphs in Fig.~2e,f can be broken into 3 subgraphs: blue and yellow nodes, blue nodes only and yellow nodes only.
This decomposition is necessary to verify the unigraphicality of each induced subgraph, that is, the unigraphicality of each subgraph in the decomposition.
As we will show later, each of these subgraphs being unigraphical is a necessary condition for the entire design set being unigraphical. which will turn out to be a necessary requirement for the unigraphicality of the full design.
First, we formalize this decomposition as follows:

\begin{definition}
    Let $\mathcal{D}$ be a fully specific design set and $G_\mathcal{D}=(V,E,f)\in\widetilde{\mathcal{F}}$ a realization. The (bi)regular decomposition $\mathcal{S}$ is given by the set of subgraphs $\mathcal{S}=\{S_{st}(V_{st},E_{st})\}_{i\leq j}$ where $$V_{st}=\theta_s\cup\theta_t$$ and $$E_{st}=\{(i,j)\in E \mid f(i)=s, f(j)=t\}$$.
\end{definition}

\begin{figure}
    \centering
    \includegraphics[width=\textwidth]{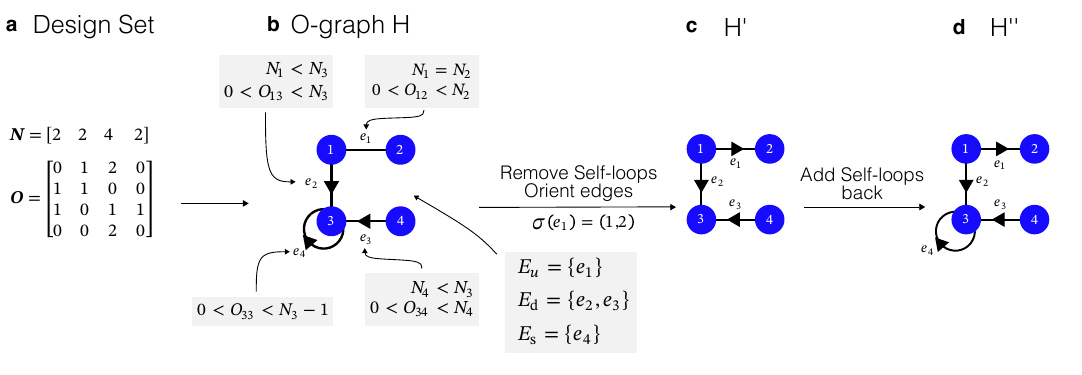}
    \caption{All auxiliary graphs (\textbf{b} $H$, \textbf{c} $H'$ and \textbf{d} $H''$) of the design set given in panel \textbf{a}.  }
    \label{fig:specificproof2}
\end{figure}

Notably, the requirement that every vertex of type $s$ have the same number of type-$t$ neighbors, for all pairs $s,t$, means that the partition of nodes by type $\{\theta_s\}_{s\in[M]}$ is an \emph{equitable partition}~\cite{godsil2013algebraic,godsil1997compact,newman2014equitable}. We do not use the machinery of equitable partitions here, and merely note the connection; it becomes relevant in Sec.~\ref{si:data-analysis}, where we give its formal definition. 

Next, we construct an auxiliary graph $H$, called the $O$-graph, that will help us determine whether or not a design set is unigraphical. Effectively, this $O$-graph encodes the information contained in the design set. Note that the mapping is not 1:1. The $O$-graph does not contain all information from the design set, and a single $O$-graph can correspond to several design sets. This is because the $O$-graph is merely a tool to identify unigraphical design sets - only information relevant for this determination is encoded. 

The vertices of the $O$-graph is merely the set of node types in the design set, as defined below. We additionally define three disjoint sets of edges - undirected and directed edges as well as self-loops, which encoded the relationships between all pairs of node types $(s,t)$, including the self-pairs $(s,s)$.

\begin{definition}
    Let $\mathcal{D}$ be a design set. We define the $O$-graph $H=(V(H),E(H))$, where $|V(H)|=M$ the amount of types. The edge set is given by the union of $E(H)=E_u\cup E_d\cup E_s$, where
    \begin{itemize}
        \item  $E_u=\{\{s,t\} \mid s,t\in V_O, s\neq t, N_{s}=N_{t},0<O_{st}<N_{t}\}$ define the undirected edges,
        \item $E_d=\{(s,t) \mid s,t\in V, s\neq t, N_{s}<N_{t},0<O_{st}<N_{t}\}$ the directed edges and
        \item $E_s=\{\{s,s\}\mid s\in V, 0<O_{ss}<N_{s}-1\}$ the self-loops.
    \end{itemize}
\end{definition}

In words, this means that each node represents a specific node type $s$. The undirected edges $\{s,t\}$ represent biregular subgraphs $S_{st}\in\mathcal{S}$ where $N_s=N_t$ for node types $s\neq t$. Next, directed edges $(s,t)$ represent subgraphs $S_{st}\in\mathcal{S}$ where $N_s<N_t$ for node types $s\neq t$. Thus, edges point from the node type with fewer vertices to the node type with more vertices. Finally, self-loops $\{s,s\}$ represent regular subgraphs $S_{ss}\in S$ between vertices of the same type. If a subgraph in $\mathcal{S}$ is complete or empty, we do not include edges between the types $s$ and $t$ or $s$ to itself as they are not relevant for determining unigraphicality, as we will show in the following sections. An example of the construction process of the $O$-graph is given in Fig.~\ref{fig:specificproof2}. Here, a design set of a network with 4 distinct types gets mapped to an O-graph $H$ with 4 vertices, one undirected edge, two directed edges and one self-loop.

Readers familiar with the theory of equitable partitions may recognize a resemblance between the $O$-graph and the quotient graph, whose nodes likewise represent parts of a partition~\cite{godsil2013algebraic}. Our $O$-graph, however, uses a non-standard edge placement, designed specifically to determine unigraphicality. A related construction appears in Ref.~\cite{arvind2017graph}, where it was used to test whether a graph is amenable to the color-refinement isomorphism check.

Using the biregular decomposition of the graph and the $O$-graph we can determine the unigraphicality of a fully specific design set.
We first explain the method at a high level before formally proving the theorem.
First, we must check that each subgraph in the biregular decomposition is unigraphical. If there exist one subgraph that is not, this subgraph will lead to multiple networks forming, therefore causing $|\mathcal{F}|>1$.
If all subgraphs are unigraphical, then we must check that the $O$-graph is given by a forest of arborescences, i.e., a collection of disconnected directed trees which point away from their root.
This particular structure of an $O$-graph guarantees that if edges between nodes are swapped while satisfying the binding matrix, the network will remain unchanged.
It does this by encoding the specifics of each possible edge connection between node types in the structure of the $O$-graph.

Using the definitions introduced above, we can now formally state the conditions under which a fully specific design set is unigraphical, i.e., the Unigraphical Design Theorem for fully specific systems.

\begin{restatable}[\textbf{Unigraphicality of Fully Specific Graphs}]{theorem}{myCoolTheorem}\label{thm:fullyspecific}
    A fully specific design set $\mathcal{D}$ is unigraphical, with associated unigraph $G_\mathcal{D}\in\widetilde{\mathcal{F}}$, if and only if the following conditions are satisfied
    \begin{enumerate}
        \item all subgraphs $S_{ij}$ in the biregular decomposition of $G_\mathcal{D}$ are unigraphical,
        \item $\exists$ $f_o:E_u(H)\rightarrow V(H)\times V(H)$, $f_o(\{s,t\})\in\{(s,t),(t,s)\}$, such that the oriented graph $H'=(V(H),E_d(H)\cup \{f_o(e):e\in E_u(H)\})$, where self-loops are removed, is given by a forest of arborescences, and
        \item self loops in $H''=(V(H),E_s\cup E(H'))$ are only present on the root vertices of the arborescences.
    \end{enumerate}
\end{restatable}

Note that in this theorem we have introduced slight variations on the $O$-graph, $H'$ and $H''$. The latter is an oriented version of original $H$ and the former takes the oriented $H'$ and removes its self-loops. We show and example of these graphs in Fig.~\ref{fig:specificproof2}. 

In Fig.~\ref{fig:specificproof3}a,b we show an example of a unigraph $G_\mathcal{D}$ and associated O-graph $H$ that satisfy the conditions in the proof. We see that $H$ consists of two disconnected trees. If vertices $4$ and $5$ are defined as the roots of these trees, then the edges $\{5,2\}$ and $\{3,4\}$ can be oriented such that the graph becomes a forest of arborescences. Note also that a disconnected $H$ graph does not necessarily imply that the corresponding graph $G_\mathcal{D}$ is disconnected. Indeed, there is no edge between vertices $1$ and $2$ in $H$, disconnecting it. This is due to the fact that the biregular subgraph between vertices of types $1$ and $2$ in $G_\mathcal{D}$ is complete, leading to a fully connected graph.    

In order to prove Theorem \ref{thm:fullyspecific}, we require machinery that we introduce in the following. Concretely, we need a way to distinguish subgraphs that can achieve unigraphicality and subgraphs that cannot. In Lemma~\ref{lemmaregularunigraphs} we give an exhaustive list of regular unigraphs and, equivalently, in Lemma.~\ref{lemmabiregularunigraphs} we give an exhaustive list of biregular unigraphs. Because the (bi)regular decomposition $\mathcal{S}$ is made up of regular and biregular graphs, this list allows us to prove condition (1) of the Theorem.~\ref{thm:fullyspecific}. 

\begin{figure}
    \centering
    \includegraphics[width=\textwidth]{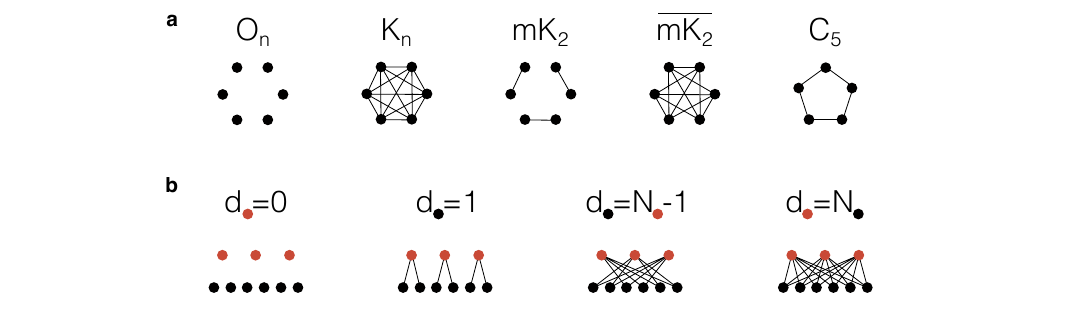}
    \caption{\textbf{a} The set of all regular unigraphs. \textbf{b} The set of all biregular unigraphs.}
    \label{fig:specificproof1}
\end{figure}

\begin{lemma}[Johnson 1975 \cite{johnson1975simple}, Koren 1976 \cite{koren1976sequences}]\label{lemmaregularunigraphs}
    The graphs $K_n,O_n,mK_2,\overline{mK_2},C_5$ yield all regular unigraphs of a single type.
\end{lemma}

\begin{lemma}[Koren 1976 \cite{koren1976sequences}] \label{lemmabiregularunigraphs}
    A biregular graph $G(V_a,V_b)$ with degree pair $(d_a,d_b)$ is a bipartite unigraph if and only if one of the following cases holds
    \begin{itemize}
        \item $d_a=d_b=0$
        \item $d_a=1$ or $d_b=1$
        \item $d_a=N_b-1$ or $d_b = N_a-1$
        \item $d_a=N_b$ and $d_b=N_a$
    \end{itemize}
\end{lemma}

Examples of these (bi)regular graphs are given in Fig.~\ref{fig:specificproof1}.

\begin{figure}
    \centering
    \includegraphics[width=\textwidth]{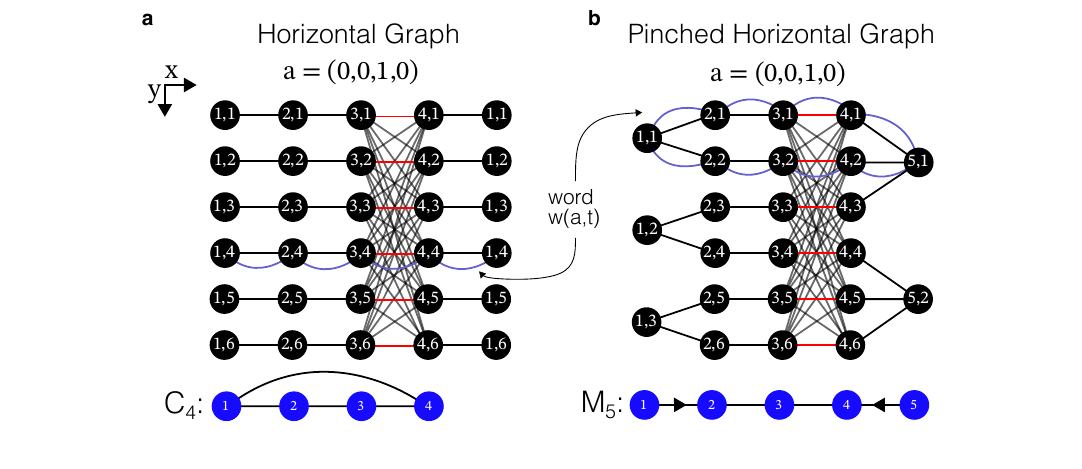}
    \caption{An example of a horizontal graph (\textbf{a}) as well as a pinched horizontal graph (\textbf{b}). Two words $w(a,t)$ are shown as blue paths in both examples. }
    \label{fig:specificproof4}
\end{figure}

In addition to these lemmas, for clarity in describing the proof, we introduce several more definitions. We begin with the \emph{horizontal and pinched horizontal graphs} as shown in Fig.~\ref{fig:specificproof4}a and Fig.~\ref{fig:specificproof4}b, respectively. We will show that these graphs can be seen as `problematic' subgraphs of a realization $G$ of a design set $\mathcal{D}$, i.e., subgraphs that lead to non-unigraphicality. Their vertices are labeled with a pair of integers, where the first encodes the node type $s$ and the second enumerates the set of $N_s$ copies vertices of this type. We then connect these vertices in a way that will become relevant in the proof of the Theorem. In Fig.~\ref{fig:specificproof4}a we show examples of this connectivity pattern. We see that edges (or non-edges) are placed horizontally between the nodes.

\begin{definition}\label{def:horizontalgraph}
    Let $F=(V,E,a)$ be a graph where each node is uniquely defined by a type index $x\in [m]$ and a copy index $y\in[n]$. The vertex set is thus defined as $V=\{(x,y)\mid x\in[m],y\in[n]\}$. We now define a sequence $a=(a_1,a_2,...,a_{m})$ s.t. $a_i\in\{0,1\} \forall i\in [m]$. The edge set is then given by 
    \begin{equation}
        E = \bigcup\limits_{x\in [m]}
        \begin{cases}
        \{\{(x,y),((x\mod m)+1, y)\}\mid y\in[n]\} &\text{ if } a_x = 0\\
        \{\{(x,y),((x\mod m)+1, y')\}\mid y,y'\in[n],y'\neq y\}&\text{ if } a_x = 1
        \end{cases}
    \end{equation}
    The graph $F=(V,E,a)$ is called the \emph{horizontal graph}. An example of a horizontal graph is given in Fig.~\ref{fig:specificproof4}a.
\end{definition}

In Fig.~\ref{fig:specificproof4}a the horizontal graph corresponds to an induced subgraph of $G$ on the set of vertices of types $s\in\{1,2,3,4\}$. Additionally, all $\theta_s$ are equally large, implying that there are the same number of copies of each type. The biregular subgraph $S_{st}$ between the node types is such that the corresponding subgraph in the $O$-graph is an undirected four cycle. This is because the $S_{s,(s\mod 4) +1}$ are neither complete nor empty, whereas all other (biregular) subgraphs between the node types considered here are empty. Specifically, $S_{12}$,$S_{23}$ and $S_{41}$ are all ladder graphs $6K_2$, whereas $S_{34}$ is the complement of the ladder graph $\overline{6K_2}$. This pattern is encoded in the sequence $a=(0,0,1,0)$, where $a_s=0$ represents $S_{s,(s\mod 4) +1}$ being a ladder graph and $a_s=1$ implying it is the complement ladder graph. We next introduce the formal definition of the pinched horizontal graph, an example of which is given in Fig.~\ref{fig:specificproof4}b. We see that the edges (or non edges), are placed horizontally between the nodes, with only the left- and rightmost edges being pinched to accommodate the different number of nodes in these columns. Again, this formalizing this connectivity pattern will help us prove Theorem~\ref{thm:fullyspecific}.   

\begin{definition}\label{def:horizontalgraph2}
    Let $F=(V,E,a)$ be a graph where the vertex set is given by 
    \begin{equation}
        V = \{(x,y)\mid x\in \{2,3,...,m-1\}, y\in [n_h]\}\cup \{(1,y)\mid y\in [n_1]\}\cup \{(m,y)\mid y\in [n_m]\},
    \end{equation}
    where $n_1<n_h>n_m$, $\frac{n_h}{n_1}\in\mathbb{N}$, $\frac{n_h}{n_m}\in \mathbb{N}$ and $n_1>n_m$. We now define a sequence $(a_i)=(a_1,a_2,...,a_{m-1})$ s.t. $a_i\in \{0,1\} \forall i\in[m-1]$. The edge set is then given by $E=E_1\cup E_h\cup E_m$, where
    \begin{equation}
        E_h = \bigcup\limits_{x\in \{2,3,...,m-2\}}
        \begin{cases}
        \{\{(x,y),(x+1, y)\}\mid ,y\in[n_h]\} &\text{ if } a_x = 0\\
        \{\{(x,y),(x+1, y')\}\mid ,y,y'\in[n_h],y'\neq y\}&\text{ if } a_x = 1
        \end{cases},
    \end{equation}
    $E_{1}=\{\{(1,y_1),(2,\frac{n_h}{n_1}y_1+y_2)\}\mid y_1\in[n_1],y_2\in[n_h/n_1]\}$ and similarly for $E_m$. \\  
    The graph $F=(V,E,a)$ is called a \emph{pinched horizontal graph}. An example of a pinched horizontal graph is given in Fig.~\ref{fig:specificproof4}b.
\end{definition}

In Fig.~\ref{fig:specificproof4}b the pinched horizontal graph again corresponds to an induced subgraph of $G$ on the set of vertices of types $s\in\{1,2,3,4,5\}$. Here, $N_1<N_2=N_3=N_4>N_5$ and $N_1>N_5$. The biregular subgraphs $S_{st}$ between the node types is such that the corresponding subgraph in the $O$-graph is line graph of length $5$, with both outermost edges directed inwards.

We next introduce formally the concept of a \emph{word}. This defines a path through the network, with some hops between nodes taking place through edges and others taking place through non-edges. These words will allow us to distinguish different topologies that are not isomorphic.  

\begin{definition}
    Let $F=(V,E)$ be a graph with vertex set $V = \{(x,y) \mid x \in [m],\; y \in [n]\}$. Fix two sequences $ a = (a_1,\dots,a_{l-1}), \quad a_i \in \{0,1\},$ and $ t = (t_1,\dots,t_{l}), \quad t_i \in [m]$. A \emph{word} $w(a,t)$ is a sequence of vertices $\big((t_1,y_1), (t_2,y_2), \dots, (t_{l},y_{l})\big)$ such that for each $i\in[l-1]$ we have $\big\{(t_i,y_i),(t_{i+1},y_{i+1})\big\} \in E
  \quad \text{if } a_i=0,$ and $\big\{(t_i,y_i),(t_{i+1},y_{i+1})\big\} \notin E \quad \text{if } a_i=1$. Additionally, we require that $(t_1,y_1)=(t_l,y_l)$, i.e., that the word describes a cycle in the graph. 
\end{definition}

Examples of such words can be found in Fig.~\ref{fig:specificproof4}a,b. Here, we see how words are cycles and how they can either visit node types once or multiple times. 

Non-isomorphic graphs can be distinguished by their word-statistics, i.e., by counting the number of words of a given type which are present in said network. For this reason in Lemma~\ref{lemma:wordcount1} we count how many distinct words there are in a network.

\begin{lemma}\label{lemma:wordcount1}
    Let $F=(V,E,a)$ be a horizontal graph (Def.~\ref{def:horizontalgraph}). For the type sequence $t = (1,2,...,m,1)$, the set of distinct words $w(a,t)$ has cardinality $n$. 
\end{lemma}

\begin{proof}
    Without loss of generality, let us start at node $(1,1)$. If $a_1=0$ then the walk must pass through an edge, and by construction, this means the only candidate node that respects the type sequence is $(2,1)$. If $a_1=1$, the same is true, as we must now pass through a non-edge. Therefore, independently of $a$, the next node in the sequence must be $(2,1)$. We can continue this argument until we loop back to $(1,1)$, leading to a single word $w_0(a,t) = ((1,1),(2,1),...,(m,1),(1,1))$. Of course, the same holds true for all starting vertices $(1,y)$ for $y\in\mathbb{Z}_n$. Therefore, the total amount of words that respect $a$ and $t$ is given by $n$. An example of the word $w((0,0,1,0),(1,2,3,4,1)) = ((1,4),(2,4),(3,4),(4,4),(1,4))$ is given in Fig.~\ref{fig:specificproof4}a.
\end{proof}

In Lemma.~\ref{lemma:wordcount2} we count the number of circular words (that start and end in the same node type) that are present in a pinched horizontal graph. 

\begin{lemma}\label{lemma:wordcount2}
    Let $F=(V,E,a)$ be a pinched horizontal graph (Def.~\ref{def:horizontalgraph2}). For the type sequence $t = (1,2,...,m-1,m,m-1,...,2,1)$, the set of distinct words $w(a,t)$ has cardinality $\sum_{i\in[n_1],j\in [n_m]}\binom{\mathcal{I}(i,j)}{2}$, where $$\mathcal{I}(i,j) = \max\left(0,\min\left(\frac{n_h}{n_1}(i+1),\frac{n_h}{n_m}(j+1)\right)-\max\left(\frac{n_h}{n_1}i,\frac{n_h}{n_m}j\right)\right).$$.
\end{lemma}

\begin{proof}
    Let us initiate our word at $(1,1)$. If $a_0=0$, we must pass through an edge and there are exactly $n_h/n_1$ candidates for the next vertex in our word. We choose one and now have $((1,1),(2,i))$, where $i\in[n_h/n_1]$. If $a_1=1$, we must pass through a non-edge, leading to exactly the same situation. Next, if $a_2=0$, there is only one option for the next step, leading to $((1,1),(2,i),(3,i))$. Once again, the same is true for $a_1=1$. This continues until we reach $((1,1),(2,i),...,(m-1,i))$. Now, if $a_{m-1}=0$, we must pass through an edge, implying that the only viable next step is $(m,1)$. If $a_{m-1}=1$, we pass through a non-edge leading to the same node. Because we cannot back-track, we only have $n_h/n_1-1$ choices for the next step, leading to $((1,1),(2,i),...,(m-1,i),(m,1),(m-1,j))$, where $j\in[n_h/n_1-1]\}$ and $j\neq i$. Following the same logic as before, we now have only one way back to $(0,0)$, regardless of $a$. Therefore, there are $(n_h/n_1)(n_h/n_1-1)$ such options. Of course, we have overcounted as the direction of the walk does not matter because $t$ is symmetric. The final amount of words $w(a,t)$ starting and ending in $(1,1)$ is then given by $\binom{n_h/n_1}{2}$. An example of such a word ($w((0,0,1,0,0,1,0,0),(1,2,3,4,5,4,3,2,1))=((1,1),(2,1),(3,1),(4,1),(5,1),(4,2),(3,2),(2,2),(1,1))$ is given in Fig.~\ref{fig:specificproof4}b. \\

    With the same set of arguments it can be shown that the amount of words that start at a node $(1,i)$, pass through $(m,j)$ and return to $(1,i)$ is given by $\binom{\mathcal{I}(i,j}{2}$, where the overlap term $\mathcal{I}(i,j)$ tells us how many stubs (non-stubs) attached to $(1,i)$ lie horizontally across from stubs (non-stubs) connected to $(m,j)$. It is given by $\mathcal{I}(i,j) = \max\left(1,\min\left(\frac{n_h}{n_1}(i+1),\frac{n_h}{n_m}(j+1)\right)-\max\left(\frac{n_h}{n_1}i,\frac{n_h}{n_m}j\right)\right)$. This can be understood by noting that the interval of vertices connected to $(1,i)$ goes from $\frac{n_h}{n_1}i$ to $\frac{n_h}{n_1}(i+1)$ and the interval of vertices connected to $(m,j)$ goes from $\frac{n_h}{n_m}j$ to $\frac{n_h}{n_m}(j+1)$. Then, $\mathcal{I}$ is just the overlap between these two intervals. \\
    
    The cardinality of the set of words $w(a,t)$ is then given by $\sum_{i\in[n_1],j\in [n_m]}\binom{\mathcal{I}(i,j}{2}$, with contributions from all starting vertices $(1,i)$ passing through all vertices $(m,j)$.

\end{proof}

With this machinery set up, we now provide the proof of Thm.~\ref{thm:fullyspecific}. For clarity, we repeat the theorem statement here.

\myCoolTheorem*

\begin{proof}

$(\Rightarrow)$
\begin{enumerate}
    \item  We prove the necessity of this condition through contradiction. Let $S_{ss}$ be a regular subgraph that is not a unigraph, i.e., it is not $K_n,O_n,mK_2,\overline{mK_2}$ nor $C_5$ (Lemma ~\ref{lemmaregularunigraphs}). We can then define a rewiring $\mathcal{R}(S_{ss})=S'_{ss}$ s.t. $S_{ss}\not\cong S'_{ss}$. We now define $G'$ as the graph associated to the biregular decomposition where $S_{ss}$ is replaced with $S'_{ss}$. We now ask if $G'(V,E,f)\cong_f G_{\mathcal{D}}(V,E',f)$. If we assume that it is, then there exists a color preserving isomorphism $\phi:V\rightarrow V$ between them. As color needs to be preserved, we know that if $\phi(i)=j$ then $f(i)=f(j)$. Thus, the different subgraphs can be studied in isolation. This means that $\phi|_{\theta_s}:\theta_s\rightarrow \theta_s$, the isomorphism restricted to the vertices of type $s$, must also be an isomorphism between $S_{ss}'$ and $S_{ss}$. However, we assumed that these were not isomorphic, so we have a contradiction and $S_{ss}$ must indeed be unigraphical. A similar proof can be given for the biregular subgraphs $S_{st}$ where $s\neq t$, using Lemma \ref{lemmabiregularunigraphs}.
    
    \item Let us say that the $H'$ consists of $c(H')$ weakly connected components $\{H'_i\}_{i=1}^{c(H')}$. We prove the necessity of condition 2 by contradiction. To this end we note that an arborescence can be defined by adding subsequent constraints to a general directed graph.  
    
    \begin{enumerate}
    \item A directed acyclic graph (DAG) is directed graph with no directed cycles. 
		
    Say that one of the $H'_i$ is not a DAG, i.e. that it contains a directed cycle of length $m$: $C_l=s_1\rightarrow s_2\rightarrow...\rightarrow s_{m}\rightarrow s_1$. Therefore, such a cycle is only possible if $N_q=N_p\equiv n$ for all $s,t \in \{s_0,s_1,...s_{m}\} $ in the cycle. Note that this implies that in $H$, this cycle consist of only undirected edges. We now invoke the fact that by condition (1), all corresponding subgraphs $S_{st}$ must be biregular unigraphs. Of the four options listed in Lemma~\ref{lemmabiregularunigraphs} (Fig.~\ref{fig:specificproof1}b), only $O_{st}=O_{ts}=n-1$ and $O_{st}=O_{ts}=1$ are possible as the empty and complete subgraphs would not lead to an edge in $H$. We now construct two graphs $F_1=(U,E_1)$ and $F_2=(U,E_2)$, where $U=\bigcup\limits_{p\in[m]}\{i\in V\mid f(i)=s_p\}=\bigcup\limits_{p\in[m]}\theta_{s_p}$, i.e. the vertices corresponding to the types in the cycle. Since $|\theta_{s_p}|=n$ $\forall p\in [m]$, we can define a labeling $\Psi:U\rightarrow [m]\times[n]$ given by $\Psi(i)=(x,y)$, where the first label refers to the type-label $p$ and the second enumerates the $N$ vertices of type $s_p$.  \\

    We place the edges $E_1$ s.t. $F_1=(U,E_1,a)$ is a horizontal graph (Def.~\ref{def:horizontalgraph} w.r.t. the labeling $\Psi$. Here, the sequence $a$ is based on the $O_{s_ps_{p+1}}$ in the cycle. When $O_{s_ps_{p+1}}=1$, $a_p=0$ and when $O_{s_ps_{p+1}}=n-1$, $a_p=1$. Without loss of generality, let $a_0=0$. If this is not the case one can look at the complement graph. For a cycle of $m=4$ with $n=6$ this would lead to the horizontal graph shown in Fig.~\ref{fig:specificproof4}a. From Lemma \ref{lemma:wordcount1} we know that the amount of words $w(a,t)$ where $t=(1,2,...,m,1)$ is given by $n$. \\

    We construct $F_2=(U,E_2)$ by performing the rewiring where we remove the edges $((1,1),(2,1))$ and $((1,2),(2,2))$ and replace them with $((1,1),(2,2))$ and $((1,2),(2,1))$. It is clear that this rewiring preserves $\boldsymbol{N}$, as no new vertices are created. It also preserves $\boldsymbol{C}$, as it is degree preserving. Finally, it preserves $\boldsymbol{O}$ because it is degree preserving and occurs within a biregular graph. This rewiring reduced the amount of words $w(a,t)$ by two because a longer cycle of length $2m$ is created. \\

    The fact that $F_1$ and $F_2$ have different cycle statistics proves that they are not isomorphic. However, both satisfy the constraints of the system, and therefore $\mathcal{D}$ cannot be unigraphical if the $H'$-graph contains a directed cycle. 

    \item A polytree is a DAG where the underlying undirected graph is a tree, i.e. where there are no cycles of any type.
		
    We prove that all $H'_i$ must be polytrees by contradiction. Say that one of the $H'_i$'s is not a polytree but is a DAG, i.e. it contains an undirected cycle. In the original $H$ graph, such an undirected cycle will always contain the motif $M_{m}=(s_1\rightarrow s_2- s_3-  ... - s_{m-2}- s_{m-1} \leftarrow s_m)$. The only other cycle that is possible in $H$ by construction is one where all edges are undirected, but we showed in (a) that this leads to non-isomorphic graphs. By construction we have that $N_{s_{1}}<N_{s_{1}}=N_{s_{2}}=...=N_{s_{m-1}}>N_{s_{m}}$. We define $N_{s_1}=n_1$, $N_{s_2}=n_h$ and $N_{s_m}=n_m$. Without loss of generality, we now assume that $n_{1}>n_{m}$ and that $O_{s_2s_1}=1$ and construct two graphs $F_1=(U,E_1)$ and $F_2=(U,E_2)$, where $U=\bigcup\limits_{p\in[m]}\theta_{s_p}$, i.e. the vertices corresponding to the types in the cycle. We label these with $\Psi'$ in a similar way to $\Psi$ in $(a)$, taking into account that now $p\in[m]$ and that $n_{1}< n_{h}>n_{m}$.

    We construct $F_1=(U,E_1,a)$ as a pinched horizontal graph as in Def.~\ref{def:horizontalgraph2} w.r.t. the labeling $\Psi'$. Here $a_p=0$ if $\min(O_{s_{p+1},s_p},O_{s_{p},s_{p+1}})=1$ and $\alpha_p=1$ otherwise $\forall j$. An example of the mapping between $M_5$ and a pinched horizontal graph with $m=5$ and $n_1=3,n_5=2$ and $n_h=6$ is given in Fig.~\ref{fig:specificproof4}b. By Lemma~\ref{lemma:wordcount2}, the amount of words $w(a,t)$, where $t=(1,2,...,m-1,m,m-1,...,1)$ is given $\sum\limits_{i\in[n_{1}],j\in[n_m]}\binom{\mathcal{I}(i,j)}{2}$. For the purposes of this proof, we only have to calculate the four combinations between $i\in\{1,n_{1}\}$ and $j\in\{1,n_{m}\}$. Plugging in these values we see that 
    \begin{alignat}{6}
        \mathcal{I}(1,1)=\mathcal{I}(n_{1},n_{m})&=\frac{n_{h}}{n_{1}}\\
        \mathcal{I}(1,n_{m})=\mathcal{I}(n_{1},1)&=0.
    \end{alignat}
    Therefore, the first two terms both contribute $\binom{n_{h}/n_{1}}{2}$ to the sum whereas the last two contribute nothing.
    \\

    We construct $F_2=(U,E_2)$ by performing the rewiring where we remove the edges $((2,1),(3,1))$ and $((2,n_{h}),(3,n_{h}))$ and replace them with $((2,1),(3,n_{h}))$ and $((2,n_{h}),(3,1))$. As before, this rewiring respects the design set. We now ask how this affects the word count. The only affected overlaps $\mathcal{I}(i,j)$ are $\mathcal{I}((1,1)$, $\mathcal{I}(n_{1},n_{m})$, $\mathcal{I}(n_{1},1)$ and $\mathcal{I}(1,n_{m})$. The former two decrease by one and the latter two increase by the same amount, leading to
    \begin{alignat}{6}
        \mathcal{I}(1,1)=\mathcal{I}(n_{1},n_{m})&=\frac{n_{h}}{n_{1}}-1\\
        \mathcal{I}(1,n_{m})=\mathcal{I}(n_{1},1)&=1.
    \end{alignat}
    The first two terms now both contribute $\binom{n_{h}/n_{1}-1}{2}\neq \binom{n_{h}/n_{1}}{2}$ to the sum, whereas the last two still contribute nothing. Therefore, the total amount of words has reduced by $2(n_{h}/n_{1}-1)$. Of course, if $n_{h}/n_{1}=1$, this evaluates to zero. However, as we assume that $n_{1}<n_{h}$ this is never the case. \\

    The fact that $F_1$ and $F_2$ have different cycle statistics proves that they are not isomorphic, implying that the $H'$-graph must be a polytree.
		
    \item An arborescence is a  polytree where there is a unique node $r$ s.t. there exists a directed path from $r$ to any other vertex $v$. \\

    Say $\exists$ $v\in V(H')$ s.t. there is no directed path between $r$ and $v$. As we have assumed that $H'_i$ is weakly connected, there must exist an undirected path $p=(r,s_1,s_2,...,s_m,v)$. For this to not also be a directed path, there are two options.\\
    
    First, in $H$, this path could contain the motif $M=(s_p\rightarrow s_{p+1}- s_{p+2} -  ... - s_{p+m-1} \leftarrow s_{p+m})$. However, we showed before that such a motif cannot exist if $\mathcal{D}$ is unigraphical. Second, $H$ could contain the motifs $s_p\rightarrow s_{p+1}-s_{p+2}$ or $s_p- s_{p+1}\rightarrow s_{p+2}$. However, because $H'$ is a polytree, one can always choose an orientation $\sigma_r$ s.t all undirected edges in $H$ are oriented away from the root. This will then remove such motifs, implying that $\exists \sigma_r$ s.t. $H'_i$ is an arborescence. Of course, this can be done for each disconnected component $H'_i$, leading to a forest of arborescences. The orientation map $\sigma_o$ in the theorem is then $$\sigma_0=\bigcup_i \sigma_{r_i},$$ where $\sigma_{r_i}:E_u(H_i)\rightarrow V(H_i)\times V(H_i)$ mapping $H_i$ to $H_i'$.
		
	\end{enumerate}
	\item We again prove the necessity of this condition by contradiction. First, we construct $H_i''$ using the orientation map $\sigma_r$ relative to the root node $r$. Assume $H_i''$ has a self loop $(u,u)$ where $u\neq r$. There are several possibilities:

    \begin{enumerate}
    \item  First, $H_i$ could contain the motif $M=(\tikz[baseline=-1.2ex] \draw[->, thick] (0,0) arc[start angle=45,end angle=315,radius=0.5em];
	u-s_p- s_p-...- s_{m-1}\leftarrow s_m)$, where, by construction, $N_s=N_t$ $\forall s,t\neq s_m$ and $N_{s_{m-1}}<N_{s_m}$. This situation is very similar to the one for the motif $(s_1\rightarrow s_2- s_3-  ... - s_{m-1} \leftarrow s_m)$ in $H$, and indeed an equivalent proof can be used to show that $M$ cannot exists if $\mathcal{D}$ is unigraphical. 
    \item If $H_i$ does not contain $M$, it could contain the motif $M'=(
    \tikz[baseline=-1.2ex] \draw[->, thick] (0,0) arc[start angle=45,end angle=315,radius=0.5em];
	u- s_1-s_2-...-s_{m-1}-s_{m}
    \tikz[baseline=0ex] \draw[->, thick] (0,0) arc[start angle=225,end angle=135+360,radius=0.5em];)$, where $n_s=n_t$ $\forall s,t$. Once again, it can be shown that this motif cannot exist if $\mathcal{D}$ is unigraphical.
    \item If $H_i$ does not contain $M$ nor $M'$, it must contain $M''=(\tikz[baseline=-1.2ex] \draw[->, thick] (0,0) arc[start angle=45,end angle=315,radius=0.5em];
	u- s_1-s_2-...-s_{m-1}-s_{m}- r)$, where $N_s=N_t$ $\forall s,t$. In this case, one can redefine the orientation mapping $\sigma_r$ s.t. $u$ becomes the new root. Then, $(u,u)$ is attached to the root node in $H_i''$ and is no longer in contradiction with the condition.
    \end{enumerate}
\end{enumerate}

\begin{figure}[h]
    \centering
    \includegraphics[width=\textwidth]{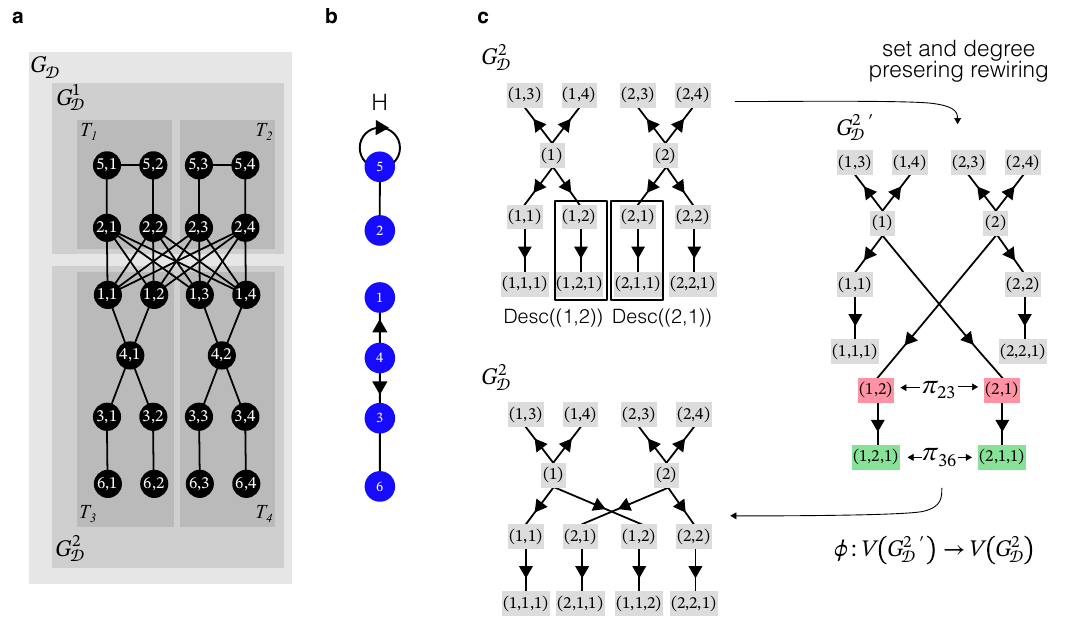}
    \caption{\textbf{a} The full graph $G_\mathcal{D}$ can decomposed into two subgraphs $G_\mathcal{D}^1$ and $G_\mathcal{D}^2$ based on whether the types of the corresponding node sets are disconnected in the O-graph $H$ (panel \textbf{b}). These subgraphs can then be decomposed into four different trees $T_1,...,T_4$. \textbf{c} For the subgraph $G_\mathcal{D}^2$, we show how the vertices can be labeled based on their relation to the root vertices $(1)$ and $(2)$. Then a rewiring between $((1),(1,2))$ and $((2),(2,1))$ can be undone using the isomorphism $\phi$ based on the perfect matchings $\pi_{23}$ at depth $2$ and the $\pi_{3,6}$ at depth $3$.}
    \label{fig:specificproof3}
\end{figure}

($\Leftarrow$)

Take a realization $G_\mathcal{D}=(V,E,f)$ of the design set $\mathcal{D}$ s.t. the previous conditions are satisfied. In the following, we will prove that any legal rewiring, i.e., one that respects $\mathcal{D}$, will lead to a graph that is isomorphic to $G_\mathcal{D}$. We first mention several observations that will make this proof easier.

First, we note again that any isomorphism between such two graphs must act within types, i.e., $\phi:V\rightarrow V$  where $\phi|_{\theta_s}:\theta_s\rightarrow \theta_s$ $\forall s\in[m]$. \\

Assume a rewiring occurs in a (bi)regular subgraph $S_{st}$. We claim that we then only have to consider types whose corresponding vertices in $H$ lie in the same weakly connected component as vertices $s$ and $t$. As mentioned, this is not the same as looking at the connected component of the graph $G_\mathcal{D}$ that $S_{st}$ is part of, as complete subgraphs in $G_\mathcal{D}$ do not map to edges in $H$. Say that in $H$, node $p$ is weakly connected to $s$ (and therefore to $t$) and that $q$ is not. Also assume that $S_{lp}$ is fully connected. This corresponds to the situation in Fig.~\ref{fig:specificproof3}a, where $p=1$, $q=2$ and $s$ and $t$ are, for example, $4$ and $3$, respectively. Any relabeling of the vertices in $\theta_p$ necessary to undo the rewiring in $S_{st}$ will not affect the neighborhoods of any of the vertices in $\theta_q$, as, in this example, all vertices in $\theta_q$ are connected to all vertices in $\theta_p$. The same type of argument can be made about any fully connected $S_{pp}$ subgraph. This allows us to only worry about subgraphs of $G_\mathcal{D}$ related to vertices in individual $H_i$'s and ignore all fully connected subgraphs $S_{pq}$. Formally, we define $G^i_\mathcal{D}$ where
\begin{alignat}{6}
V(G^i_\mathcal{D})&=\{v\in V(G_\mathcal{D})\mid f(v)\in V(H)\}\\
E(G^i_\mathcal{D})&=\{\{u,v\}\in E(G_\mathcal{D})\mid  f(u)\neq f(v)\in V(H),O_{f(u)f(v)}\neq N_{f(v)}\}\notag\\
&\cup \{\{u,v\}\in E(G_\mathcal{D})\mid  f(u)=f(v)\in V(H),O_{f(u)f(u)}\neq N_{f(u)}-1\}.
\end{alignat}
The definition of the edge set reflects the relation between edges in $H$ and the binding matrix and component vector. For example, in Fig.~\ref{fig:specificproof3}a, the graph $G_\mathcal{D}$ can be divided into two parts, corresponding to vertices of types $5$ and $2$ and vertices of types $1,4,3$ and $6$.

Finally, we note that for all $u,v\in V(G_\mathcal{D}^i)$ where $f(u)\neq f(v)$ and $N_{f(u)}\leq N_{f(v)}$, $O_{f(v)f(u)}=1$ or $O_{f(v)f(u)}=N_{f(u)}-1$. This is the result of condition $(1)$ which tells us that all edges in $H$ must correspond to unigraphical biregular subgraphs in $G_\mathcal{D}$. Now, because proving that a mapping $\phi:V\rightarrow V$ is an isomorphism only requires showing that neighbor relations are maintained if the the mapping does not change types, we can actually replace all subgraphs relating to $O_{f(v)f(u)}=N_{f(u)}-1$ with their complements. This is because maintaining non-neighbor relations is equivalent to maintaining neighbor relations. Similar arguments hold for the self-loop vertices where $f(u)=f(v)$. We then redefine $E(G_\mathcal{D}^i)$ as 

\begin{alignat}{6}
E(G^i_\mathcal{D})&=\{\{u,v\}\in E(G_\mathcal{D})\mid  f(u)\neq f(v)\in V(H),\min(O_{f(u)f(v)},O_{f(v)f(u)})= 1\}\notag\\
&\cup\{\{u,v\}\not\in E(G_\mathcal{D})\mid  f(u)\neq f(v)\in V(H),O_{f(u)f(v)}= N_{f(v)}-1\}\notag\\
&\cup \{\{u,v\}\in E(G_\mathcal{D})\mid  f(u)=f(v)\in V(H),O_{f(u)f(u)}= 1\}\notag\\
&\cup \{\{u,v\}\not\in E(G_\mathcal{D})\mid  f(u)=f(v)\in V(H),O_{f(u)f(u)}= N_{f(u)}-2\}.
\end{alignat}

The fact that we know that the corresponding $H_i'$ is an arborescence induces a natural labeling $\Psi$ on the vertices of $V(G_\mathcal{D}^i)$. Our construction of $G_\mathcal{D}^i$ implies that it is given by a set of $n_{r}$ (the number of vertices related of root-type) trees $\{T_i\}$. In Fig.~\ref{fig:specificproof3}c, $G_\mathcal{D}^2$ has $n_r=2$ and so consists of two trees. In general, these trees are connected pairwise at the roots if $H$ contains a self-loop and are disconnected if it does not. For this proof we focus on the latter case. The proof for the former is equivalent.

First, we orient the trees $T_i$ with respect to their root node $r_i$. This provides each node $v$ with a path towards the root. This fact allows us to define a labeling function $\Phi:V\rightarrow \mathbb{Z}^{<\omega}$, where $\mathbb{Z}^{<\omega}$ denotes the set of all finite sequences of integers. The function acts on a node $v$ in tree $T_i$ as 
\begin{equation}\label{eq:treelabeling}
    \Psi(v)=
    \begin{cases}
        (i) &\text{ if } v=r_i\\
        \Psi(p(v))\smallfrown i_v &\text{ else}
    \end{cases},
\end{equation}
where $p(v)$ is the parent of $v$, $\smallfrown$ symbolizes concatenation and and $i_v\in[|\{u\in V(T_i)\mid p(u)=p(v)\}|]$ is a unique index distinguishing the children of $p(v)$. Thus, each node is labeled with a unique sequence on integers, where the first entry marks the tree the node is part of. We show in Fig.~\ref{fig:specificproof3}c an example of this labeling. Here, the trees have depth 2, such that node labels are sequences of length 1 (the roots), 2 (the children of the roots) and 3 (the grandchildren of the roots). \\

We now perform an arbitrary, design set preserving, rewiring, where we remove the edges $\{u,p(u)\}$ and $\{v,p(v)\}$ and replace them with $\{u,p(v)\}$ and $\{v,p(u)\}$. Note that such a rewiring must always take place within a subgraph of the (bi)regular decomposition. It is known that the spaces of bipartite and simple graphs are both connected through these degree preserving edge swaps (switchings)~\cite{erdHos2018efficiently}. Therefore, if any such switching can be undone with a type preserving isomorphism, any graph compatible with the design set must be isomorphic, making the design set unigraphical. In Fig.~\ref{fig:specificproof4}c, we perform the rewiring $\{((1),(1,2)),((2),(2,1))\}\rightarrow\{((1),(2,1)),((2),(1,2))\}$. Because this rewiring respects the design set, we know that $f(u)=f(v)$ and $f(p(u))=f(p(v))$. By construction, this also implies that $|\Psi(u)|=|\Psi(v)|=|\Psi(p(u))|+1=|\Psi(p(v))|+1$, i.e., the two parents and the two children have the same depth within the trees. This rewiring results in the graph $G_\mathcal{D}^{\prime i}$. \\

We now construct an isomorphism between the two graphs. To this end we first define the \emph{descendants} of $v$ as $\text{Desc}(v)=\{u\in V\mid \exists i\in[|\Psi(u)|]\,s.t.\Psi(u)[0:i]=\Psi(v)$\}, i.e., the set of vertices $u$ where $\Psi(v)$ is a \emph{prefix} of $\Psi(u)$ (Fig~\ref{fig:specificproof3}c). Note that $v$ is also included in this set. Because the design set is fully specific and vertices $u$ and $v$ are of the same type, the induced sub-trees on the sets $\text{Desc}(v)$ and $\text{Desc}(u)$ are isomorphic. This is because they must share the same number of children, with the same distribution of types. This is then also true for these children, and so on until the leaves of the sub-trees. \\

For each tree depth $y$ and node type $\tau$, we now define two sets of vertices $\mathcal{A}_{y\tau}=\{w\in \text{Desc}(v)\mid f(w)=\tau,|\Psi(w)|=y\}$ and $\mathcal{B}_{y\tau}=\{w\in \text{Desc}(u)\mid f(w)=\tau,|\Psi(w)|=y\}$. Because the two descendant sub-trees of $u$ and $v$ are isomorphic, these two sets have the same size, and we can thus define a perfect matching $\pi_{y\tau}:\mathcal{A}_{y\tau}\rightarrow\mathcal{B}_{y\tau}$ between them. In Fig.~\ref{fig:specificproof3}c we define the two matchings $\pi_{13}$ and $\pi_{26}$, respectively for vertices of type $3$ at depth $1$ and vertices of type $6$ at depth $2$. The isomorphism $\phi:V(G_\mathcal{D}^{\prime i})\rightarrow V(G_\mathcal{D}^{i})$ is then given by
\begin{equation}
    \phi(w)=
    \begin{cases}
        w&\quad\text{if}\quad w\not\in\text{Desc}(u)\cup \text{Desc}(v)\\
        \pi_{|\Psi(w)|f(w)}(w)&\quad\text{if}\quad w\in\text{Desc}(v)\\
        \pi^{-1}_{|\Psi(w)|f(w)}(w)&\quad\text{if}\quad w\in \text{Desc}(u).
    \end{cases}
\end{equation}
This mapping is effectively exchanging the labels on the two sub-trees induced by the descendants of $u$ and $v$, where node type and node depth is respected. This means that vertices are mapped together with their parents and children, maintaining the neighbor relations, making this mapping an isomorphism. In Fig.~\ref{fig:specificproof3}c this corresponds to $\phi((1,2))=(2,1)$, $\phi((2,1))=(1,2)$, $\phi((1,2,1))=(2,1,1)$ and $\phi((2,1,1))=(1,2,1).$ \\

We can also convince ourselves that this mapping leads to the original graph $G_\mathcal{D}^i$. After the rewiring, node $u$ is no longer connected to its parent $p(u)$ but rather to the node $p(v)$. The mapping $\phi$ interchanges $u$ and $v$, such that once again $p(u)$ is connected to $u$. This by itself would not be enough, as now a child $w$ of node $u$ would be connected to $v$ and not $u$. However, the mapping interchanges $w$ with an equivalent child $w'$ of node $v$. Then $w$ is once again connected to $u$. This logic can then be continued until the leaves of the two sub-trees.
\end{proof}

We have now proved Theorem~\ref{thm:fullyspecific}, allowing us to determine exactly and efficiently whether a fully specific design set is unigraphical, by examining at its (bi)regular subgraphs and how they are connected to one another. 

Using this theorem, we show that in the extreme case where all node types are different (the fully diverse case $\varphi=1$) and the system is fully specific, unigraphical assembly is always guaranteed.

\newcounter{tempMaster}
\setcounter{tempMaster}{\value{theorem}} 

\setcounter{theorem}{2}

\begin{corollary}\label{cor:fullydiverse}
    Let $\mathcal{D}$ be a fully specific design set that is maximally diverse. Then, $\mathcal{D}$ is unigraphical.
\end{corollary}

\setcounter{theorem}{\value{tempMaster}}

\begin{proof}
    Because $\mathcal{D}$ is maximally diverse, each biregular subgraph $S_{st}$ is trivially unigraphical as all these subgraphs are $K_2$, the dimer. These dimer subgraphs represent complete graphs, meaning that the O-graph $H$ is empty. An empty graph is trivially an arborescence and there are no self-loops. Therefore, by Theorem \ref{thm:fullyspecific}, $\mathcal{D}$ is unigraphical. 
\end{proof}

\subsubsection{The Unigraphical Design Theorem for Semi-specific Design Sets}

For semi-specific design sets, we can employ similar tools used for non-specific systems, where we have to take into account that the binding matrix imposes additional rules for semi-specific systems. These rules only affect the second condition of the UDT for non-specific previously described. Because the binding matrix must now be considered, not all nodes with the same degree can be interchanged with one another. Thus, rather than checking that all nodes can be switched as in the non-specific case, we only switch nodes which can switch without violating the binding matrix. If a every such pair nodes can be swapped without changing the global structure and the degree sequence is unigraphical, then the design set must be unigraphical.

While the conditions for non-specific systems always sufficient, they are not always necessary. Specifically, if the degree sequence is not unigraphical, the full design set may still be unigraphical due to the extra constraints encoded in the binding matrix. In these cases, we turn to the numerical methods presented in Sec.~\ref{si:milp}. It must be remarked, however, that this is only necessary for about 8\% of datasets.
We present the proof of the UDT for semi-specific systems here:

\begin{theorem}[\textbf{Unigraphicality of Semi-specific Graphs}]
    Let $\mathcal{D}$ be a semi-specific design set and let $\boldsymbol{d}=(d_1,d_2,...,d_N)$ be the corresponding degree sequence. Then $\mathcal{D}$ is unigraphical if 
    \begin{enumerate}
        \item the degree sequence $\boldsymbol{d}$ is unigraphical and
        \item
        for all $i,j\in V$ such that $d_i=d_j$, $|\{k\in nb(i)\mid f(k)=t\}|\leq O_{f(i)t}$ $\forall t\in\{1,2,...,M\}$ and $|\{k\in nb(j)\mid f(k)=t\}|\leq O_{f(j)t}$ $\forall t\in\{1,2,...,M\}$, if we define $f'$ such that $f'(i)=f(j)$, $f'(j)=f(i)$ and $f'(k)=f(k)$ $\forall k\neq i,j$, then $\exists\pi\in Aut(G_\mathcal{D})$ such that $f\circ \pi=f'$.
    \end{enumerate}

    If condition (2) does not hold, then $\mathcal{D}$ is not unigraphical. 
\end{theorem}

We see that conditions (1) and (2) are sufficient for unigraphicality, but that (1) is not necessary. 

\begin{proof}
    $(\Leftarrow)$\\
    The proof for the sufficiency of conditions (1) and (2) is nearly identical to the one for the non-specific case. Again, we note that any type function $f'$ can be related to the original type function $f$ through a product of transpositions. Then, the transpositions as defined in the theorem are the only ones that satisfy the conditions set by the capacity vector $\boldsymbol{C}$ and binding matrix $\boldsymbol{O}$. The rest of the proof is then identical. 
    
    ($\Rightarrow$ condition (2))\\
    This again directly follows from the definition of the color preserving isomorphism. 

    ($\not\Rightarrow$ condition (1))\\
    To show that condition (1) is not always necessary we give a counter example. Let $\boldsymbol{N}=[3,2]^T$, $\boldsymbol{C}=[2,1]^T$ and $\boldsymbol{O}=[[2,1],[1,0]]$ define the design set $\mathcal{D}$. Because $\sum_i O_{i1}\neq C_1$ and $O_{12}\neq C_1$, this design set is semi-specific. The corresponding degree sequence is $d=(2,2,2,1,1)$. As can be seen in Fig.3b of the main text, this sequence is not unigraphical as it can be realized by two unlabeled graphs. Therefore, condition (1) is not satisfied. However, one of the two graphs, which consists of a triangle together with a dimer, would require two type $2$ vertices to connect. However, this is not allowed by the binding matrix. Therefore, this graph is not part of $\mathcal{F}$, and the cardinality of this set reduces to 1, making the design set unigraphical. 
\end{proof}

In summary, we have stated and proved the Unigraphical Design Theorem. This theorem is broken into three pieces--non-specific, fully specific and semi-specific--to account for differing specificities. Next, we introduce complementary numerical methods that will allow us to (1) verify the analytic results of the UDT and (2) handle the few cases where the conditions for semi-specific sets are not enough. Together, these tools allow us to determine the unigraphicality of the real design sets in our data set, leading to the results shown in Fig.~2k in the main text.

\subsection{Mixed Integer Linear Programming Unigraphical Assembly}
\label{si:milp}
To numerically determine whether a design set can unigraphically assemble, we restate our problem statement as a mixed integer linear program.
In general, our goal is to find the set of networks from a given design set with the highest number of edges. In many cases, this implies that all vertices must be at capacity. However, this method will also work if the degree sequence derived from the capacity vector is not graphical, i.e., when there exists at least one node under capacity. This is especially important for guided assembly, where subsets of vertices are brought together sequentially, implying that generally there will not be enough potential binding partners to ensure filled capacity.
If the set of compatible networks has cardinality one, then the network can unigraphically assemble.
To find the set of most stable networks, we define a vector $\mathbf{a}$ which is a vectorized version of the upper triangular quadrant of the adjacency matrix $\mathbf{A}$ for an undirected graph.
Since the most stable configuration of the design set is that with the maximum number of edges, we aim to minimize $-\mathbf{a}^T\mathbf{1}$.

The constraints of our linear program are defined by the design matrix.
For ease of notation, we write our constraints in terms of the adjacency matrix.
These constraints can be easily transformed into vectorized notation.
To enforce the constraints of the binding matrix, we require that $\sum_{j\in\theta_v}\text{vec}(A_{ij})\leq O_{uv}$ for all vertices $i$ of type $\theta_u$.
Similarly, we require that $\sum_{j}\text{vec}(A_{ij})\leq C_u$ for all node $i$ of type $\theta_u$.
Because we do not allow weighted edges, we also require that $a_{ij}\in\{0,1\}$ for all $i,j$.

With these constraints, we can identify one possible stable network by solving the linear program
\begin{align}
    \text{min}&-\mathbf{a}^T\mathbf{1}\\
    \text{such that }&\sum_{j\in\theta_v}\text{vec}(A_{ij})\leq O_{uv},\forall i\in\theta_u\\
    &\sum_{j}\text{vec}(A_{ij})\leq C_u, i\in\theta_u\\
    &a_{ij}\in\{0,1\}.
\end{align}
We use the python SciPy package to solve this mixed-integer linear program and find a solution \cite{2020SciPy-NMeth}.

Solving this program only finds one stable solution of the design set.
We call this solution $\mathbf{a}^*$.
A second solution can be found by adding an additional set of constraints requiring that at least one edge is different from our previous solution.
Therefore our linear program becomes
\begin{align}
    \text{min}&-\mathbf{a}^T\mathbf{1}\\
    \text{such that }&\sum_{j\in\theta_v}\text{vec}(A_{ij})\leq O_{uv},\forall i\in\theta_u\\
    &\sum_{j}\text{vec}(A_{ij})\leq C_u, i\in\theta_u\\
    &a_{ij}\in\{0,1\}\\
    &\sum_{i,j}a_{ij}(1-a_{ij}^*) + (1-a_{ij})a_{ij}^*\geq 1.
\end{align}
We will call this new solution $a^+$.
If $\sum_{i,j}a_{ij}>\sum_{ij}a^+_{ij}$, then we know the set of stable graphs is one and the network can be unigraphically assembled.
If $\sum_{i,j}a_{ij}=\sum_{i,j}a^+{ij}$, then we first need to check whether the associated networks, $G^*$ and $G^+$ are isomorphic.
If they are not, then we know that there are at least two stable graphs and the resultant network cannot be unigraphically assembled.
If they are isomorphic, we add a new set of constraints to our problem, $\sum_{i,j}a_{ij}(1-a^+_{ij}) + (1-a_{ij})a^+_{ij}\geq 1$ and repeat the same procedure.
We continue until a solution results in a non-isomorphic network.
If this solution has less edges than all other solutions, then we have found a network which is not in $\mathcal{F}$ and the design set must be unigraphical.

\clearpage
\section{Guided Assembly}
In the main text we apply the tools introduced in the previous section to determine that of the 3,618 networks studied, 813 unigraphically assemble. To understand how the remaining 2,805 networks can be built reproducibly, we introduce guided assembly, where nodes are interact sequentially. In the following we introduce a set of propositions that allow us to determine how nodes need to be brought together to guarantee a single final outcome for a set of network topologies. Next, we provide numerical tools that allow us to determine the guiding protocol for networks not amenable to theoretical conditions. 

\subsection{The Guided Assembly Propositions}
An assembly protocol is visualized by an assembly tree $T$ where the vertices of the graph are distributed over the leaves of this assembly tree, such that each leaf $T_\alpha$ contains a disjoint subset $V_\alpha\in V$ of the total node set.
The parent of a leaf contains the union of the subsets of vertices contained in its children. Note here that our assembly trees have a non-standard orientation \emph{towards} the root, meaning that children point towards parents instead of the other way around.
Consequently, the root of any assembly tree will contain all the vertices of the network. 
For clarity, we will call vertices $T_i$ in the assembly tree  $T$ \emph{partitions} and refer to vertices in the network as \emph{vertices}.

For each partition $T_i$ of the design protocol, we consider a design set of a subset of vertices. Formally, the design set of a subset of vertices $U\subseteq V$ is given by $\mathcal{D}_U=\{\boldsymbol{N}_U,\boldsymbol{C_U},\boldsymbol{O_U}\}$, where $\boldsymbol{C_U}=\boldsymbol{C}$, $\boldsymbol{O_U}=\boldsymbol{O}$ and where $\boldsymbol{N_U}=[|{i\in U|f(i)=s}|]_{s\in[M]}$ counts the number of vertices of each type that are present in the subset. 

Here it is important to note a subtlety in the definition of the set of final states $\mathcal{F}$. Previously, we have only seen design sets where it was possible for all vertices to reach capacity simultaneously. Using graph theoretical language, we might say they are \emph{graphical}. However, \emph{a priori}, a design set does not need to be graphical. Indeed, when working with subsets of vertices it will often happen that there are fewer vertices available than would be necessary to satisfy every node's capacity. In these cases, $\mathcal{F}$, as defined in the main text, would be empty. To still be able to have a concept of final states, we slightly alter the second condition on $\mathcal{F}$ to state that $G\in \mathcal{F}$ if there is no other graph compatible with the design set that contains more edges. Note that this generalization reduces to the definition in the main text when a design set is graphical - if all vertices are at capacity, no additional edges can be placed. 

We start this section by formalizing the design protocols introduced in the main text. This formal language is necessary to be able to state and prove the Guided Assembly Propositions as well as for describing the numerical tools needed for networks not amenable to these propositions. 

\begin{definition}\label{def:designprotocol}
A design protocol $\mathcal{P}$ defines the order of interactions between vertices of the target graph $G$ under the constraints set by the design set $\mathcal{D}$. A protocol is represented visually using a rooted assembly tree $T$, whose partitions $T_\alpha$ represent assembly steps. Each partition $T_\alpha$ is associated to a subset $V_\alpha\subseteq V(G)$ of vertices of the target graph. Edges are directed towards the root, indicating which subsets are merged. An assembly tree
\begin{enumerate}
    \item assigns each node $i\in V(G)$ of the target graph to a unique leaf $T_\alpha\in L(T)$ of the assembly tree, where $L(T)=\{T_\alpha\in V(T)\mid d(T_\alpha)=1\}$, such that $\bigcup\limits_{T_\alpha\in L(T)}V_\alpha=V(G)$ and $V_\alpha\cap V_\beta=\emptyset$ $\forall T_{\alpha}\neq T_\beta\in L(T)$,
    
    \item defines the set of unlabeled stable networks $\mathcal{F}_\alpha$ for each leaf $T_\alpha\in L(T)$, as the set of stable networks formed by the vertices $V_\alpha$, and 
    
    \item defines the set of stable networks for any non-leaf partition $T_\beta\in V(T)\setminus L(T)$ of the assembly tree with children $\mathcal{B}_\beta=\{T_\alpha\in V(T)|(T_\alpha,T_\beta)\in E(T)\}$ iteratively. Here, given the set of stable networks $\mathcal{F}_\alpha$ for $T_\alpha\in\mathcal{B}_\beta$, we define the set of stable networks for partition $T_\beta$ as  $\mathcal{F}_\beta=g\left(\mathcal{G}\right)$, where $\mathcal{G}\in\Pi_\alpha\mathcal{F}_\alpha$, i.e., an element of the cartesian product of the final states of the children, and where $g$ is a function that returns all possible networks that can be realized from the vertices in $V_\beta$ under the rules of the design set, given the edges already present in the graphs in $\mathcal{G}$.
\end{enumerate}
\end{definition}

\begin{figure}[h]
    \centering
    \includegraphics[width=.75\linewidth]{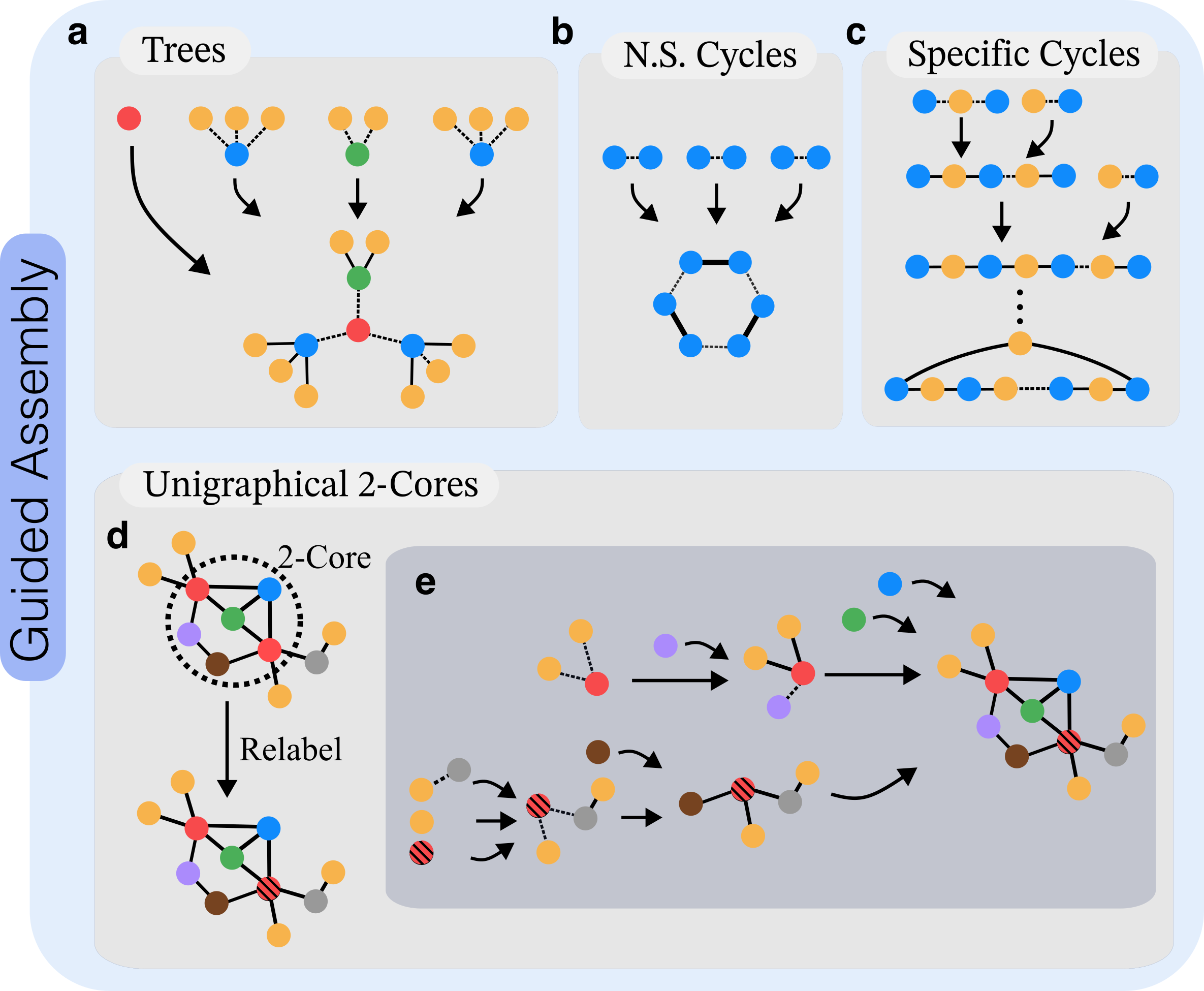}
    \caption{\textbf{a} Every tree has a design protocol obtained by selecting a root and adding vertices sequentially from the leaves to the root. \textbf{b} Non-specific cycles have a non-trivial design protocol only if $N=6$. \textbf{c} Fully specific cycles have a design protocol for two node types by add pairs of vertices sequentially. \textbf{d-e} Any graph with tree-like appendages can be checked for guided assembly by assessing whether it's 2-core is a unigraph after relabeling vertices \textbf{d} based on their type and their appendages. Then a protocol is made by first assembling each appendage separately and then combining all vertices at once and letting the unigraph form \textbf{e}.}
    \label{fig:si-guided}
\end{figure}

In the case of guided assembly, we can prove the existence of a design protocol for specific network structures and specificity classes. We begin by showing how all trees can be built through guided assembly without consideration of specificity (Fig.~\ref{fig:si-guided}a).
We then show how unigraphical networks with tree-like appendages can be built through guided assembly (Fig.~\ref{fig:si-guided}d,e). Finally, we show when cycles can be built for networks that contain only a single type as well as for certain fully specific design sets (Fig.~\ref{fig:si-guided}b,c).

For trees (networks without cycles) we can always find a protocol by assigning an arbitrary root to the tree and then building up the tree from the leaves to the root. This procedure is illustrated in Fig.~\ref{fig:si-guided}a, where the root node is given in red. Note that this procedure is not necessarily the most efficient assembly protocol, i.e., the one with fewest steps. However, its existence means that guided assembly is possible. We formalize this in the following proposition. 

\begin{proposition}[\textbf{Guided Assembly for Trees}]\label{prep:trees}
    Let $G=(V,E,f)$ be a target tree with design set $\mathcal{D}$.
    Then there exists a protocol $\mathcal{P}$ such that $G$ can be built through guided assembly.
\end{proposition}

\begin{proof}
    We work by induction on the number of vertices $N$ in the tree. Assume in all cases that the tree consists of one connected component.
    The case for $N=2$ is trivial as there must be one edge to connect the graph and there is one location to place the edge. Because $G$ is a realization of $\mathcal{D}$, the rules of $\mathcal{D}$ must be satisfied and therefore the protocol $\mathcal{P}$ has one step: combining all vertices at once.

    Assume $N=3$. The only tree of size 3 is a path of length 2. By the same argument of $N=2$, $G$ must have a design protocol $\mathcal{P}$ where all vertices combine at once.

    Assume $N=k+1$. Choose an arbitrary node $r$ to be the root of the tree.
    Consider the set of sub-trees $\mathcal{S}$ rooted at the children of the root of $G$. Such a sub-tree $S_i\in \mathcal{S}$ is the induced subgraph on the set of descendants $Desc(i)$ of a child $i\in nbs(r)$ of the root $r$. These subgraphs must be of size $|S_i|\leq k$ $\forall i\in nbs(r)$. By induction, we assume there exists a design protocol to build any such subgraph $S_i$ by guided assembly.
    We then apply this protocol to each subgraph, resulting in $|\mathcal{S}|=|nbs(r)|$ trees with roots that are one connection away from reaching capacity.
    Because the root $r$ of the tree $G$ has no connections and each child of the root has only one vacant connection, the only realization which will maximize the edges in the network is connecting all children to the root, thereby realizing the original graph $G$.
\end{proof}

Next, we use the result for trees to determine a assembly protocol for networks with unigraphical 2-cores. The 2-core of a network is given by removing, iteratively, all vertices of degree 1. Visually this can be thought of as removing all tree-like appendages from an irreducible core containing cycles (Fig.~\ref{fig:si-guided}d,e). In the following we will first define what it means for a 2-core to be unigraphical and then show that networks whose 2-core is unigraphical can be built through guided assembly.

\begin{definition}\label{def:reducedgraph}
    Let $G=(V,E,f)$ be a realization of $\mathcal{D}$ and let $G'=(V^{(2)},E^{(2)})$ be the associated $2$-core. Let $\mathcal{H}^{(1)}=\{H^{(1)}_i\}_{i\in V^{(2)}}$ be the collection of oriented tree-like appendages rooted at vertices in the 2-core, let $\mathcal{H}^{(1)}/\cong_f$ be the set of isomorphism classes and let $I_{\mathcal{H}^{(1)}}$ be its index set. We then define $f^{(2)}:V^{(2)}\rightarrow I_{\mathcal{H}^{(1)}}$ the coloring function on the 2-core, where vertices are labeled based on the isomorphism class of their associated tree-like appendage. The graph $G^{(2)}=(V^{(2)},E^{(2)},f^{(2)})$ is then called the reduced graph. 
\end{definition}

The idea behind Def.~\ref{def:reducedgraph} is that two vertices of the same type might be attached to tree-like appendages with different topologies. If we now ignore the appendages, one might be tempted to say that the two vertices can be interchanged without changing the topology as they are of the same type. However, this is of course not true as interchanging vertices with different appendages could potentially lead to a different overall topology. To account for this fact, we generate a new labeling that takes into account both the anchor node's type as well as the topology of its corresponding appendage (Fig.~\ref{fig:si-guided}d). This requires us to define the \emph{reduced design set} as follows:

\begin{definition}\label{def:reduced_D}
    Let $G=(V,E,f)$ be a realization of $\mathcal{D}$ and let $G^{(2)}=(V^{(2)},E^{(2)},f^{(2)})$ be the associated reduced graph, which contains $M^{(2)}$ distinct types. Additionally, define $\mathcal{H}^{(1)}=\{H^{(1)}_i\}_{i\in V^{(2)}}$ as before. 
    We then define the reduced design set $\mathcal{D}^{(2)}=\{\boldsymbol{N}^{(2)},\boldsymbol{C}^{(2)},\boldsymbol{O}^{(2)}\}$ as follows:
    \begin{itemize}
        \item For all $p\in[M^{(2)}]$, $N_p^{(2)}$ is the amount of vertices of type $p$,
        \item for all $p\in[M^{(2)}]$, $C_p^{(2)}$ is the degree of vertices of type $p$, and 
        \item for all $i,j\in V^{(2)}\subseteq V$, $O_{f^{(2)}(i)f^{(2)}(j)}=O_{f(i)f(j)}-|\{k\in V\mid f(k)=f(j),k\in H^{(1)}_i,k\in nb(i)\}|$, i.e., a node in the reduced graph can have the same amount of neighbors of a certain type as in the original graph, subtracting those connections that lead to the tree-like appendages.
    \end{itemize}
\end{definition}

We can now use Defs.~\ref{def:reducedgraph},\ref{def:reduced_D} to state the second Guided Assembly Proposition, where we show that graphs with tree-like appendages whose reduced 2-core is unigraphical will always have an assembly protocol.
This is done by first building each individual  tree-like appendages using Prop.~\ref{prep:trees}, where the root is pre-defined as the node connecting the appendage to the 2-core of the network. Finally, all these appendages are combined together with the remaining nodes of the 2-core, which unigraphically assemble into the final structure. An example of such a procedure is given in Fig.~\ref{fig:si-guided}d. 

\begin{proposition}[\textbf{Guided Assembly for Unigraphical 2-cores}]
    Let $G=(V,E,f)$ be a realization of $\mathcal{D}$ and let $G^{(2)}=(V^{(2)},E^{(2)},f^{(2)})$ be the reduced graph with associated reduced design set $\mathcal{D}^{(2)}$. Then, $G$ can be constructed using guided assembly if $\mathcal{D}^{(2)}$ is unigraphical.
\end{proposition}

\begin{proof}
    Because we only claim sufficiency, it suffices to find a design protocol $\mathcal{P}$ such that $G$ is created consistently. By Proposition \ref{prep:trees} we can generate all $H_i^{(1)}$ through guided assembly, where we note that these tree-like appendages are rooted at the node $i$. We combine the assembly protocols of all these trees $|V^{(2)}|$, each containing a complete tree with a rooted node that has $C_{f^{(2)}}$ free stubs. We then combine all these trees into a single assembly node. Because the vertices that are not part of the $2$-core are already at full capacity, they can be ignored in the dynamics. However, their presence does effectively lead to a different coloring as switching two distinct appendages between vertices of the same type will generally lead to a different final graph. Additionally, a link towards a node of a certain type in the appendage $H_i^{(1)}$ means that the node $i$ in the 2-core will have one less binding site available to bind vertices of that type. This situation is therefore equivalent to that of the formation of the reduced graph $G^{(2)}$ with associated design set $\mathcal{D}^{(2)}$. Therefore, if $G^{(2)}$ unigraphically assembles, i.e., if $\mathcal{D}^{(2)}$ is unigraphical, so do the collection of trees $\mathcal{H}^{(1)}$ present in the final assembly step. 
\end{proof}

Finally, we turn to cyclic graphs, specifically 2-regular graphs, which necessarily form into disjoint unions of cycles (Fig.~\ref{fig:si-guided}b,c). Here, we focus on target graphs where all vertices are contained in a single cycle. 

\begin{proposition}
    Let $\mathcal{D}$ be a design set corresponding to $N>5$ vertices and $M$ types, where $\boldsymbol{C}_i=2$ $\forall i\in[M]$, i.e, where all vertices are of degree 2. Then there exists a design protocol $\mathcal{P}$ such that $C_N$, the $N$-cycle, is the unique final state
    \begin{enumerate}
        \item when $M=1$ if and only if $N=6$, and
        \item when $M=2$, if $N\mod2=0$ and $\boldsymbol{O}=[[0,2],[2,0]]$ or $\boldsymbol{O}=[[1,1],[1,1]]$.
    \end{enumerate}
\end{proposition}

\begin{proof}
    \begin{enumerate}
        \item ($\Leftarrow$) We first show the sufficiency of these conditions, i.e., that for $M=1$ guided assembly is possible if $N=6$. 
        In the first step of the protocol, i.e., the leaves of the assembly tree $T$, we subdivide the vertices into three subsets: (1) $\{v_1,v_2\}$, (2) $\{v_3,v_4\}$ and (3) $\{v_5,v_6\}$.
        Dimers are unigraphical, and, thus, the edges $e_{12}, e_{34}$ and $e_{56}$ will form.
        We then combine all three of these subsets at once. There are now two scenarios possible. First, edges are created between only subsets. For example, (1) and (2) could combine to form a 4-cycle. However, as subsets cannot interact with themselves as this would lead to double edges, this scenario leaves all vertices in (3) under-capacity meaning that a stable final state is not produced. Second, all three subsets interact and a 6-cycle is created. Now, all vertices are at capacity and the final state is stable. Of course, there are many different ways each scenario can play out. For example, $v_1$ can connect to any of the vertices $v_3,v_4,v_5,v_6$. However, because all vertices are of the same type ($M=1$) this does not lead to non-isomorphic final states. 
        Thus $C_6$ is the only network at full capacity and thus uniquely designable with guided assembly.

        ($\Rightarrow$) Next, we prove the necessity of this condition, i.e., that, when $M=1$ and guided assembly is possible, that $N= 6$. We prove this by contraposition, showing that if $N\geq 7$, guided assembly is not possible when $M=1$. First, note that $C_N$ with $N\geq 6$ does not unigraphically assemble (Lemma~\ref{lemmaregularunigraphs}), and having a trivial design protocol is thus not viable. This is due to the fact that for any $N\geq 6$, we can find a partition of integers $N_1,N_2\geq 3$ such that $N_1+N_2=N$, implying the possibility of a $C_{N_1}\cup C_{N_2}$ final state. We also know that any $n>2$ is compatible with a $C_n$ graph. This implies that any leaf node of the assembly $T$ that contains $2<n<N$ vertices can produce an intermediate state given by $C_n$. Because edges placed during an assembly step are fixed in our assembly protocols, such small cycles would inevitably lead to a disconnected final states, which is incompatible with fully connected cyclic final state.  
        Thus, each leaf of the assembly tree can contain no more than 2 vertices, i.e, $|V_\alpha|\leq 2$ for $\forall T_\alpha\in L(T)$. For any $N$, the most general node partition is then given by $p$ leaves with $2$ vertices and $N-2p$ leaves with $1$ vertices. The leaves with $2$ vertices will then produce dimers whereas the leaves will $1$ node will produce isolated vertices. Combining two single node leaves will lead to a dimer, meaning that this is equivalent to starting off with $p+1$ leaves with $2$ vertices and $N-2p-2$ leaves with $1$ node. 
        
        We now take $0\leq q_2\leq p$ dimers and $0\leq q_1\leq N-2p$ isolated vertices and combine them into a single assembly node. For this assembly step to be non trivial, at $2q_2+q_1\geq 3$. For any such $q_1,q_2$, this leads to cycles of length $n=2q_2'+q_1'$, where $q_1'\in [q_1]$ and $q_2'\in [q_2]$. Even if $2q_2+q_1=N$, i.e., all vertices are combined in the assembly step, the $C_N$-cycle is only one of many cycles produced. For example, if $q_1\neq 0$, the cycle of length $n=2q_2+q_1-1\leq N$ will also be produced.  Therefore, any such step will inevitably lead to smaller stable cycles being produced, making a fully connected final state impossible.

        \item This condition is only sufficient, not necessary, and thus we only prove the $(\Leftarrow)$ direction. We first investigate the case $\boldsymbol{O}=[[0,2],[2,0]]$.
        
        Because type 1 can only connect to type 2 and vice versa, the sequence in the cycle must be alternating. 
        This is always possible as $N$ is even and gives $N_1=N_2=N/2$. For readability, we call vertices of type 1 $a$-vertices and vertices of type 2 $b$-vertices.
        Create an assembly leaf containing one $a$-node and two $b$-vertices. This will lead to a unique 3-path where
        the $a$-node is at capacity.
        Create another leaf with a single $a$ node.
        Pair the other $N_1-2$ $a$ vertices and $N_2-2$ $b$ vertices into subsets of size 2, each containing one node of each type.
        Each of these will form a single link with neither node at capacity.
    
        With all leaves defined, sequentially combine the subset containing the chain of size 3 with a pair of $a$ and $b$ vertices.
        In the first combination, this will create a chain of size 5 ending in two $b$ vertices.
        The next combination will combine into a chain of size 7, etc.
        Once all pairs have been added sequentially, one subset will have one $a$ node and the other will have a chain of size $N-1$ which ends in one $b$ node on both ends.
        Combine these subsets, closing the chain and creating a cycle of size $N$.

        When $\boldsymbol{O}=[[1,1],[1,1]]$, the final cycles must consist of alternating pairs of $a$ and $b$ vertices, i.e., $...aabbaabb...$. Again, this is possible because $N$ is even and $N_1=N_2=N/2$. We now create an assembly leaf made out of two $a$'s and an $b$. This leads to a unique 3-path. We then create $N_1-2$ leaves containing an $a$ and a $b$, leading to $ab$ dimers. Finally, we create a leaf containing a single $b$ node. 
        
        Next, we add the dimer leaves sequentially to the 3-path. This will lengthen it in a unique way, because no dimer can connect to the $aa$ terminus of the chain. The path can also not close in on itself as it could only connect to a $bb$ terminus, which adding the dimers never creates. Finally, when all dimers are added we add the single $b$ node, which closes the cycle, leading to the unique cycle of length $N$.  
    \end{enumerate}
\end{proof}

Together, these propositions prove the existence of a guided assembly protocol for various specific network topologies. Next, we investigate how numerical results are able to identify assembly protocols for networks which are not encompassed by the Guided Assembly Propositions.  

\subsection{Numerical Search for Assembly Protocols}
\label{si:assembly-trees}
The three Guided Assembly Propositions cover large sub-families of target graphs, but are not exhaustive. For this reason we turn to numerical methods to search for assembly protocols for general target graphs. We do so by searching the space of assembly trees using a Markov Chain Monte Carlo algorithm. First, we discuss the size of the space of possible assembly trees and then demonstrate an approach to explore this space for small networks. Future work should explore more efficient methods for identifying design protocols, capable of handling larger graphs.

\subsubsection{Space of Possible Assembly Trees}
\label{si:space-of-trees}

As discussed, the first step in constructing an assembly protocol is distributing the vertices of the target graph over the leaves of the assembly tree. This is equivalent to partitioning the \emph{multiset} of vertices, where a multiset is a set where elements can be repeated. While there is no closed form formula to calculate the amount of distinct partitions given a multiset, asymptotic results are known~\cite{bender1974partitions,bender1984partitions}. Assume that $\max_s N_s\equiv r$ is finite, which in turn implies that $M\propto N$. Then, as $N\rightarrow\infty$, the number of partitions $v^*(\boldsymbol{N})$ for a component vector $\boldsymbol{N}$ scales as
\begin{equation}
    v^*(\boldsymbol{N})\sim \frac{B(N)}{\prod\limits_{s=1}^M N_s!}\exp\left(\sum\limits_{s=1}^M\binom{N_s}{2}/n\right),
\end{equation}

where $n$ is defined through $N=n\log n$, and where $B(N)$ denotes the Bell number, which is the number of partitions of a set of cardinality $N$. In order to find how this scale with $N$, let us assume that the $N_s$ are randomly distributed i.i.d. around some average $\overline{N_s}$. Using the central limit theorem, this then allows us to write 
\begin{alignat}{6}
    v^*(\boldsymbol{N})&\sim B(N) \exp\left(\frac{1}{n}\sum_s\left(\overline{\ln N_s!}+\overline{N_s}!-(\overline{N_s}-1)!-2!\right)\right)\notag\\
    &\sim B(N)/n^c\sim B(N)\left(\ln N/N\right)^c,
\end{alignat}

where we have used that $\overline{N_s}\sim N^0$ such that $c$ is some constant. We also use that $n\sim N/\ln N$. It is known that the Bell number scales super-exponentially with $N$ \cite{bell1927partition} (faster than $b^{N/a}$ for any constants $a$ and $b$), and the correction coming from working with multisets instead of regular sets does not change this. 

Now, this is of course only a lower bound on the total amount of assembly trees that can be made for a certain component vector $\boldsymbol{N}$ as it only tells us how many different ways we can distribute the vertices over the leaves, and does not take into account that for a certain number of leaves, many trees are possible. 

To calculate an upper bound, we assume that all vertices are different, i.e., $N_s=1$ $\forall s$. Given a certain number of leaves $Q$, the number of assembly trees $R(Q)$ we can make is given by the number of series-reduced rooted trees with Q labeled leaves. We also know that there are exactly $S(N,Q)$ ways to divide a set of $N$ distinct vertices into $Q$ leaves, where $S(N,Q)$ is a Stirling number of the second kind. Therefore, the total amount of possible assembly trees is given by
\begin{equation}
    N_T(N)=\sum_{Q=1}^N S(N,Q)R(Q).
\end{equation}
We now evoke Chebyshev's sum inequality, the fact that $B(N)=\sum_Q S(N,Q)$ and that $R(Q)$ is an increasing function of $Q$ to upper bound the number of trees as
\begin{equation}
    N_T(N)\leq B(N)R(N).
\end{equation}
It is known that $R(N)$ grows super-exponentially with $N$~\cite{foulds2006determining,oeisA000311}, and so does $B(N)R(N)$. Note that this calculation was done for a fully diverse system. However, repeated vertices just mean that some trees are identical, making $B(N)R(N)$ also an upper bound of the general case. Thus, both the lower and upper bound of $N_T(N)$ grow super-exponentially, meaning that $N_T(N)$ must too, independently of the precise component vector $\boldsymbol{N}$. 

Because of the super-exponential growth of the space of possible assembly trees, it is infeasible to exhaustively check all possible design protocols.
While future work should explore more robust numerical methods, we propose a Markov Chain Monte Carlo based approach to search the space of possible assembly trees in order to find one that allows for guided assembly.
While this method is able to correctly identify design protocols for small networks (like the C1-q complex), it is not robust enough nor fully optimized to identify design protocols for large networks. 
We leave the development of improved methods as future work.

\subsubsection{Markov Chain Monte Carlo Search}
We develop a Bayesian pipeline to find the assembly tree that maximizes the probability of reproducing the target network $G^*$.
To do this, we must define the probability that a given design protocol $\mathcal{P}$ assembles the target graph $G$.
In line with definition~\ref{def:designprotocol}, we note that $\mathcal{P}$ is equivalent to its rooted assembly tree $T$.
Therefore, we define $P(G\mid T)$ as the probability the target graph assembles from a given design protocol.

To calculate $P(G\mid T)$, we need to define the probability distribution of networks associated with each leaf partition $T_\alpha\in L(T)$.
Let $V_\alpha\subseteq V(G)$ be the vertices of the network $G$ contained in $T_\alpha$.
Using the numerical methods described in Section \ref{si:milp}, we can generate the set $\mathcal{S}$ of possible networks which can be formed by the vertices in $V_\alpha$ according to the design set $\mathcal{D}$.
Note that this set may contain isomorphic networks.
We then define the distribution $P(H=S\mid V_{\alpha})=|\{H'\in \mathcal{S}\mid H'\cong_f S\}|/|\mathcal{S}|$ indicating the probability that the leaf partition $T_\alpha$ generates the graph $S$.

Let $T_\beta$ be a non-leaf partition.
We further let $T_{\alpha_1},T_{\alpha_2},...,T_{\alpha_r}$ be the set of all children of $T_\beta$.
Additionally, we define the sample space of $P(H\mid T_{\alpha_i})$ as $\mathcal{S}_i$.
We define the distribution $P(H\mid T_\beta)$ as the convolution of the distributions over the leaf vertices.
That is,
$P(H=S\mid T_{\alpha_i}) = \sum_{(H_1,H_2,...,H_r)\in\prod_i\mathcal{S}_i}P(H=S\mid H_1,...,H_r)\prod_{i=1}^rP(H_i\mid T_{\alpha_i})$
where again $P(H\mid H_1,...,H_r)$ is calculated empirically using the methods in Section \ref{si:milp}.
Specifically, we alter the MILP mechanism to include the constraints that the edges in $H_1$, $H_2$, up to $H_r$, must be present in $H$.

By repeating this same process recursively, we can calculate the probability distribution over the final networks produced by the design protocol, i.e., the networks produced in the root of $T$.
If the target graph $G$ can be formed by $T$, then we have $P(G\mid T)=P(H=G\mid T_0)$ where $T_0$ is the root of the assembly tree.
If $G$ is not in the sample space, we define $P(G\mid T)=0$.

Using a Metropolis-Hastings' (MH) inspired approach, we develop an algorithm to search the space of possible assembly trees $T$ to identify an assembly tree which maximizes $P(G\mid T)$.
The algorithm works as follows.
We begin by initializing the algorithm with a tree with one node, i.e.~the design protocol for unigraphical assembly.
This will bias our result naturally towards shallower trees, which is desirable as they represent simpler design protocols.
At each step of the algorithm, we propose a new assembly tree $T'$ and calculate $P(G\mid T')$.
The proposed tree is then accepted or rejected based on the ratio $\tau = P(G\mid T')/P(G\mid T)$.
We repeat this process for either 100,000 time steps or until $P(G\mid T')=1$. 
Because our goal is not to estimate the distribution $P(H\mid T)$ but rather find a $T$ where $P(S\mid T)=1$, we use the standard MH ratio for acceptance.

\begin{algorithm}
\caption{Metropolis-Hastings for Finding Assembly Trees}\label{alg:mh}
\begin{algorithmic}
\State $G \gets G^*$ \Comment{Set target graph}
\State $P_0 \gets P(G^*\mid T_0)$ \Comment{Set probability of trivial tree}
\State $T \gets T_0$ \Comment{Set trivial tree as initial tree}
\State $r \leftarrow 0$
\If{$P_0>0$} \Comment{Do not search if $G$ is not in sample space}
\While{$P_0 < 1$}
\If{$r>10^5$} \Comment{End at $10^5$ time steps}
\State break
\EndIf
\State $T'\gets$ModifyTree($T$) \Comment{Propose new tree}
\State $P_1\gets P(G^*\mid T')$ 
\State $\tau \gets P_1/P_0$
\State $u \sim U[0,1]$ \Comment{Draw random number}
\If{$u<\tau$} \Comment{Accept or reject proposal}
    \State $P_0\gets P_1$
    \State $T\gets T'$
\EndIf
\State $r \leftarrow r+1$
\EndWhile
\EndIf
\end{algorithmic}
\end{algorithm}


New trees $T'$ are proposed by modifying the current $T$ through one of two modification steps:
\begin{enumerate}
    \item Branching, and
    \item Intermediate partition addition.
\end{enumerate}

\emph{Branching} is when a leaf branches into multiple children.
If a leaf partition contains $n$ vertices, this is done by randomly sampling an integer from $\{2,3,...,n-1\}$ and creating that many children partitions.
The $N$ vertices are then randomly divided between these children.
Branching is reversible by selecting the parent of a leaf partition and removing all the parent's children partitions.

\emph{Intermediate partition addition} adds a new partition between a parent and its children.
This is done by first selecting a random non-leaf partition $A$ and then selecting a subset $S$ of its children, not including the set of all children.
A new partition is then added as the parent of the partitions in the subset $S$ and the parent of the new partition is the original partition $A$.
This step can be reversed by randomly selecting a non-leaf partition and connecting its children to its parent and then removing the non-leaf partition.

These steps produce a Markov Chain that is irreducible, i.e., where any vertices can be reached from any other. 

To propose a new tree in our modified Metropolis-Hastings' algorithm, we modify the current tree with either a branching or intermediate partition step or one of their reverse actions.
Each of these actions are equally likely to occur.
Naturally, this will bias the search to the initial tree, which in our case is the trivial tree. However preliminary investigations show that the entire space of possible trees can be reached over time. 
We accept this bias as we prefer assembly pathways that are as short as possible to minimize the number of specific assembly steps that must take place.

It is important to note that when proposing a new tree $P(G\mid T')$ we generally do not recalculate $P(G\mid T')$ entirely. Because the modifications that transform $T$ into $T'$ are local, many of the $P(H\mid T_\beta)$ terms used to calculate $P(G\mid T)$ can be reused to calculate $P(G\mid T')$. In fact, only those $P(S\mid T_\beta)$ that lie between the point of intervention and the root of $T$ need to be recalculated. 

\clearpage
\section{Data Analysis}
\label{si:data-analysis}

Using the theoretical and numerical tools presented, we have built a large dataset containing the networks described in Sec.~\ref{si:datasets}, labeling each network as unigraphical assembly or not unigraphical assembly.
The goal of our analysis presented in this section is to identify key network and design set features that can predict whether a given target network unigraphically assembles.

\subsection{Network and Design Set Features.} For each network, we measure a collection of common network statistics: number of vertices $(N)$, number of edges $(E)$, average degree ($\langle k\rangle$), degree second moment ($\langle k^2\rangle)$, degree variance $(\langle k^2\rangle-\langle k\rangle^2)$, degree standard deviation $(\sqrt{\langle k^2\rangle-\langle k\rangle^2})$, degree coefficient of variance $(\sqrt{\langle k^2\rangle-\langle k\rangle^2}/\langle k\rangle)$, clustering coefficient $(\langle c\rangle$) and number of connected components ($C$).
Notably we transform the feature $N$ to $\log N$.
We also measure less common network statistics relevant to the design problem.
The first is number of orbits $N_O$ and redundancy $1-\frac{N_O-1}{N-1}$. An orbit is a set of vertices that be mapped to one another through an automorphism. Here, we look only at type preserving automorphisms. Redundancy normalizes the number of orbits to a value between 0 and 1~\cite{macarthur2008symmetry,ball2018symmetric}.
When redundancy is 0, all vertices are in different orbits and when redundancy is 1 all vertices are in the same orbit.
The next measure is the maximum chordless cycle size as well as cyclicity $\frac{(E-N+C)}{E}$, where $C$ is the number of connected components.
In addition to network statistics, we measure statistics relevant to the design set.
We measure diversity ($\varphi)$ which is the number of node types $M$ normalized by the number of vertices $\frac{M-1}{N-1}$ as well as specificity (see SI II).

\subsection{Logistic Regression Analysis.}
\begin{figure}
    \centering
    \includegraphics[width=\linewidth]{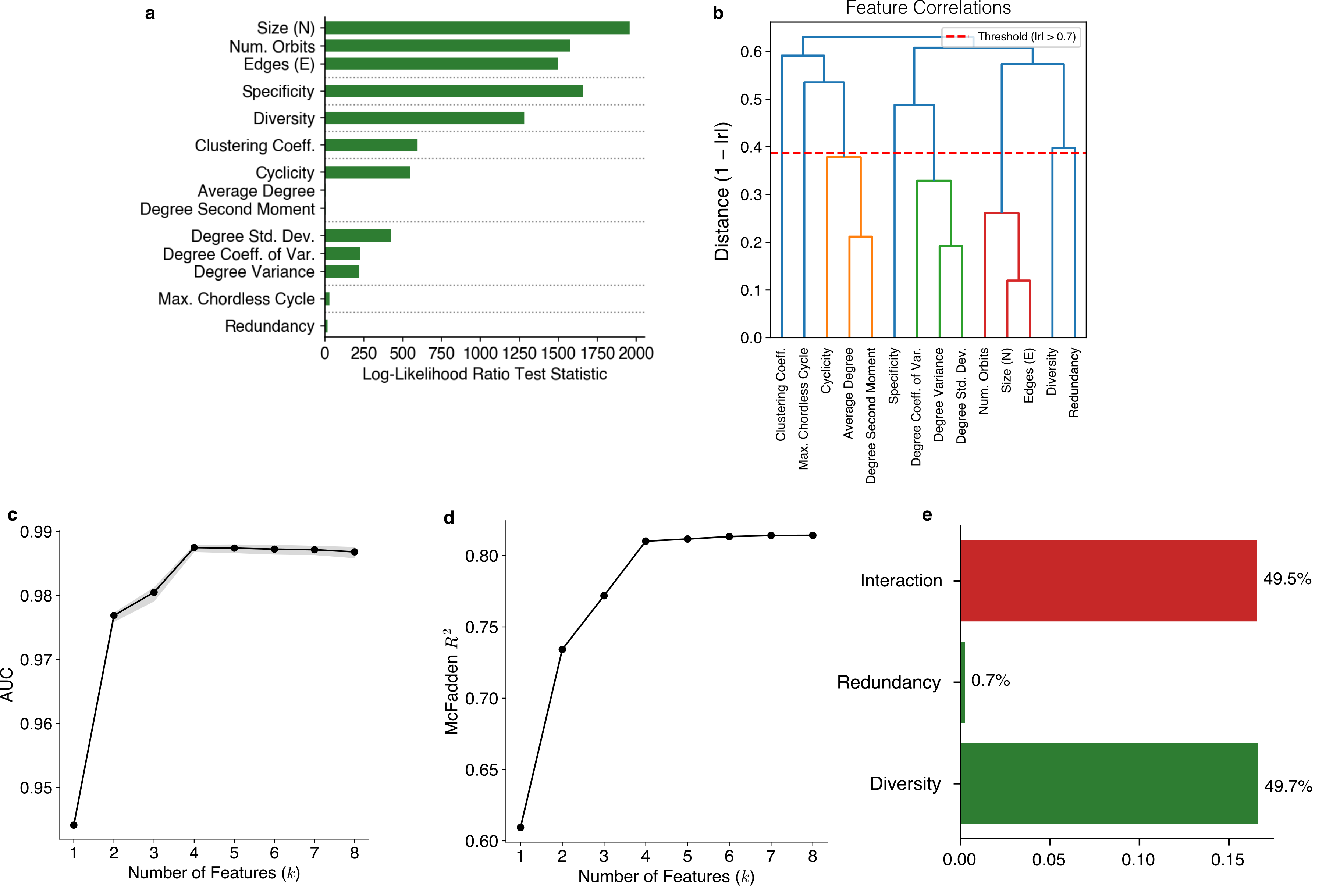}
    \caption{\textbf{Logistic Regression Analysis.} \textbf{a} Logistic regression predicts the impact of 15 network characteristics on unigraphical assembly by measuring the log-likelihood ratio. \textbf{b} Network characteristics are grouped by correlations where $|r|>0.7$. \textbf{c,d} We see that performance significantly increases for both AUC and McFadden $R^2$ metrics when $d=2$ and plateaus at $d=4$. \textbf{e} The SHAP contribution of diversity, redundancy and their interaction points to the predictive power of the combination of both variables.}
    \label{fig:si-analysis}
\end{figure}

To determine the predictive power of each feature, we use logistic regression to predict unigraphical assembly using each feature individually.
Then for each model, we measure the log-likelihood ratio for each feature where a larger value indicates greater predictive power (Fig.~\ref{fig:si-analysis}a).

Features are then correlated using a standard Pearson correlation coefficient and features are clustered together if their correlation is greater than $|r|>0.7$ (Fig.~\ref{fig:si-analysis}b).
We take the feature with highest log-likelihood ratio test score as the representative for each cluster (Fig.~\ref{fig:si-analysis}a). 
This leaves nine features to consider in the following analyses.

For $d\in[1,9]$, we generate every combination of $d$ distinct features and fit a logistic regression for each combination.
For each model, we generate multiple resamples of the data using bootstrapping to determine the model's AUC and determine whether the given combination is statistically significant as compared to a random model.
We additionally measure the McFadden $R^2$ value.

As seen in Fig.~\ref{fig:si-analysis}c,d, we see that the model performance begins to indicate diminishing returns at $d=2$, with a full elbow at $d=4$. 
While the top performing model (AUC=0.98, $R^2=0.73$) for $k=2$ uses both $N$ and specificity, this is likely overfit to our data which has many small networks. Furthermore, the second top performing model has comparable performance (AUC=0.97, $R^2=0.67$) using diversity and redundancy, with a decision boundary at $12.2339\varphi+9.1530r-9.5329=0$.
Lastly, we see that when $d=4$ and performance has plateaued, all four features aforementioned, diversity, redundancy, $N$ and specificity are used.

To determine how much diversity and redundancy contribute to the predictive power individually, we compute the exact Shapley value of each feature by comparing all 4 possible models involving just diversity and redundancy or neither (model 1 - neither, model 2 - diversity only, model 3 - redundancy only, model 4 - diversity and redundancy) and obtaining the weighted marginal contribution of each feature.
These values are reported in Fig.~\ref{fig:si-analysis}e.
In Fig.~\ref{fig:si-analysis}, we see that the power of this prediction comes heavily from the interaction of diversity and redundancy (49.5\%). 

\subsection{The Diversity-Redundancy Boundary}
In this section we explore the diversity-redundancy boundary ($\varphi+r=1$). In the main text, we find that a large majority of unigraphical networks lie on or close to this boundary. Our aim is to understand the mechanism that leads to this effect.

First, we note that the concept of redundancy, which can be thought of as a measure of symmetry in the graph, is intimately related to type preserving automorphisms, which are type preserving isomorphisms (Def.~\ref{def:type_preserving_isomorphism}) from a graph to itself ($\sigma:V\rightarrow V$). Given a colored graph $G(V,E,f)$, we denote the set of type preserving automorphisms by $Aut_f(G)$. The automorphisms allow us to define the set of orbits - groups of nodes that play structurally equivalent roles. 

\begin{definition}
    Let $G(V,E,f)$ be a colored graph and let $Aut_f(G)$ be the set of automorphisms on it. Then, two vertices $u$ and $v$ are in the same orbit if $\exists\sigma\in Aut_f(G)$ s.t. $\sigma(u)=v$. 
\end{definition}
The redundancy of a colored graph is then formally defined as 
\begin{equation}
    r = 1-\frac{N_O-1}{N-1},
\end{equation}
where $N$ is the number of vertices in the graph and $N_O$ is the number of orbits~\cite{ball2018symmetric}. When $N_O=N$, each node is in its own orbit, and there are no structurally equivalent nodes. There is therefore no redundancy in the graph, and $r=0$. When $N_O=1$, all nodes are in the same orbit, and are thus all structurally identical. In this case, there is maximal redundancy and $r=1$. 

We now formalize the concept of equitable partitions, which will help link our work to the mathematical literature. 

\begin{definition}
    Let $G$ be a graph with $N$ vertices and let $\pi$ be a partition of $V(G)$ with cells $\theta_1,...,\theta_M$. We call $\pi$ equitable if for any ordered pair of cells $(\theta_i,\theta_j)$, the number of vertices in $\theta_i$ adjacent to a fixed vertex in $\theta_j$ only depends on $i$ and $j$. 
\end{definition}

As mentioned in Sec.~\ref{sec:fullyspecificUDT}, equitable partitions arise in fully specific systems where, for each pair of node types $s,t$, all nodes of type $s$ is connected to the same number of nodes of type $t$. We formalize this claim in the following Lemma:

\begin{lemma}
    Let $\mathcal{D}$ be a fully specific design set and let $G$ be a realization. The partition of $V(G)$ defined by the set of types related to $\mathcal{D}$ is equitable.
\end{lemma}

We now introduce the concept of a group of automorphisms. Say $Aut(G)$ is the set of automorphisms on the uncolored graph. Then, under the composition operation (which applies functions sequentially), this set defines a group $\Gamma$. Using the language of group theory is useful, as it allows us to talk about subsets of automorphisms that are closed under the group operation, i.e., where you can never compose two elements of the subset to create a group action that is not in the subset.

\begin{lemma}[Godsil 1997~\cite{godsil1997compact}]
    Let $G(V,E)$ be a graph and let $\Gamma'\leq \Gamma$ be any subgroup of the automorphisms $\Gamma$ on $G$. The partition of $G$ defined by the orbits of this group is equitable. 
\end{lemma}

It can be shown that for any coloring $f$ of the graph, the group $\Gamma_f$ related to $Aut_f(G)$ is a subgroup of $\Gamma$. The question is now if the coloring resulting from the equitable, orbit partition of $\Gamma_f$ coincides with the coloring $f$. In this case, $N_O=M$, i.e., the number of orbits is equal to the number of types, and so $r+\varphi=1$. Graphs where $N_O=M$ for every equitable partition is called \emph{Godsil}, as defined in the following definition.

\begin{definition}[Arvind 2017~\cite{arvind2017graph}]
    The class of graphs $G$ where any equitable partition must be generated by a group of automorphisms $\Gamma$ on $G$ is called Godsil. 
\end{definition}

If a graph is Godsil, then any equitable partition, so also the one defined by $f$, must be related to at a least one automorphism group $\Gamma'$. Because $\Gamma_f$ is the largest subgroup that preserves type, i.e., it maps between equivalent vertices in the partition, the orbits of $\Gamma_f$ coincide with the type partitions if the graph is Godsil. Then, for $\varphi+r=1$ to hold for all graphs, Godsil must be equal to class of all graphs. However, it was proven that this is not the case. 

\begin{lemma}[Godsil 1997~\cite{godsil1997compact}]
    Not all graphs are Godsil.
\end{lemma}

\begin{proof}
    For a given $n\geq 7$, it can be proven (Godsil 1997~\cite{godsil1997compact}) that one can define an equitable partition on the line graphs of the complete graph $L(K_n)$ such that the partition is not generated by any group of automorphisms on $\Gamma$.
\end{proof}

This means that not all fully-specific graphs lie on the diversity-redundancy boundary, even though fully specific graphs realize equitable partitions. The line graph $L(K_7)$ used in the above proof, for example, permits a design set that does not lie on this boundary. 

We can however, show that all fully specific graphs that unigraphically assemble must lie on the boundary, i.e., must be Godsil. The following lemma can be seen as a corollary of the UDT for fully specific systems, as it uses very similar tools for its proof.

\begin{lemma}
    Let $\mathcal{D}$ be a unigraphical, fully-specific design set realized by $G$. Then, the partition $\pi$ defined by the node types is generated by the colored automorphism group $\text{Aut}_f(G)$. 
\end{lemma}

\begin{proof}
    Take any pair of vertices $u,v\in V$ s.t. $f(u)=f(v)$. We now need to find a colored automorphism $\sigma:V\rightarrow V$ s.t. $\sigma(u)=v$ and $\sigma(v)=u$. 

    If $u=v$, $\sigma=e\in Aut_f(G)$, the identity. 

    If $u\neq v$ we construct the mapping $\sigma$ using similar steps as were used in the proof of the Unigraphical Design Theorem. Because $\mathcal{D}$ is unigraphical, we know that the $O$-graph $H$ is given by a forest of arborescences where the only self-loops are located on the roots. To prove that a certain $\sigma$ is a colored automorphism, we need to show that it preserves local structure, i.e., that it preserves neighbor relations. As explained in the proof of the Unigraphical Design Theorem, this implies we only need to look at the arborescence that contains type $f(v)$. Additionally, because preserving neighbor relations is equivalent to preserving non-neighbor relations, we can assume that for any pair of types $s,t$, where $N_s\leq N_t$, $O_{ts}=1$. Finally, assume that the root of the arborescence does not contain a self-loop (the proof where the self-loop is present is equivalent). This implies that $u$ and $v$ will live in a forest of isomorphic trees, like the ones shown in Fig.~\ref{fig:specificproof3}c. These trees are rooted by the vertices of with the type that correspond to the root node of the arborescence in the corresponding $O$-graph. We first endow these trees with the labeling on their vertices $V'$ defined through Eq.~\ref{eq:treelabeling}
    
    Let $w$ be the lowest common ancestor of vertices $u$ and $v$, i.e., the common ancestor that has the greatest depth with respect to the root of the tree. Let $w_u$ be a child of $w$ and an ancestor or $u$ and let $w_v$ be a child of $w$ and an ancestor of $v$. For each tree depth $y$ and node type $\tau$, we now define two sets of vertices $\mathcal{A}_{y\tau}=\{a\in \text{Desc}(w_u)\mid f(a)=\tau,|\Psi(a)|=y\}$ and $\mathcal{B}_{y\tau}=\{a\in \text{Desc}(w_v)\mid f(a)=\tau,|\Psi(a)|=y\}$. Because the two descendant sub-trees of $w_u$ and $w_v$ are isomorphic, these two sets have the same size, and we can thus define a perfect matching $\kappa_{y\tau}:\mathcal{A}_{y\tau}\rightarrow\mathcal{B}_{y\tau}$ between them. Because $u$ and $v$ have the same depth ($|\Psi(u)|=|\Psi(v)|$) and the same type ($f(u)=f(v)$), we can define this matching such that $\kappa_{|\Psi(u)|f(u)}(u)=v$ and $\kappa^{-1}_{|\Psi(u)|f(u)}(v)=u$. The colored automorphism $\sigma:V\rightarrow V$ is then given by
    \begin{equation}
        \sigma(a)=
        \begin{cases}
            a&\quad\text{if}\quad a\not\in\text{Desc}(w_u)\cup \text{Desc}(w_v)\\
            \kappa_{|\Psi(a)|f(a)}(a)&\quad\text{if}\quad a\in\text{Desc}(w_u)\\
            \kappa^{-1}_{|\Psi(a)|f(a)}(a)&\quad\text{if}\quad a\in \text{Desc}(w_v).
        \end{cases}
    \end{equation}
    In words, this function interchanges the labels of the two sub-trees corresponding to the vertices $w_v$ and $w_u$, making sure node type is preserved. By definition, this leads to $\sigma(u)=v$ and $\sigma(v)=u$. Because we are interchanging the entire sub-tree and because $w_u$ and $w_v$ have the same parent $w$, the local neighborhoods are preserved and $\sigma$ is an automorphism.
\end{proof}

The final question we ask is if lying on the diversity-redundancy boundary ensures unigraphical assembly. We prove that this is not the case, as stated in the following proposition.

\begin{proposition}
    Not all fully-specific design sets where $\varphi+r=1$ are unigraphical. 
\end{proposition}

\begin{proof}
    As a counterexample we take the Petersen graph. In Ref.~\cite{arvind2017graph} it is shown that this graph is Godsil, and so the related fully-specific design sets, defined by any equitable partition, must lie on the $\varphi+r=1$ boundary. Because the Petersen is 3-regular, we can, for example, take all vertices to be of the same type. In this case, $\varphi=0$ and $r=1$ (it is vertex-transitive). The biregular decomposition contains only the entire, 3-regular graph, which is not unigraphical, implying that the first condition of the UDT is violated, making the design set not unigraphical. 
\end{proof}

Arvind \textit{et al.}~\cite{arvind2017graph} do, however, identify a subclass of the Godsil graphs (they lie on the diversity-redundancy boundary) that will necessarily unigraphically assemble. These are the graphs that are \textit{amenable} to the color refinement algorithm, which finds the coarsest equitable partition of a graph. A graph is amenable to the algorithm if it can be used to test if the graph is isomorphic to another. Arvind \textit{et al.}~identify a way to test if a graph is amenable, by providing conditions on the found coarsest partition. These coincide exactly with the conditions of the fully specific Unigraphical Design Theorem (Sec.~\ref{sec:fullyspecificUDT}). This implies that all amenable graphs where the design set corresponds to the coarsest partition unigraphically assemble. 

We can summarize the results of this section as follows:

\begin{itemize}
    \item All unigraphical, fully specific design sets lie on the $\varphi+r=1$ boundary.
    \item Not all design sets that lie on the $\varphi+r=1$ are unigraphical.
    \item Not all fully-specific design sets lie on the $\varphi+r=1$ boundary. 
\end{itemize}

\clearpage
\section{Experiment}

In the previous sections, we described how local interactions rules can be leveraged to determine whether a system can be consistently be reproduced. In the following we address the inverse problem, where we want to design the building blocks and local interaction rules such that only one final state is possible. We do so by relying on construction sets, specifically Duplo and K'nex, which, as we discussed in the paper, do not encourage the design of unique outcomes, but are rather purposefully built to generate multiple structures and are therefore compatible with only a few trivial structures capable of unigraphical assembly. By increasing the diversity of these sets by adding custom 3D-printed building blocks, we are able to drastically increase the number of unigraphical sets. 

\subsection{Standard Construction Sets}

We first look at the standard building blocks used in the main text: Duplos and K'nex. We examine the design set of each construction set separately.

\subsubsection{Duplo}
We only consider $2\times 2$ and $4\times 2$ Duplo blocks.
For simplicity, we assume that no partial connection between blocks are possible. For example, two $2\times2$-blocks must connect through 4-pips, and two $4\times2$ blocks must connect through 8. 
Because different block shapes connect with different coverage (i.e.~a $4\times 2$ Duplo block can only place one $4\times 2$ on top of it while it can fit two $2\times 2$ blocks on top, we choose the capacity of a block to be the maximum number of connections it can make with any combination of blocks.

We additionally introduce two new blocks, a ground and ceiling block. A ground block can connect to four pips on the bottom of any Duplo and is flat on the bottom. A ceiling block can connect to four pips on the top of any Duplo and is flat on top. This combination of blocks gives the following capacity vector and binding matrix:
\begin{align}
O=\begin{bmatrix}2&2&1&1\\4&2&2&2\\1&1&0&1\\1&1&1&0\end{bmatrix}&&C=\begin{bmatrix}2\\4\\1\\1\end{bmatrix}
\end{align}
where the rows refer to $2\times 2$ blocks, $4\times 2$ blocks, floor blocks and ceiling blocks, respectively.

In general, many different structures can be built with these four block types by building vertically, horizontally, and turning $4\times 2$ blocks at different angles. Therefore, to determine a lower bound on the possible number of Duplo structures we consider the specific case where there are at most two ground and two ceiling blocks.
This restriction predetermines that structures can only be built vertically due to the blocks physical constraints and consequently allows us to combinatorially calculate the number of possible vertical structures. We call the number of blocks of dimensions $2\times 2$ and $4\times 2$ $N_2$ and $N_4$, respectively.

First, let us assume we have only one top and one bottom block. In this case, we can have no $4\times 2$ blocks, and the only structure we can make with $N$ blocks is a tower made of $N_2=N-2$ $2\times2$ blocks. This is then a unigraph. 

Next, we take two tops and two bottoms. In this case, we will generally start with a tower of size $2\times 2$, followed by a part of size $4\times 2$ ending in a tower of size $2 \times 2$. Let us call these different parts $B_1$, $B_2$ and $B_3$ respectively. Note that for the topology of the network it does not matter if the $B_1$ tower is on the same side as the $B_3$ tower or not. Blocks of size $4\times 2$ can only live in part $B_2$, whereas $2\times 2$ blocks can live anywhere. Let us say that there are $x$ $2\times 2$ blocks in $B_1$, $y$ in $B_2$ and $N_2-x-y$ in $B_3$. Note that $y$ must be even. Now, there is only one way to build towers $B_1$ and $B_3$. However, depending on the number of $4\times 2$ blocks, there are many ways to build tower $B_2$. Noting that the height of tower $B_2$ is $y/2+N_4$, the exact number can be calculated by finding the number of ways to distribute $N_4$ blocks over $y/2+N_4$ urns, which is given by
\begin{equation}
    \binom{y/2+N_4}{N_4}
\end{equation}
Of course, given a certain $N_2$, there are many ways to distribute the small blocks over the three towers. To account for this, we multiply the above value by the number of ways ($N_2-y+1$) to distribute the remaining $N_2-y$ blocks over towers $B_1$ and $B_3$ and sum over all possible values of $y$, noting that it must be even. We also use that given a certain $N$, $N_2\in[0,N-4]$. However, here we assume that there is at least one $4\times 2$ block, as the case where there are only small blocks needs to be treated differently. Taking this into account the number of constructions possible with $N$ blocks, where $N_4\geq1$, is given by
\begin{alignat}{6}
    &\sum_{N_2=0}^{N-5}\sum_{y'=0}^{\lfloor N_2/2\rfloor} (N_2-2y'+1)\binom{y'+N_4}{N_4}\sim 3\,{}_2F_1(-N/2,-N/2,-N,-4),
\end{alignat}
where $2F_1(-N/2,-N/2,-N,-4)$ is the hypergeometric function, which grows exponentially in $N$. Note that only one of these structures, the one where $N_2=0$, is a unigraph.

If there are only $2\times 2$ blocks, then we obtain 2 disconnected towers. There are $\lfloor(N-4)/2\rfloor+1$ different height combinations of these towers. This number grows polynomially, and can thus be neglected. 

To find the average number of networks per $N$, we just need to divide by the number of design sets. There are $N-2$ construction sets in total. This leads to the final average number of construction sets per $N$ 
\begin{equation}
    \langle |\mathcal{F}|\rangle \sim \frac{2}{N}{}_2F_1(-N/2,-N/2,-N,-4)
\end{equation}

\subsubsection{K'Nex}
We only consider K'Nex of two types: a wheel with 8 possible connections and a rod which can connect two wheels.
The wheel has 8 connections around the center of the wheel.
A wheel cannot connect to a wheel and a rod cannot connect to a rod. This is because wheels consist of notches and rods end in pegs.
Thus any network formed by K'Nex will be a bipartite network where one part has degree 2 and one part has degree 8.
We also introduce stopper pieces that have degree one. These pieces can be used to fill empty wheel or rod pieces which have not yet reached capacity.

A K'Nex design set $\mathcal{D}_{knex}$ has $O$ matrix and capacity vector
\begin{align}
    O&=\begin{bmatrix}0&8&0&8\\2&0&2&0\\0&1&0&1\\1&0&1&0\end{bmatrix}&\text{and}&&\mathbf{C}=\begin{bmatrix}8\\2\\1\\1\end{bmatrix}.
\end{align}

We estimate the average number of possible graphs given a fixed $N$ numerically. For this system, physical constraints are important, as our graph must (1) be planar and (2) rods have fixed lengths and fit into the wheels at certain angles. We simulate the K'Nex system as a growing network. We start with a representation where vertices are wheels and edges are links. We start with a single node at the origin $(0,0)$. We then place a second node at $(\sin\theta,\cos\theta)$ and attach it to the first. Here, $\theta$ is a randomly selected angle $\theta\in\{0,\pi/4,...,7/4\pi\}$. We then choose an existing node at random and place a new node at some angle relative to it. We continue this procedure iteratively. If a proposed node location leads to edges intersecting we reject it. If a proposed node location coincides with an already existing vertices, only an edge gets placed. We continue this procedure until a certain number of vertices have been placed, leading to final, connected, physical network. Note that this network only consists of rods and wheels, and that many wheels are not at full capacity. We therefore perform two post-processing steps. First, we take all degree one wheels. Say there are $n$ such degree one wheels. As we want to find all possible topologies, we generate $2^n$ possible final graphs by either filling a wheel's empty binding sites with degree one stopper rods or replacing it with a degree one cap. 

We iteratively generate such networks, checking if the topology of a candidate network is isomorphic to any previously generated graph. If so, the new topology is rejected. If not, we add the network to the set of possible topologies. We continue this process until saturation, i.e., until no new graphs have been added to the set during a sufficient number of attempts. Of course, our process generates networks of different sizes. We therefore sort the generated networks based on the total size $N=\sum_s N_s$ and count the number of topologically distinct graphs produced for each $N$. We obtain the average number of graphs by dividing this number by the amount of possible sets given a certain $N$. We once again observe in Fig.~5f in the main text that the trend of the number of graphs grows exponentially with $N$. It must be noted that symmetries provoked by the physical nature of the system imply this growth is not monotonous. 

Again, the number of unigraphs is $\mathcal{O}(1)$. We can calculate this number numerically by counting the number of possible graphs for a specific $\boldsymbol{N}$. The results are shown in Fig.~5m.

\subsection{Adapted Construction Sets}
We now ask how we can adapt our construction sets such that we produce unigraphical design sets. 

\subsubsection{Enumerating Unigraphs}\label{si:enumerating_unigraphs}
We first consider adding 4 adapters that connect the two standard sets. The first two connect the bottom of a $2\times 2$ Duplo block to either a K'Nex notch or peg. The second two connect the top of a $2\times2$ Duplo block to either a K'Nex notch or peg. The $4\times 2$ Duplo block will not be included.

These adapters raise the total diversity of our systems, and we thus expect to find a larger number of unigraphs. We again investigate this numerically.

To encode the physicality of the system we make our graphs directed. For the Duplo blocks, the tops correspond to the out degree and the bottoms to the in degrees. So, for example, the $2\times2$ block would have $k_{in}(2\times2 \text{ block})=1$ and $k_{out}(2\times2 \text{ block})=1$. For the K'Nex blocks, the number of notches corresponds to in-degree and the number of pegs correspond to the out-degree. Then, the wheels have $k_{in}(\text{wheel})=1$ and rods have $k_{out}(\text{rod})=2$. Going to a directed system, for example, allows us to demand that graphs cannot contain directed cycles, thus making cycles of Duplo blocks impossible. 

Next, we split our graph into a multiplex, where one layer corresponds to `Duplo edges' (pips) and the other to `K'Nex edges' (notches and pegs). Note that connectors have connections in both layers, whereas pure building blocks only make connections within a single layer. Within a layer, no information from the $O$-matrix is needed as all binding rules are already encoded in the directionality of the edges. For example, one no longer needs to know the $O$-matrix to know that two wheels cannot connect as both have $k_{out}=0$. This allows us to use standard graph-theoretical tools that are optimized for degree sequences, not binding rules. 

For a given number of vertices $N$, we first check all combinations of building blocks for graphicality, i.e., if the resulting set of building blocks can generate a network at all. Note that we do this for both of the layers of the multiplex. We also check if it is possible for the set to produce fully connected graphs, as we are only interested in fully connected unigraphs. 

For each of these graphical design sets, we exhaustively check all possible graphs for each of the layers of the multiplex through stub-matching. If more than two graphs are possible, the design set is discarded. If both layers are unigraphical, we combine them through the connectors as follows: Say our design set contains a single connector with a Duplo top and a wheel bottom. This means it has $k_{out}^{Duplo}=1$ and $k_{in}^{K'Nex}=1$. We can then select a node with in degree 1 in the K'Nex layer of the multiplex and a node with out degree 1 in the Duplo layer and identify them. Of course, if there are multiple of such vertices in each of the layers, different vertices can be identified. If this leads to different topologies in the combined graph, the design set is discarded as it is not unigraphical. If only one topology is possible, the design set is unigraphical. 

The advantage of this algorithm is that it is exhaustive - all possible unigraphs of a certain size can be found. However, the stub-matching on which the central step of the algorithm is based is slow. Additionally, for large $N$, the amount of possible component vectors grows fast. Therefore, this algorithm can only be used to identify relatively small unigraphs. 

\subsubsection{Physicality's Role in Unigraphical Assembly}

\begin{figure}[h]
    \centering
    \includegraphics[width=0.6\textwidth]{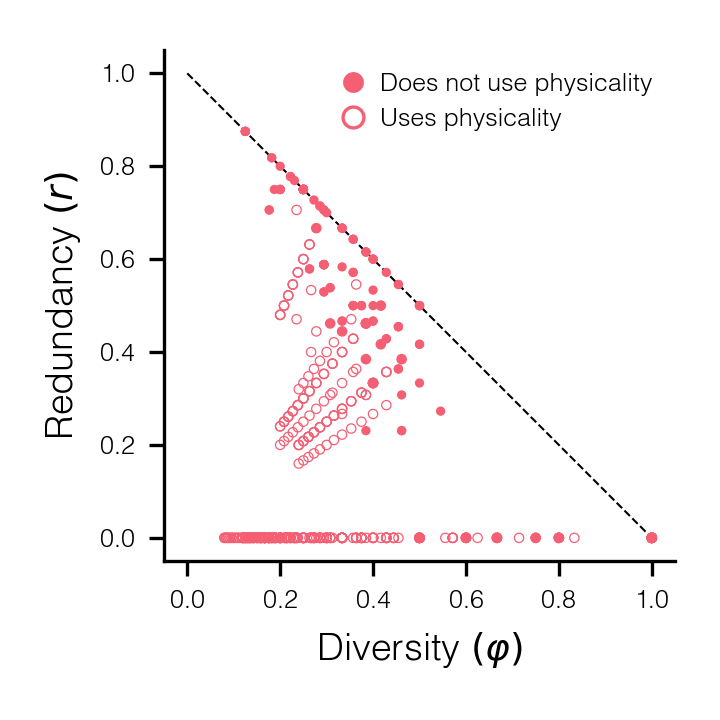}
    \caption{The diversity $\varphi$ and redundancy $r$ of all structures described in Sec.~\ref{si:enumerating_unigraphs}. Structures that require physical constraints to be unigraphical are represented by empty circles, whereas structures that can assemble regardless of physicality are given by full circles.  }
    \label{fig:roleofphysicality}
\end{figure}

For each structure, we measure the diversity and redundancy. In Fig.~\ref{fig:roleofphysicality} we see that many structures fall well below the diversity-redundancy line. 
This is because most of these structures are not fully specific, only achieving full specificity if the correct proportions of toys are combined.
This raises the question of how so many networks with low $\varphi+r$ achieve unigraphical assembly.

For each structure, we can determine whether physicality was used in order to constrain the space of possible networks. This is done by inputting the design set of particular construction into the MILP algorithm defined in Sec.~\ref{si:milp}.
If the MILP algorithm outputs one possible network realization, then physicality was not needed in order to induce unigraphical assembly.
In Fig.~\ref{fig:roleofphysicality}, we use full circles to indicate unigraphical structures that do not rely on physicality and empty circles to indicate structures which rely on physicality to unigraphically assemble.
Notably, networks that do not rely on physicality tend to fall near the $\varphi+r=1$ boundary as previously observed.
On the other hand, networks which rely on physicality to unigraphically assemble fall much farther from the diversity-redundancy boundary.
This provides preliminary evidence that invoking system-specific constraints such as physicality can allow design sets far from the diversity-redundancy boundary to unigraphically assemble.

\subsubsection{Designing Reproducible Structures.}

In order to create the bridge seen in Fig.~5l, we implement the additional constraint that pieces may only connect to other pieces of the same color.
For example, a blue Duplo can only connect to either another blue Duplo or a blue Duplo adapter.
It is important to note that K'Nex rods are not links in this network representation, rather they are vertices that can be colored.
Thus, a red K'Nex rod can only connect to a red K'Nex wheel or red adapter.
We additionally introduce new custom made adapters which connect a Duplo brick with three K'Nex rods.

Because high diversity is needed in order to make the bridge unigraphical, we introduce multiple coloring patterns onto single vertices.
For example, a single K'Nex rod will have two colors, one on each end.
In order for pieces to come together, colored patterns must match precisely.
If they only match partially, it is deemed an invalid connection.
The resultant $O$ matrix can be found in Fig.~5k and the capacity vector is trivially extracted. Because this graph is both fully diverse $(\varphi=1)$ and fully specific ($\psi=1)$, Cor.~\ref{cor:fullydiverse} predicts that it must be unigraphical.

While unigraphical assembly is possible due to the bridge's high diversity, it should be noted that the diversity can be significantly reduced and unigraphcal assembly is maintained due to the physical constraints enforced by each node's shape.
For example, if color was ignored there would still only be one possible way to connect all the pieces. This is because K'Nex rods must connect at right angles to Duplos, Duplos must stack directly on top of one another, and not all K'Nex rods are of the same length.
However, these physical constraints are not encoded into the design set and therefore their influence on the system's physicality cannot be verified using the UDT.

\clearpage
\bibliography{ref.bib}